\documentclass[11pt]{amsart}
\usepackage{latexsym,amsmath,amssymb,amscd, eucal,verbatim}
\usepackage[all]{xy}
\usepackage{graphicx}
\DeclareMathOperator{\K}{\sf K}
\DeclareMathOperator{\sT}{\sf T}

\DeclareMathOperator{\KK}{\sf KK}
\DeclareMathOperator{\Hom}{Hom}

\newcommand{\HE}{{\sf HL}}

\DeclareMathOperator{\HP}{\sf HP}
\DeclareMathOperator{\HA}{\sf HA}
\DeclareMathOperator{\End}{End}
\DeclareMathOperator{\Aut}{Aut}
\DeclareMathOperator{\Ext}{Ext}
\DeclareMathOperator{\Res}{Res}

\DeclareMathOperator{\ch}{ch}
\DeclareMathOperator{\Ind}{index}
\DeclareMathOperator{\KR}{\sf KR}

\newcommand{\bbr}{\mathbb{R}}
\newcommand{\bbR}{\mathbb{R}}
\newcommand{\bbc}{\mathbb{C}}
\newcommand{\bbC}{\mathbb{C}}
\newcommand{\bbS}{\mathbb{S}}
\newcommand{\cC}{\mathcal{C}}

\newcommand{\cD}{\mathcal{D}}
\newcommand{\xr}{\xrightarrow}
\newcommand{\call}{\mathcal{L}}
\newcommand{\cale}{\mathcal{E}}
\newcommand{\fk}{\mathfrak{K}}
\newcommand{\fC}{\mathfrak{C}}
\newcommand{\fB}{\mathfrak{B}}

\newcommand{\newsection}{\setcounter{equation}{0}\section}

\newcommand{\eq}{\begin{equation}}
\newcommand{\eqend}{\end{equation}}

\newbox\ncintdbox \newbox\ncinttbox
\setbox0=\hbox{$-$} \setbox2=\hbox{$\displaystyle\int$}
\setbox\ncintdbox=\hbox{\rlap{\hbox to
\wd2{\hskip-.125em\box2\relax \hfil}}\box0\kern.1em}
\setbox0=\hbox{$\vcenter{\hrule width 4pt}$}
\setbox2=\hbox{$\textstyle\int$}
\setbox\ncinttbox=\hbox{\rlap{\hbox to
\wd2{\hskip-.175em\box2\relax \hfil}}\box0\kern.1em}
\newcommand{\ncint}{\mathop{\mathchoice{\copy\ncintdbox}%
{\copy\ncinttbox}{\copy\ncinttbox}{\copy\ncinttbox}}\nolimits}

\def\Dirac{{D\!\!\!\!/\,}} 

\newcommand{\cR}{\mathcal R}
\newcommand{\twga}{C_r^*(\Gamma_g,\sigma_\theta)}

\newcommand{\Aosc}{{\sf A}}

\newcommand{\complex}{{\mathbb C}} 
\newcommand{\zed}{{\mathbb Z}} 
\newcommand{\bbZ}{{\mathbb Z}}
\newcommand{\bbT}{{\mathbb T}}
\newcommand{\nat}{{\mathbb N}} 
\newcommand{\real}{{\mathbb R}} 
\newcommand{\rat}{{\mathbb Q}} 
\newcommand{\mat}{{\mathbb M}} 
\newcommand{\id}{{1\!\!1}} 
\def\alg{{\mathcal A}}
\def\balg{{\mathcal B}}
\def\calg{{\mathcal C}}
\def\dalg{{\mathcal D}}
\def\ealg{{\mathcal E}}
\def\hil{{\mathcal H}}
\def\bun{{\mathcal E}}
\def\cE{{\mathcal E}}
\def\cK{{\mathcal K}}

\def\K{{\sf K}}
\def\KO{{\sf KO}}
\def\H{{\sf H}}
\def\P{{\sf P}}

\def\E{{\sf E}}
\def\b{{\rm b}}
\def\B{{\rm B}}

\def\W{{\sf W}}
\def\HP{{\H\P}}

\def\op{{\rm o}}
\def\Id{{\rm id}}
\def\Cl{{\sf Cliff}}
\def\ch{{\sf ch}}
\def\Ch{{\sf Ch}}
\def\PD{{\sf Pd}}

\def\Aroof{{\widehat{A}}}
\def\Todd{{\sf Todd}}
\def\frh{{\mathfrak{h}}}
\def\frf{{\mathfrak{f}}}
\newcommand{\bbz}{{\mathbb Z}}

\def\nn{\nonumber}

\newcommand{\Tr}[1]{\:{\rm Tr}\,#1}
\def\e{{\,\rm e}\,}

\def\be{\begin{equation}}
\def\ee{\end{equation}}
\def\bea{\begin{eqnarray}}
\def\eea{\end{eqnarray}}
\def\bd{\begin{displaymath}}
\def\ed{\end{displaymath}}

\def\dd{{\rm d}}

\def\ii{{\,{\rm i}\,}}

\makeatletter
\newdimen\normalarrayskip              
\newdimen\minarrayskip                 
\normalarrayskip\baselineskip
\minarrayskip\jot
\newif\ifold             \oldtrue            
\def\arraymode{\ifold\relax\else\displaystyle\fi} 
\def\@arrayskip{\ifold\baselineskip\z@\lineskip\z@
     \else
     \baselineskip\minarrayskip\lineskip2\minarrayskip\fi}
\def\@arrayclassz{\ifcase \@lastchclass \@acolampacol \or
\@ampacol \or \or \or \@addamp \or
   \@acolampacol \or \@firstampfalse \@acol \fi
\edef\@preamble{\@preamble
  \ifcase \@chnum
     \hfil$\relax\arraymode\@sharp$\hfil
     \or $\relax\arraymode\@sharp$\hfil
     \or \hfil$\relax\arraymode\@sharp$\fi}}
\def\@array[#1]#2{\setbox\@arstrutbox=\hbox{\vrule
     height\arraystretch \ht\strutbox
     depth\arraystretch \dp\strutbox
     width\z@}\@mkpream{#2}\edef\@preamble{\halign \noexpand\@halignto
\bgroup \tabskip\z@ \@arstrut \@preamble \tabskip\z@ \cr}%
\let\@startpbox\@@startpbox \let\@endpbox\@@endpbox
  \if #1t\vtop \else \if#1b\vbox \else \vcenter \fi\fi
  \bgroup \let\par\relax
  \let\@sharp##\let\protect\relax
  \@arrayskip\@preamble}
\makeatother

\newcommand{\beq}{\begin{eqnarray}}
\newcommand{\eeq}{\end{eqnarray}}

\newcommand{\G}{\Gamma}

\def\appendix#1{\addtocounter{section}{1}\setcounter{equation}{0}
\renewcommand{\thesection}{\Alph{section}}
\section*{Appendix \thesection. #1}
\protect\indent \parbox[t]{11.715cm}
}

\def\ii{{\,{\rm i}\,}}

\newtheorem{theorem}{Theorem}[section]
\newtheorem{lemma}[theorem]{Lemma}

\newtheorem{corollary}[theorem]{Corollary}
\newtheorem{proposition}[theorem]{Proposition}
\theoremstyle{definition}
\newtheorem{definition}[theorem]{Definition}
\theoremstyle{remark}
\newtheorem{example}[theorem]{Example}
\newtheorem{remark}[theorem]{Remark}

\numberwithin{equation}{section}

\begin{document}

\title[D-Branes on Noncommutative Manifolds]{D-Branes, RR-Fields and
  duality \\ on noncommutative manifolds}

\author{Jacek Brodzki}

\address{Faculty of Mathematical Studies, University of Southampton,
Southampton SO17 1BJ, UK}

\email{j.brodzki@soton.ac.uk}

\author{Varghese Mathai}

\address{Department of Pure Mathematics, University of Adelaide,
Adelaide 5005, Australia}

\email{vmathai@maths.adelaide.edu.au }

\author{Jonathan Rosenberg}

\address{Department of Mathematics, University of Maryland,
College Park, MD 20742, USA}

\email{jmr@math.umd.edu}

\author{Richard J. Szabo}

\address{Department of Mathematics and Maxwell Institute for
  Mathematical Sciences, Heriot-Watt University, Riccarton, Edinburgh
  EH14 4AS, UK}

\email{R.J.Szabo@ma.hw.ac.uk}

\begin{abstract}
We develop some of the ingredients needed for
string theory on noncommutative spacetimes,
proposing an axiomatic formulation of T-duality as well as
establishing a very general formula for D-brane charges.
This formula is closely related to a noncommutative
Grothendieck-Riemann-Roch theorem that is proved here.
Our approach relies on a very general form of Poincar\'e duality, which is
studied here in detail.  Among the technical tools employed are
calculations with iterated
products in bivariant $\K$-theory and cyclic theory, which are
simplified using a novel
diagram calculus reminiscent of Feynman diagrams.
\end{abstract}
 

\maketitle

{\baselineskip=11pt
\tableofcontents
}

\newsection*{Introduction}

As  proposed by  \cite{MM} and elaborated in 
 \cite{W,FW,Horava,OS99,MW},  D-brane charges and RR-fields in
string theory are 
classified by the $\K$-theory of spacetime $X$, or
equivalently by the $\K$-theory of the $C^*$-algebra $C_0(X)$ of
continuous functions on $X$ vanishing at infinity.
Recently, in a far-sighted suggestion at KITP, I.M.~Singer
suggested working out string theory and duality on spacetimes that are
general noncommutative $C^*$-algebras, with some minimal
assumptions. This paper can be viewed as a preliminary step towards
this goal.  Some of our main results are a formula for the charges of
D-branes in noncommutative spacetime and
a fairly complete treatment of a general framework for T-duality.
The main technical tools are a study of
Poincar\'e duality in both $\KK$-theory
and bivariant cyclic theories, a definition of Gysin (``wrong-way'') maps,
and a version of the Grothendieck-Riemann-Roch theorem.

Previous work ( \cite{BEM,MR}, among many other references) 
already showed that a good formulation of T-duality requires the use of 
noncommutative algebras. We develop a formalism for dealing with
T-duality in the context of 
general separable $C^*$-algebras and in Section \ref{sect:T-duality}
we give an axiomatic definition. 
This includes the requirement that the RR-fields and D-brane charges of $\alg$
should be in bijective correspondence with the RR-fields and D-brane charges 
of the T-dual ${\sT}(\alg)$, and that T-duality applied twice yields a
$C^*$-algebra  
which is physically equivalent to the $C^*$-algebra that we started out with. 
This general T-duality formalism can be viewed as a noncommutative
version of the [topological aspects of the] Fourier-Mukai transform. 

In the classical case, 
D-brane charges are expressed in terms of the Chern character in
$\K$-homology (see  \cite{ReisSz}), as formulated topologically 
by the Baum-Douglas construction of  \cite{BD}. 
In formulating the notions of D-brane charges and RR-fields on arbitrary
$C^*$-algebras, one is faced with the problem of developing Poincar\'e
duality and constructing characteristic classes in this general
setting. In  \cite{Connes:CMP}, Connes initiated this study, pointing out 
that the analogue of a spin$^c$
structure for a $C^*$-algebra $\alg$ is a fundamental class $\Delta$
for its $\K$-theory, whereas the analogue of a spin 
structure is a fundamental class for its $\KO$-theory. In  \cite{Moscovici97}, 
Moscovici gives an elegant application of Poincar\'e duality, deriving an 
analogue of the Vafa-Witten inequalities for spectral triples that implement
Poincar\'e duality, under a finite topological type hypothesis.
One of the goals of this paper is to define
the Todd class and Todd genus for a spin$^c$ $C^*$-algebra $\alg$, which 
generalize the notion of the classical Todd class and Todd genus
of a compact spin$^c$ manifold $X$.
 If $\Delta$ is a fundamental class for the $\K$-theory
of the spin$^c$ $C^*$-algebra $\alg$ and $\Xi$ is a fundamental class in 
bivariant cyclic homology of  $\alg$
(which is the analogue of an orientation for a smooth manifold),
then we define in Section \ref{sect:Todd}, the Todd class of $\alg$ to
be the invertible element 
\[
{\sf Todd}(\alg) =  \Xi^\vee\otimes_{\alg^\op}
{\sf ch}(\Delta)\,.
\]
in bivariant cyclic homology of the algebra $\alg$. (The notations are
explained below; $\Xi^\vee$ is the dual fundamental class to $\Xi$
and $\alg^\op$ is the opposite algebra to $\alg$.)
In the special case when $\alg$ is a spin $C^*$-algebra and 
the $\K$-theory fundamental class $\Delta$ comes from
a fundamental class in  $\KO$-theory, the characteristic class as
defined above is called the Atiyah-Hirzebruch class $\widehat
A(\alg)$. One of our main results, Theorem \ref{thm:GRR-strong}, shows that
the Todd class as defined above is exactly the correction factor
needed in the noncommutative Grothendieck-Riemann-Roch formula.  

Our final main result, the D-brane charge formula
of Section \ref{GenConstr}, is a noncommutative
analogue of the well-known formula (1.1) in  \cite{MM} (cf.\
 \cite{W,Horava,OS99,MW}). It takes the
familiar form,
\[
{\sf Q}_\xi= \ch(f_!(\xi))  \otimes_\alg \sqrt{{\sf Todd}(\alg) }\,,
\]
for a D-brane $\balg$ in a noncommutative spacetime $\alg$
with given weakly $\K$-oriented morphism $f\colon \alg \to \balg$ and
Chan-Paton bundle  
$\xi \in \K_\bullet(\balg)$, where $f_!$ denotes the Gysin map
associated to $f$. 
With this modification of the Chern character, 
one obtains an isometry between the natural intersection pairings in
$\K$-theory and cyclic theory of $\alg$. There is also a similar dual
formula for the charge of a D-brane given by a Fredholm module,
representing the Chern-Simons coupling of D-branes with RR-fields.

The central mathematical technique of the paper is the development of a novel 
diagram calculus for $\KK$ theory and the analogous diagram calculus  for 
bivariant cyclic theory, in Appendix B. 
The rules of this diagram calculus are reminiscent of those for the calculus 
of Feynman diagrams, and are likely to become an important tool for simplifying
iterated sequences of intersection products in $\KK$-theory and in
cyclic theory, and for establishing identities in these theories.

\subsection*{Acknowledgments}

We would like to thank P. Bouwknegt, K. Hannabuss, 
J.~Kaminker, R.~Plymen, R.~Reis and I.M.~Singer for helpful
discussions. J.B. and R.J.S.\
were supported in part by the London Mathematical Society. 
V.M.\ was supported by the Australian Research Council.
J.R.\ was supported in part by the USA National Science Foundation,
grant number DMS-0504212. R.J.S.\ was supported in part by PPARC Grant
PPA/G/S/2002/00478 and by the EU-RTN Network Grant
MRTN-CT-2004-005104. J.B., V.M., and J.R.\ all thank the
Erwin Schr\"odinger International Institute for Mathematical Physics
for its hospitality under the auspices of the programme in 
Gerbes, Groupoids, and Quantum Field Theory, which made part of this
collaboration possible.

\newsection{D-Branes and Ramond-Ramond Charges\label{ChargeDef}}

In this section we give a detailed mathematical description of brane
charges in the language of topological $\K$-homology and singular
cohomology. Our aim later on is then to generalize these constructions
to analytic $\K$-homology and cyclic cohomology suitable to generic
noncommutative settings, some examples of which we describe below. For
a description of D-branes in terms of $\K$-theory
see \cite{MM,MW,OS99,W}, and in terms of $\K$-homology
see \cite{AST,ReisSz,Szabo1}.

\subsection{Flat D-Branes\label{FlatD}}

Let $X$ be a spin manifold of dimension $d=10$ with
metric. In Type~II superstring theory, $X$ is called the {\it
  spacetime}. If $X$ is non-compact, appropriate compact support
conditions are always implicitly understood throughout. In our later
applications we can typically relax some of these requirements and
only assume that $X$ is a finite-dimensional Hausdorff space which has
the homotopy type of a finite CW-complex.

\begin{definition}
  A {\it flat D-brane} in $X$ is a triple $(W,E,\phi)$,
  where $\phi:W\hookrightarrow X$ is a closed, embedded spin$^c$
  submanifold and $E\in\K^0(W)$. The submanifold $W$ is called the
  {\it worldvolume} and the class $E$ the {\it Chan-Paton bundle} of the
  D-brane.
\label{defflatD}\end{definition}

When $E$ is the stable isomorphism class of a complex vector bundle
over $W$, we assume that it is equipped with a connection and
refer to the triple as a {\it brane system}. When $E$ is only a
virtual bundle, say $E=E^+-E^-$, we can loosely regard it as the class
of a complex $\zed_2$-graded bundle $E^+\oplus E^-$ equipped with a
superconnection and the triple is called a {\it
  brane-antibrane system}. The requirement that a D-brane $(W,E,\phi)$
be invariant under processes involving brane-antibrane creation and
annihilation is the statement of stable isomorphism of Chan-Paton
bundles. Physical quantities which are invariant under deformations of
$E$ thereby depend only on its $\K$-theory class in $\K^0(W)$  \cite{W}.
Deformation invariance, gauge symmetry enhancement and the possibility
of branes within branes then imply that any D-brane $(W,E,\phi)$
should be subjected to the usual equivalence relations of topological
$\K$-homology \cite{BD}, i.e., bordism, direct sum and vector bundle
modification, respectively  \cite{ReisSz}.
We will not distinguish between a D-brane and
its $\K$-homology class in $\K_\bullet(X)$, nor between the
Chan-Paton bundle and its isomorphism class in $\K^0(W)$.

To define the charge of a flat D-brane in the
spacetime manifold $X$, we begin by introducing a natural
bilinear pairing on the $\K$-theory of $X$,
\beq
\langle-,-\rangle\,:\,\K^0(X)\times\K^0(X)~\stackrel{\otimes}
{\longrightarrow}~\K^0(X)~\xr{{\rm index}(\Dirac_{(-)})}~\zed \ ,
\label{Kthpairing}\eeq
where $\Dirac_N:C^\infty(X,{\sf S}_X^+\otimes N)\to C^\infty(X,{\sf S}_X^-\otimes
N)$ is the twisted Dirac operator on $X$, with respect to a chosen
connection on the complex vector bundle $N\to X$, and ${\sf
  S}_X^\pm\to X$ are the two half-spinor
bundles over $X$. When tensored over $\rat$ the pairing (\ref{Kthpairing})
is nondegenerate, which is equivalent to Poincar\'e duality in
rational $\K$-theory. In the topological setting, Poincar\'e duality
is generically determined by the bilinear cap product
\beq
\K^0(X)\otimes\K_\bullet(X)~\stackrel{\cap}{\longrightarrow}~
\K_\bullet(X)
\label{captop}\eeq
defined for any complex vector bundle $F\to X$ and any D-brane
$(W,E,\phi)$ in $X$ by
\beq
F~\cap~(W,E,\phi)~=~(W,E\otimes\phi^*F,\phi) \ .
\label{FcapDbrane}\eeq
The index pairing $\K^0(X)\otimes\K_\bullet(X)\to\zed$ is then
provided by the Dirac operator on $W$ as
\beq
F\otimes(W,E,\phi)~\longmapsto~{\rm index}(\Dirac_{E\otimes\phi^*F}) \
{}.
\label{K0indpair}\eeq

On the other hand, in cohomology the natural
bilinear pairing is given by the intersection form
\beq
(-,-)\,:\,\H^p(X,\zed)\times\H^{d-p}(X,\zed)~\stackrel{\cup}
{\longrightarrow}~\H^d(X,\zed)~\xr{(-)[X]}~\zed \ .
\label{cohpairing}\eeq
Again nondegeneracy of this pairing over $\rat$ is equivalent to the
Poincar\'{e} duality theorem of classical rational cohomology
theory \cite[p. 44]{BT}. For a compact oriented manifold $X$, the
pairing between cohomology groups of complementary degrees leads to
the duality
\beq
\H^p(X,\rat) \cong \big(\H^{d-p}(X,\rat)\big)^\vee
\cong\H_{d-p}(X,\rat) \ .
\eeq
It is important to realize that this pairing is determined purely in
terms of the topology of $X$, while the index pairing between
$\K$-theory and $\K$-homology uses both the knowledge of the topology
of $X$ and the analysis of the Dirac operator $\Dirac$. This
difference will become important when we compare the two pairings
using the Chern character below. The statement of cohomological
Poincar\'{e} duality in the non-oriented case requires the use of
twisted coefficients, while in $\K$-theory the Poincar\'e pairing
involves twisting whenever $X$ is not spin$^c$. This links very
importantly with twisted $\K$-theory \cite{AS}.

Recall that the Atiyah-Hirzebruch class $\widehat{A}(X)$ of the
manifold $X$ is invertible with respect to the cup product on
cohomology \cite[p.~257]{ML}. An application of the Atiyah-Singer
index theorem (recalling that $X$ is spin) then immediately gives the
following fundamental result.

\begin{proposition}\label{modifiedChern}
The modified Chern isomorphism
\beq
{\sf Ch}\,:\,\K^0(X)\otimes\rat~\longrightarrow~\H^{\rm
  even}(X,\rat)=\bigoplus_{n\geq0}\,\H^{2n}(X,\rat)
\label{calQiso}\eeq
defined by
\beq
{\sf Ch}(N)={\sf ch}(N)\cup\widehat{A}(X)^{1/2}
\label{chernmod}\eeq
is an isometry with respect to the natural inner products
\eqref{Kthpairing} and \eqref{cohpairing},
\beq
\big\langle N\,,\,N'\,\big\rangle=\big({\sf Ch}(N)\,,\,
{\sf Ch}(N'\,)\big) \ .
\label{Chisom}\eeq
\label{propmodCh}\end{proposition}
\noindent
Note that the ordinary Chern character $\ch$ preserves the addition
and multiplication on $\K$-theory and cohomology, but not the bilinear
forms. The modified Chern character $\Ch$ preserves addition but not
the cup products. A similar statement is also true for the Chern
character on $\K^{-1}(X)\otimes\rat\to\H^{\rm
  odd}(X,\rat)=\bigoplus_{n\geq0}\,\H^{2n+1}(X,\rat)$. However,
because of the suspension isomorphism
$\K^{-1}(X)\cong\K^0(X\times\real)$ it will suffice to work explicitly
with $\K^0$ groups alone in the following. In string theory terms this
means that we work only with Type~IIB D-branes, the analogous results
for Type~IIA branes being obtainable by T-duality (See
Section~\ref{sect:T-duality}).

There is an elementary but useful alternative interpretation of
Proposition~\ref{modifiedChern}. Since the $\widehat{A}$-class is an
even degree (inhomogeneous) class in the cohomology ring
$\H^{\text{even}}(X$, $\rat)$, with non-zero constant term,
its square root $\widehat{A}(X)^{1/2}$ is
also invertible. It follows that taking products with this class
produces an isomorphism $\frh: \H^{\text{even}}(X,\rat) \rightarrow
\H^{\text{even}}(X,\rat)$ which is given explicitly by 
\beq
\omega ~\longmapsto~ \omega \cup \widehat{A}(X)^{1/2} \ . 
\label{hisosqrt}\eeq
When we combine this isomorphism with the pairing given by
Poincar\'{e} duality, we obtain a new nondegenerate pairing
$(-,-)_\frh:\H^{\rm even}(X,\rat)\times\H^{\rm even}(X,\rat)\to\real$
defined by
\bea
\big(\alpha\,,\, \alpha'\,\big)_\frh &  :=& 
\big(\alpha \cup \widehat{A}(X)^{1/2}\,,\, \alpha'
\cup\widehat{A}(X)^{1/2}\big)\nonumber\\
& =& \big(\alpha \cup \widehat{A}(X)^{1/2}\cup \alpha' 
\cup\widehat{A}(X)^{1/2}\big)[X] \nonumber\\
& = &\big(\alpha \cup  \alpha' \cup\widehat{A}(X)\big)[X]
\label{hpairing}\eea
where we have used commutativity of the cup product. It is now easy to
see that the classical Chern character is an isometry with respect to
the two pairings (\ref{Kthpairing}) and (\ref{hpairing}),
\beq
\big\langle N\,,\, N'\, \big\rangle =\big( \ch (N)\,,\,
\ch(N'\,)\big)_\frh \ . 
\label{classisoch}\eeq
{}From this point of view the isomorphism $\frh$ transforms the purely
topological pairing $(-,-)$ to the ``index'' pairing $(-,-)_\frh$, where
the latter contains the information about the  extra piece of index
machinery given by the Atiyah-Hirzebruch class. 

For any closed oriented embedding $\phi:W\hookrightarrow X$ of
dimension $p$, we denote by $[W]$ its orientation cycle in
$\H_p(X,\zed)$, by ${\sf Pd}_X(W)={\sf Pd}_{W\hookrightarrow
  X}=([X]\cap\,\underline{\phantom{X}}\,)^{-1}[W]$ its Poincar\'e dual
in $\H^{d-p}(X,\zed)$, and by
$\phi_!:\K^\bullet(W)\to\K^{\bullet+d-p}(X)$ the corresponding
$\K$-theoretic Gysin homomorphism. Recall that on cohomology, the
Gysin map is given explicitly by $\phi_!={\sf
  Pd}^{\phantom{1}}_X\circ\phi_*\circ{\sf
  Pd}_W^{-1}:\H^\bullet(W,\zed)\to\H^{\bullet+d-p}(X,\zed)$.

\begin{definition}
The {\it Ramond-Ramond charge} ({\it RR-charge} for short) of a D-brane
  $(W,E$, $\phi)$ in $X$ is the modified Chern characteristic class ${\sf
  Ch}(\phi_!E)\in\H^\bullet(X,\rat)$. If $(W',E',\phi'\,)$ is any
  other D-brane in $X$, then the {\it $(W',E',\phi'\,)$-charge} of
  $(W,E,\phi)$ is the integer
\beq
Q_{W',E',\phi'}(W,E,\phi)=\big(\PD_X(W'\,)\,,\,\Ch(\phi_!E)\big)
=\phi^{\prime\,*}\,\Ch(\phi_!E)[W'\,] \ .
\label{RRchargeint}\eeq
\label{defRRcharge}\end{definition}
\noindent
When $(W',E',\phi'\,)=(W,E,\phi)$, we write simply
$Q_{W,E,\phi}=Q_{W,E,\phi}(W,E,\phi)$ for the charge of the D-brane
$(W,E,\phi)$ itself. Note that this charge formula for a D-brane is
written entirely in terms of spacetime quantities.

Let us momentarily assume, for simplicity, that the spacetime manifold
$X$ is compact. Let $C(X)$ denote the $C^*$-algebra of continuous
complex-valued functions on $X$. A standard construction in
$\K$-homology then provides the following result.

\begin{proposition}
  There is a one-to-one correspondence between flat D-branes in $X$,
  modulo Baum-Douglas equivalence,
  and stable homotopy classes of Fredholm modules over the algebra
  $C(X)$.
\label{propflatDFred}\end{proposition}
\begin{proof}
Consider a D-brane $(W,E,\phi)$ such that
$\dim(W)$ is odd. The worldvolume $W$ inherits a metric from $X$ and
its Chan-Paton bundle $E$ is equipped with a (super)connection $\nabla$. Let
${\sf S}_W\to W$ be the spinor bundle over $W$, and
consider the usual twisted Dirac operator
$\Dirac_E:C^\infty(W,{\sf S}_W\otimes E)\to C^\infty(W,{\sf S}_W\otimes E)$ with
respect to the chosen connection $\nabla$. Using the metric we can
complete the vector space of smooth sections $C^\infty(W,{\sf S}_W\otimes E)$
of the twisted spinor bundle and view $\Dirac_E:{\mathcal H}\to{\mathcal
  H}$ as an unbounded self-adjoint Fredholm operator on the separable
Hilbert space
\beq
{\mathcal H}=L^2(W,{\sf S}_W\otimes E) \ .
\label{L2WSE}\eeq
Let us now define a unital algebra $*$-homomorphism $\rho:C(X)\to{\mathcal
B}({\mathcal
  H})$ as pointwise multiplication on $\mathcal H$ via
\beq
\rho(f)=m_{f\circ\phi}^{~}\otimes\id^{~}_{{\sf S}_W\otimes E} \ , 
\quad \forall\,f\in C(X) \ ,
\label{Phif}\eeq
where $m^{~}_g:C(W)\to C(W)$ is the pointwise multiplication operator
$h\mapsto g\,h$. Since the Dirac operator is closable, we can thereby form
an odd Fredholm module $({\mathcal H},\rho,F)$ over the algebra $C(X)$,
where $F=\Dirac_E/|\Dirac_E|$ is the partial isometry in the
polar decomposition of $\Dirac_E$. Similarly, when $W$ is
even-dimensional, we can form an even Fredholm module $({\mathcal
  H},\rho,F)$, where the $\zed_2$-grading $\hil=\hil^+\oplus\hil^-$ on
the Hilbert space (\ref{L2WSE}) is given by
\beq
{\mathcal H}^\pm=L^2(W,{\sf S}_W^\pm\otimes E)
\label{L2WSEpm}\eeq
with ${\sf S}_W^\pm\to W$ the two half-spinor bundles over $W$, the
odd bounded Fredholm operator $F$
constructed as above from the corresponding twisted Dirac
operator $\Dirac_E:C^\infty(W,{\sf S}_W^+\otimes E)\to C^\infty(W,{\sf
  S}_W^-\otimes E)$, and the even $*$-homomorphism
$\rho:C(X)\to{\mathcal B}({\mathcal H}^\pm)$ defined as in
(\ref{Phif}). The clases of the
Fredholm modules built in this way are independent
of the choice of metric on $X$ and connection $\nabla$ on $E$.

Conversely, allow arbitrary coefficient classes in $\K$-theory (this
requires certain care with the defining equivalence
relations \cite{Jakob,ReisSz}). Then the $\K$-homology class of the
cycle $(X,E,\Id_X)$ is the Poincar\'e dual of $E$, which can be any class in
$\K_\bullet(X)$. We conclude that all classes in the $\K$-homology
$\K^\bullet({\mathcal A})=\K\K_\bullet(\alg,\complex)$ of the algebra
${\mathcal A}=C(X)$ can be obtained by using an appropriate D-brane.
\end{proof}

Proposition~\ref{propflatDFred} of course simply establishes the
equivalence between the analytic and topological descriptions of
$\K$-homology. Any Fredholm module over the $C^*$-algebra $C(X)$ is
therefore a flat D-brane in the spacetime $X$. The usefulness of this
point of view is that it can be extended to more general brane
configurations (that we describe in the following) which are
represented by noncommutative algebras. Namely, a D-brane may be
generically regarded as the homotopy class of a suitable Fredholm
module over an algebra $\alg$. In what follows we will reformulate the
description of D-brane charge in the language of cyclic cocycles. This
will require, in particular, an analytic reformulation of the natural
pairings introduced above. More precisely, one of our main goals in
this paper is to provide a generic, noncommutative version of the
result (\ref{classisoch}).

\subsection{Ramond-Ramond Fields\label{RRFields}}

Closely related to the definition of D-brane charge given above is the
notion of a Ramond-Ramond field. In what follows
we use the cup product $\cup$ when multiplying
together cohomology classes, and exterior products $\wedge$ when
multiplying arbitrary differential forms. In a similar vein to what we
have done before, we will not distinguish between (co)homology classes
and their explicit representatives.

Let ${\rm Fred}={\rm Fred}(\hil)$ be the space of Fredholm
operators on a separable Hilbert space $\hil$. Then ${\rm Fred}$ is
a classifying space for $\K$-theory of $X$ and any vector bundle $E\to
X$ can be obtained as the index bundle of a map into ${\rm Fred}$. Let
${\sf c}$ be a choice of cocycle representative for the universal
Chern character. If $f_E:X\to{\rm Fred}$ is the classifying map
of a bundle $E\to X$, then $\ch(E)=f_E^*\,{\sf c}\in\H^{\rm
  even}(X,\real)$.

Consider triples $(f,C,\omega)$, where $f\in[X,{\rm Fred}]$,
$\omega$ is an inhomogeneous form of even degree, and $C$ is an
inhomogeneous cochain of odd degree satisfying
\beq
\dd C=\omega-f^*{\sf c} \ .
\label{catconddef}\eeq
The collection of all such triples is denoted
$\fk^0(X)$. Two elements $(f_0,C_0,\omega_0)$ and $(f_1,C_1,\omega_1)$
of $\fk^0(X)$ are called {\it equivalent} if there is a triple
$(f,C,\omega)$ on $X\times[0,1]$, with $\omega$ constant on
$\{x\}\times[0,1]$ for each $x\in X$, such that
$(f,C,\omega)|_{X\times\{0\}}=(f_0,C_0,\omega_0)$ and
$(f,C,\omega)|_{X\times\{1\}}=(f_1,C_1,\omega_1)$. The set of
equivalence classes forms an abelian group under addition of triples
called the {\it differential $\K$-theory group}
$\breve{\K}{}^0(X)$, cf.\ \S4 in  \cite{HopSing}. It fits into the short exact sequence
\beq
0~\longrightarrow~\K^{-1}(X)\otimes\real/\zed~\longrightarrow~
\breve{\K}{}^0(X)~\longrightarrow~\Aosc^0(X)~\longrightarrow~0 \ ,
\label{diffexactseq}\eeq
where $\Aosc^0(X)$ is defined by the pullback square
\beq
\xymatrix{
\Aosc^0(X)\ar[d] \ar[r]&
\Omega_{\rm cl}^{\rm even}(X)=\bigoplus\limits_{n\geq0}\,
\Omega_{\rm cl}^{2n}(X)\ar[d] \\
\K^0(X) \ar[r]_{\!\!\!\!\!\!\!\!\ch} & \H^{\rm even}(X,\real)}
\eeq
with $\Omega_{\rm cl}^{2n}(X)$ the space of closed $2n$-forms on
$X$.

The RR-fields (of Type~IIA superstring theory) are closed even degree
forms associated to elements of $\fk^0(X)$  \cite{MW,FH}.
\begin{definition}
The {\it Ramond-Ramond field} ({\it RR-field} for short) $G$
associated to an element of $\fk^0(X)$ which maps to
$(E,\omega)\in\Aosc^0(X)$ under (\ref{diffexactseq}) is the closed
differential form
\beq
G(E,\omega)=\omega\wedge\widehat{A}(X)^{1/2} \ .
\label{GEomegadef}\eeq
\label{RRfielddef}\end{definition}
\noindent
The topological equivalence class of the RR-field is the D-brane
charge regarded as an element of the appropriate $\K$-theory
group. The D-branes ``couple'' to RR-fields, and another way to define
D-brane charge is through the pairings of their characteristic classes
with these differential forms.
\begin{definition}
The {\it Chern-Simons coupling} of a D-brane
$(W,E,\phi)$ to an RR-field corresponding to the element
$(f,C,\omega)\in\fk^0(X)$ is the spacetime integral
\beq
S_{\rm CS}(W,E,\phi|C)=\int\limits_XC\wedge\Ch(\phi_!E) \ .
\label{SCSdef}\eeq
\label{defCScoupl}\end{definition}
\noindent
Given this notion, we can now formulate an alternative homological
definition of D-brane charge.
\begin{definition}
The {\it dual Ramond-Ramond charge} ({\it dual RR-charge} for short)
of a D-brane $(W,E,\phi)$ in $X$ is the rational homology class
$\Ch(W,E,\phi)\in\H_\bullet(X,\rat)$ such that
\beq
S_{\rm CS}(W,E,\phi|C)=\int\limits_{\Ch(W,E,\phi)}C
\label{SCSdualcharge}\eeq
for all RR-fields on $X$.
\label{defdualRR}\end{definition}

Evidently, the natural framework for the Chern-Simons
couplings of D-branes is $\K$-homology. The Chern character in
topological $\K$-homology is the isomorphism
\beq
\ch\,:\,\K_\bullet(X)\otimes\rat~\longrightarrow~
\H_\bullet(X,\rat)
\label{chbulletiso}\eeq
defined by
\beq
\ch(W,E,\phi)=\phi_*\circ{\sf
  Pd}_W^{-1}\big(\ch(E)\cup\Todd(W)\big)
\label{chbulletbrane}\eeq
for any D-brane $(W,E,\phi)$ in $X$. The Todd class is related to the
Atiyah-Hirzebruch class by
\beq
\Todd(W)=\e^{-{\sf d}(W)}\cup\widehat{A}(W) \ ,
\eeq
where ${\sf d}(W)\in\H^2(W,\zed)$
is a characteristic class whose reduction modulo~$2$ is the second
Stiefel-Whitney class ${\sf w}_2(W)\in\H^2(W,\zed_2)$. This specifies
the spin$^c$ structure on the brane worldvolume $W$ as follows. The
spin$^c$ groups $Spin^c(n)=Spin(n)\times_{\zed_2}U(1)$ fit into a
commutative diagram
\beq
\xymatrix{
 & & 1\ar[d] & & \\ & & U(1)\ar[d]^\jmath\ar[rd]^{~z~\mapsto~z^2}& & \\
1~\ar[r] & ~Spin(n)~\ar[r]^{\!\!\!\imath}\ar[rd]_\lambda &
~Spin^c(n)~\ar[r]^{\,\,\,\,\,\,l}\ar[d]^\lambda & ~U(1)~\ar[r] & ~1 \\
 & & SO(n)\ar[d] & & \\ & & 1 & & }
\label{Spincncommdiag}\eeq
whose row and column are exact sequences. The map $\lambda:Spin(n)\to
SO(n)$ is the universal cover of the group $SO(n)$, while
$\jmath:U(1)\hookrightarrow Spin^c(n)$ and
$\imath:Spin(n)\hookrightarrow Spin^c(n)$ are natural inclusions. The
homomorphism $l:Spin^c(n)\to U(1)$ is defined by $(g,z)\mapsto
z^2$. It induces a map $\H^1(W,Spin^c(n))\to\H^1(W,U(1))$ and thus we
may associate a complex line bundle $L\to W$ with the worldvolume
$W$. The corresponding Chern class is the characteristic class ${\sf
  d}(W):={\sf c}_1(L)$.

The homological Chern character preserves sums, as well as the cap
product in the sense that
\beq
\ch\bigl(F\cap(W,E,\phi)\bigr)=\ch(F)\cap\ch
(W,E,\phi)
\label{chcappres}\eeq
for any complex vector bundle $F\to X$. This follows from its
definition (\ref{chbulletbrane}), the multiplicativity of the
cohomological Chern character, the index theorem, and the
Atiyah-Hirzebruch version of the Riemann-Roch theorem,
\beq
\phi_!\big(\ch(E)\cup\Todd(W)\big)=\ch
(\phi_!E)\cup\Aroof(X) \ ,
\label{AHRR}\eeq
which together give
\bea
\ch(F)\diamond\ch(W,E,\phi)&=&\ch(F)\cup\phi_!\big(
\ch(E)\cup\Todd(W)\big)[X]\nn\\ &=&
\ch(F\otimes\phi_!E)\cup\widehat{A}(X)[X]\nn\\ &=&{\rm index}(
\Dirac_{F\otimes\phi_!E}) 
\label{chcohopair}\eea
with $\diamond:\H^\bullet(X,\zed)\times\H_\bullet(X,\zed)$ the pairing
between cohomology and homology. This is just the index pairing
(\ref{K0indpair}).

As the notation suggests, the dual charge
of a D-brane is a modification of the homological Chern character
analogous to the modification in the case of cohomology.
\begin{proposition}
The dual RR-charge
$\Ch(W,E,\phi)\in\H_\bullet(X,\rat)$ of a D-brane $(W,E,\phi)$ in
$X$ can be represented by
\beq
\Ch(W,E,\phi)=\PD_X^{-1}\big(\PD^{\phantom{1}}_X\circ
\ch(W,E,\phi)\cup
\widehat{A}(X)^{-1/2}\bigr) \ .
\label{dualchargeexpl}\eeq
\label{propdualRR}\end{proposition}
\begin{proof}
We use (\ref{AHRR}) along with (\ref{chernmod}) to rewrite the
D-brane charge as
\beq
\Ch(\phi_!E)=\phi_!\big(\ch
(E)\cup\Todd(W)\big)\cup\Aroof(X)^{-1/2} \ .
\label{ChDrewrite}\eeq
Along with the definition (\ref{chbulletbrane}), we can use
(\ref{ChDrewrite}) to rewrite the Chern-Simons coupling (\ref{SCSdef})
in the form
\beq
S_{\rm CS}(W,E,\phi|C)=\int\limits_XC\wedge\big(\PD_X\circ
\ch(W,E,\phi)\cup\Aroof(X)^{-1/2}\big) \ .
\label{SCSrewrite}\eeq
By comparing this with the definition (\ref{SCSdualcharge}) of the
dual charge, (\ref{dualchargeexpl}) follows.
\end{proof}

\subsection{Noncommutative D-Branes}

There are many sorts of noncommutative D-branes, i.e., D-branes
modeled as Fredholm modules over a noncommutative algebra, and here
we will discuss only a few special instances. To motivate the first
generalization of our definition of a D-brane given above, we look at
an alternative way of regarding the embedding
$\phi:W\hookrightarrow X$ of a flat D-brane into spacetime. Consider a
tubular neighbourhood $W'$ of $W$ in $X$. For any point $u\in W$,
there is an isomorphism $T_uX\cong T_uW\oplus N_u(X/W)$, where $N(X/W)\to W$
is the normal bundle, which can be identified with
$\phi^*(TX)/TW$, of the proper differentiable map
$\phi:W\hookrightarrow X$. Let $\Psi:W'\to N(X/W)$ be the diffeomorphism which
identifies the normal bundle $N(X/W)$ with the tubular neighbourhood
$W'$. Then $\widehat{\phi}:=\Psi\circ\phi$ is the zero section of
$N(X/W)\to W$, and in this way we may identify the embedding of the
worldvolume into spacetime as a smooth section of the corresponding
normal bundle, $\widehat{\phi}\in C^\infty(W,N(X/W))$.

\begin{definition}
  A {\it flat nonabelian D-brane} in $X$ is a quadruple
  $(W,E,\phi,\widehat{\phi}\,)$, where $\phi:W\hookrightarrow X$ is a closed,
embedded spin$^c$ submanifold, $E\in\K^0(W)$, and $\widehat{\phi}\in
C^\infty(W,N(X/W)\otimes\End(E))$.
\label{defflatnonab}\end{definition}

When $E$ is a complex line bundle, we identify $\widehat{\phi}=\phi$ and in
this case the nonabelian D-brane is the same object that we defined above.
Nonabelian D-branes are classified by the same $\K$-theory as abelian
ones. In general, the algebra ${\mathcal A}=C(X)\otimes C^\infty(W,
\End(E))$ acts on the Hilbert spaces (\ref{L2WSE}) or
(\ref{L2WSEpm}), and so one can formulate this
definition in the language of Fredholm modules over an algebra which
is Morita equivalent to $C(X)$.

A related notion arises within the framework of
Fredholm modules when one replaces the algebra of functions on
spacetime with an appropriate noncommutative algebra.

\begin{definition}
A {\it flat noncommutative D-brane} in $X$ is a Fredholm
module over a deformation ${\mathcal A}_\theta$ of the algebra ${\mathcal
  A}=C(X)$.
\label{defflatncD}\end{definition}
\noindent
For the most part, noncommutative D-branes are classified by the same
$\K$-theory as commutative ones. However, this assumes that
$\K$-theory is preserved under deformation \cite{Ros}, which is not
always the case. See \cite{flabby} for an interesting counterexample.

\begin{example}
Consider $X=\real^{2n}$ (with compactly-supported
cohomology groups), and let $\mathcal{S}(\real^{2n})$ be
the space of complex Schwartz functions on $\real^{2n}$. Let
$\theta=(\theta^{ij})$ be a real, invertible skew-symmetric $2n\times
2n$ matrix. For $f,g\in\mathcal{S}(\real^{2n})$, we define the
corresponding twisted product
\beq
f\star_\theta
g(x):=(2\pi)^{-2n}\,\int\!\!\!\int f\big(\mbox{$x-\frac12\,\theta\,u$}
\big)\,
g\big(x+v\big)~\e^{-\ii u\cdot v}~\dd^{2n}u~\dd^{2n}v
\label{starproddef}\eeq
where $\dd^{2n}u$ is the Lebesgue measure on
$\real^{2n}$. The deformed algebra ${\mathcal A}_\theta$ is then
defined as
\beq
\alg_\theta=\bigl(\mathcal{S}(\real^{2n})\,,\,\star_\theta
\bigr) \ .
\label{Athetadef}\eeq
This is an associative Fr\'echet algebra which defines a
noncommutative space that is often called the {\it
  Moyal $n$-plane} or {\it noncommutative Euclidean space}. D-branes
may be constructed analogously to the commutative case. For instance,
for $f\in\alg_\theta$ let $m^\theta_f:\alg_\theta\to\alg_\theta$ denote the left
multiplication operator $g\mapsto f\star_\theta g$, and let
$\hil=L^2(\real^{2n})\otimes\complex^{2^n}$ be the Hilbert space of
ordinary square-integrable spinors on $\real^{2n}$. Let $\Dirac$ be
the ordinary Euclidean Dirac operator, and define a $*$-representation
$\rho^\theta:\alg_\theta\to\mathcal{B}(\hil)$ by
$\rho^\theta(f)=m^\theta_f\otimes\id_{2^n}$. Then
$(\hil,\rho^\theta,F)$, with $F=\Dirac/|\Dirac|$, is a Fredholm module
over the algebra (\ref{Athetadef}).
\label{exflatncD}\end{example}

\begin{example}
Let $X$ be a closed Riemannian spin manifold equipped
with a smooth isometric action of a $2n$-torus $\mathbb{T}^{2n}$. The
periodic action of $\bbT^{2n}$ on $X$ induces by
pullback an action of $\bbT^{2n}$ by automorphisms $\tau$ on the
algebra $\alg=C^\infty(X)$ of smooth functions on $X$. The orbits on
which $\bbT^{2n}$ acts freely determine maps $\sigma_s:C^\infty(X)\to
C^\infty(\bbT^{2n})$. Let
$\mathcal{T}^{2n}_\theta:=(C^\infty(\mathbb{T}^{2n}),\star_\theta)$ be the
noncommutative torus defined as the algebra of smooth functions on the
ordinary torus endowed with the periodised version of the twisted
product (\ref{starproddef}). Pulling back this deformation by the maps
$\sigma_s$ gives rise to an algebra
$\alg_\theta:=(C^\infty(X),\times_\theta)$. This defines a broad class
of noncommutative spaces known as {\it toric noncommutative
  manifolds}. The product $f\times_\theta g$ is given by a periodic
twisted product just like (\ref{starproddef}), with the non-periodic
translations replaced by the periodic
$\bbT^{2n}$-action. Alternatively, $\alg_\theta$ may be defined as the
fixed point subalgebra of $C^\infty(X)\,\hat\otimes\,\mathcal{T}^{2n}_\theta$
under the action of the automorphism $\tau\otimes\tau^{-1}$ (with
$\hat\otimes$ the projective tensor product of Fr\'echet
algebras). The construction of D-branes in these cases again parallels
that of the commutative case and Example~\ref{exflatncD} above.
\label{extoricncD}\end{example}
\noindent
The special classes of noncommutative branes given by
Examples~\ref{exflatncD} and~\ref{extoricncD} above will be referred
to as {\it isospectral deformations} of flat D-branes.
Other interesting examples may be found in
 \cite{ConnesLandi} and  \cite{ConnesDBV}.

\subsection{Twisted D-Branes\label{TwistedD}}

A very important instance in which noncommutative D-branes arise is
through the formulation of the notion of a curved D-brane. These arise
when the spacetime manifold $X$ carries certain topologically
non-trivial characteristics in the following sense.
Recall that a gerbe over $X$ is an infinite rank principal bundle over
$X$ with projective unitary structure group and characteristic class
$H$. A gerbe connection is a Deligne cohomology class on $X$ with top
form $H$.

\begin{definition}
A {\it $B$-field} $(X,H)$ is a gerbe with one-connection over $X$ and
characteristic class $H\in\H^3(X,\zed)$ called an {\it NS--NS
  $H$-flux}.
\label{defBfield}\end{definition}

For any oriented submanifold $W\subset X$, we denote by
$\W_3(W)\in\H^3(W,\zed)$ the third integer Stiefel-Whitney class of
its normal bundle $N(X/W)$. It is the obstruction to a spin$^c$
structure on~$W$.

\begin{definition}
A {\it curved} or {\it twisted D-brane} in a
$B$-field $(X,H)$ is a triple $(W,E,\phi)$, where
$\phi:W\hookrightarrow X$ is a closed, embedded oriented submanifold
with $\phi^*H=\W_3(W)$, and $E\in\K^0(W)$.
\label{defcurvedD}\end{definition}
\noindent
The condition on the brane embedding is required to cancel the
global Freed-Witten anomalies \cite{FW} arising in the worldsheet functional
integral. Suitable equivalence classes of curved D-branes take
values in the twisted topological $\K$-homology
$\K_\bullet(X,H)$ \cite{Szabo1}. For $H=0$, the worldvolume $W$ is spin$^c$
and the definition reduces to that of the flat case. One should also
require that the brane worldvolume $W$ carry a certain projective structure
that reduces for $H=0$ to the usual characteristic class
${\sf d}(W)\in\H^2(W,\zed)$ specifying a spin$^c$ structure on $W$.

A $B$-field can be realized by a bundle of algebras over $X$ whose
sections define a noncommutative $C^*$-algebra. When $H\in{\rm
  Tor}(\H^3(X,\zed))$ is a torsion class, this is known as an Azumaya
algebra bundle \cite{BM1}. Via the Sen-Witten construction, D-branes
in $(X,H)$ may then be realized in terms of $n$ D9 brane-antibrane
pairs carrying a principal $SU(n)/\zed_n=U(n)/U(1)$ Chan-Paton
bundle. Cancellation of anomalies then requires $n\,H=0$. To
accommodate non-torsion characteristic classes, one must consider a
certain $n\to\infty$ limit which can be realized as follows.

Let us fix a separable Hilbert space $\hil$, and denote by
$PU(\hil)=U(\hil)/U(1)$ the group of projective unitary automorphisms
of $\hil$. Let ${\mathcal K}(\hil)$ be the $C^*$-algebra
of compact operators on $\hil$. For any $g\in U(\hil)$, the map ${\rm
  Ad}_g:{\mathcal K}(\hil)\to{\mathcal K}(\hil)$ defined by ${\rm
  Ad}_g(T)=g\,T\,g^{-1}$ is an automorphism. The assignment
$g\mapsto{\rm Ad}_g$ defines a continuous epimorphism ${\rm
  Ad}:U(\hil)\to\Aut({\mathcal K}(\hil))$ with respect to the
strong operator topology on $U(\hil)$ and the point-norm topology on
$\Aut({\mathcal K}(\hil))$ with $\ker({\rm Ad})=U(1)$. It follows
that one can identify the group $PU(\hil)$ with $\Aut({\mathcal
  K}(\hil))$ under this homomorphism. 

The exact sequence of sheaves of germs of continuous functions
on $X$ given by
\beq
1~\longrightarrow~\underline{U(1)}_{\,X}~\longrightarrow~
\underline{U(\hil)}_{\,X}~\longrightarrow~\underline{PU(\hil)}_{\,X}~
\longrightarrow~1
\eeq
induces a long exact sequence of sheaf cohomology groups as
\bea
&&~\longrightarrow~\H^1\bigl(X\,,\,\underline{U(\hil)}_{\,X}\bigr)~
\longrightarrow~\H^1\bigl(X\,,\,\underline{PU(\hil)}_{\,X}\bigr)~
\stackrel{\delta_1}{\longrightarrow}~\H^2\bigl(X\,,\,\underline{U(1)}_{\,X}
\bigr)~\longrightarrow~ \ .\nonumber\\
\eea
Since the unitary group $U(\hil)$ is contractible with respect to the
strong operator topology, the sheaf $\underline{U(\hil)}_{\,X}$ is soft and so
$\H^j(X\,,\,\underline{U(\hil)}_{\,X})=0$ for all $j\geq1$. It follows
that the map $\delta_1$ is an isomorphism. From the exact sequence of
groups
\beq
1~\longrightarrow~\zed~\longrightarrow~\real~\longrightarrow~U(1)~
\longrightarrow~1
\label{gpexactseq}\eeq
we obtain the long exact cohomology sequence
\beq
\longrightarrow~\H^2\big(X\,,\,\underline{\real}_{\,X}\big)~
\longrightarrow~\H^2\big(X\,,\,\underline{U(1)}_{\,X}\big)~
\stackrel{\delta_2}{\longrightarrow}~\H^3\big(X\,,\,\zed\big)~
\longrightarrow~\H^3\big(X\,,\,\underline{\real}_{\,X}\big)~
\longrightarrow\nonumber
\eeq
Agaia, since $\underline{\real}_{\,X}$ is a fine sheaf, one has
$\H^j(X\,,\,\underline{\real}_{\,X})=0$ for all $j\geq1$ and so
the map $\delta_2$ is an isomorphism.

The map
\beq
\delta_X=\delta_2\circ\delta_1\,:\,\H^1\bigl(X\,,\,
\underline{PU(\hil)}_{\,X}\bigr)~\longrightarrow~
\H^3\big(X\,,\,\zed\big)
\label{deltaXdef}\eeq
is thus an isomorphism on stable equivalence classes of principal
$PU(\hil)$-bundles over the spacetime $X$. If $P\to X$ is a
$PU(\hil)$-bundle and $[P] \in
\H^1(X\,,\,\underline{PU(\hil)}_{\,X})$ is its isomorphism class, then
$\delta_X(P):=\delta_X([P]) \in \H^3(X,\mathbb{Z})$ is called
the \textit{Dixmier-Douady invariant} of $P$ \cite{BM1,13,14}. The set
of isomorphism classes of locally trivial bundles over $X$ with
structure group $\Aut(\mathcal{K}(\hil))$ and fibre
$\mathcal{K}(\hil)$ form a group ${\rm Br}^{\infty}(X)$ under
tensor product called the \textit{infinite Brauer group} of $X$. Using
the identification $PU(\hil)\cong\Aut(\mathcal{K}(\hil))$, it
follows that such algebra bundles are also classified by
$\H^3(X,\mathbb{Z})$. If $\mathcal{E}$ is a bundle of this kind, then
the corresponding element of $\H^3(X,\mathbb{Z})$ is also called the
Dixmier-Douady invariant of $\mathcal{E}$ \cite{13} and denoted
$\delta_X(\mathcal{E})$.

Given a $B$-field $(X,H)$, there corresponds
a unique, locally trivial $C^*$-algebra bundle $\bun_H\to X$ with
fibre ${\mathcal K}(\hil)$ and structure group $PU(\hil)$ whose
Dixmier-Douady invariant is
\beq
\delta_X(\bun_H)=H \ .
\label{deltaXH}\eeq
Let $C_0(X,\bun_H)$ be the $C^*$-algebra of continuous sections,
vanishing at infinity, of this algebra bundle. The twisted $\K$-theory
$\K^\bullet(X,H)=\K_\bullet(C_0(X,\bun_H))$ \cite{AS,15} may then be computed as
the set of stable homotopy classes of sections of an associated algebra
bundle $P_H\times_{PU(\hil)}{\rm Fred}(\hil)$, where $P_H$ is a
principal $PU(\hil)$-bundle over $X$ and ${\rm Fred}(\hil)$ is the
algebra of (self-adjoint) Fredholm operators on $\hil$ with $PU(\hil)$
acting by conjugation. On the other hand, one can define
Dixmier-Douady classes over any D-brane worldvolume $W$ in
complete analogy with (\ref{deltaXdef}) and show that \cite{Plymen1}
\beq
\W_3\big(W\big)=\delta_W\big(\Cl(N(X/W))\big)
\label{W3Cliff}\eeq
where $\Cl(N(X/W))\to W$ is the Clifford algebra bundle of the
normal bundle $N(X/W)$. The Dixmier-Douady class $\delta_W(\Cl(N(X/W)))$ is
the global obstruction to existence of a spinor bundle
${\sf S}_W$ with 
\beq
\Cl_p\big(N(X/W)\big)\cong\End_p\big({\sf S}_W\big)
\eeq
for $p\in W$. This observation leads to the following
result.

\begin{proposition}
There is a one-to-one correspondence between twisted D-branes in
$(X,H)$ and stable homotopy classes of Fredholm modules over the
algebra $C_0(X,\bun_H)$.
\label{proptwistD}\end{proposition}

The proof of Proposition~\ref{proptwistD} fixes the appropriate equivalence
relations required for twisted topological $\K$-homology. One of these
equivalence relations (in addition to the appropriate twisted analogs
of bordism, direct sum and vector bundle modification) is based on the
observation \cite{MMS1} that while a triple
$(W',E',\phi'\,)$ may violate the embedding condition of a curved
D-brane, one can still cancel the Freed-Witten anomalies on the
submanifold $W'$ by adding a ``source'' in $W'$ corresponding to
a twisted D-brane. This D-brane can be unstable and decay due to the
configuration $(W',E',\phi'\,)$. This physical process can be stated
more precisely as follows.

\begin{lemma}[{\bf Stabilization}]
  Let $(W,E,\phi)$ be a twisted
  D-brane in $(X,H)$ whose orientation cycle $[W]$ is non-trivial in
  $\H_\bullet(X,\zed)$. Suppose that there exists a closed, embedded
  oriented submanifold $\phi':W'\hookrightarrow X$ such that $W$ is a
  codimension~$3$ submanifold of $W'$ and its Poincar\'e dual ${\sf
    Pd}_{W\hookrightarrow W'}$ satisfies the equation
\beq
\phi^{\prime\,*}H=\W_3(W'\,)+{\sf Pd}_{W\hookrightarrow W'}
\label{stabilitylem}\eeq
in $\H^3(W',\zed)$. Then $(W,E,\phi)$ is trivial in $\K_\bullet(X,H)$
\textup{(}up to twisted vector bundle modification\textup{)}.
\label{lemmastab}\end{lemma}

The structure of D-branes in torsion $B$-fields simplifies
drastically. When $H\in {\rm Tor}(\H^3(X,\mathbb{Z}))$ the
algebra $C_{0}(X,\mathcal{E}_{H})$ is Morita equivalent to an Azumaya
algebra bundle over $X$, i.e., a bundle whose fibres are Azumaya
algebras with local trivializations reducing them to $n\times n$
matrix algebras $\mat_n(\complex)$. Two Azumaya bundles
$\mathcal{E,F}$ over $X$ are called {\it equivalent} if there are
vector bundles $E,F$ over $X$ such that $\mathcal{E} \otimes 
\End(E)$ is isomorphic to $\mathcal{F} \otimes \End(F)$. The
set of equivalence classes is a group ${\rm Br}(X)$ under tensor
product called the \textit{Brauer group} of $X$. There is also a
notion of Dixmier-Douady invariant $\delta'_X$ for Azumaya bundles
over $X$, which is constructed using the same local description as
above but now with $\hil$ a finite-dimensional complex vector
space. By Serre's theorem one has ${\rm Br}(X)\cong{\rm
  Tor}(\H^3(X,\mathbb{Z}))$. This gives two descriptions of ${\rm
  Tor}(\H^3(X,\mathbb{Z}))$, one in terms of locally trivial bundles
over $X$ with fibre $\mathcal{K}(\hil)$ and structure group ${\rm
  Aut}(\mathcal{K}(\hil))$, and the other in terms
of Azumaya bundles. They are related by the following
result from \cite{13}.

\begin{proposition}
If $X$ is a compact manifold and $\mathcal{E}$ is a locally trivial
bundle over $X$ with fibre $\mathcal{K}(\hil)$ and structure group
$\Aut(\mathcal{K}(\hil))$, then the algebra
$C(X,\mathcal{E})$ is stably unital if and only if its Dixmier-Douady
invariant is a torsion element in $\H^3(X,\mathbb{Z})$.
\end{proposition}

These constructions allow us to describe the $\K$-theory of the
noncommutative $C^*$-algebra $C(X,\mathcal{E}_H)$ \cite{13}. Morita
equivalence induces an isomorphism between the $\K$-theories of
$C_{0}(X,\mathcal{E}_{H})$ and $C_{0}(X,\mathcal{A}_{H})$, where
$\mathcal{A}_{H}$ is an Azumaya bundle associated to $\mathcal{E}_{H}$
via the Dixmier-Douady invariant \cite{BM1}. A geometric description of
this $\K$-theory is provided by the notion of projective vector
bundle \cite{13}, while in the infinite-dimensional setting of a
non-torsion $B$-field one needs to introduce the notions of bundle
gerbes and bundle gerbe modules \cite{16}.

\newsection{Poincar\'{e} Duality \label{Pairings}}

A crucial point of our construction of flat D-brane charges in 
Section~\ref{FlatD} was the role played by Poincar\'e duality. With an
eye to generalizing the construction to the more general settings
described above, in this section we
will explore how and to what extent this classical notion of topology
can be generalized to generic $C^*$-algebras in the context of
$\K\K$-theory. We will describe various criteria which guarantee the
duality. There are several natural inequivalent versions of Poincar\'e
duality, which we define and study, giving many
purely noncommutative examples. Our examples range from those of classical
spaces to noncommutative deformations of spin$^c$ manifolds, and also
the more general examples of Poincar\'e duality spaces such as those
arising in the case of the free group acting on its boundary or, more
generally, for hyperbolic groups acting on their Gromov boundaries. 

\subsection{Exterior Products in $\K$-Theory\label{ExtProdsK}}

To describe Poincar\'e duality generically in $\K$-theory, we first need
to make some important remarks concerning the product structure. Let
$\alg_1$ and $\alg_2$ be unital $C^*$-algebras. If
$p_1\in\mat_{k}(\alg_1)$ is a projection representing a
Murray-von~Neumann equivalence class in
$\K_0(\alg_1)$ and a projection $p_2\in\mat_{l}(\alg_2)$ represents a
class in $\K_0(\alg_2)$, then the tensor product $p_1\otimes p_2 $ is
a projection in $\mat_k(\alg_1)
\otimes\mat_l(\alg_2)\cong\mat_{k\,l}(\alg_1\otimes\alg_2)$ for any
$C^*$-tensor product and so it represents a class in
$\K_0(\alg_1\otimes\alg_2)$ (in this section we will work mostly with
the maximal tensor product). In this way we obtain a map
\beq
\K_0(\alg_1) \times \K_0(\alg_2) ~\longrightarrow~
\K_0(\alg_1\otimes\alg_2) \ .
\label{Kextprodgen}\eeq
This definition extends to non-unital algebras in a standard
way \cite[p. 104]{HR}. In the special case $\alg_1  =\alg_2 =\alg$ we
obtain a map
\beq
\K_0(\alg) \times \K_0(\alg) ~\longrightarrow~ \K_0(\alg\otimes\alg)  \
.
\label{KextprodA}\eeq

It is important to note that, in contrast to the topological case, it
is not possible in general to make $\K_0(\alg)$ into a ring. We recall
that for a compact topological space $X$ there is an exterior product
map
\beq
\K^0(X) \times \K^0(X) ~\longrightarrow~ \K^0(X\times X)
\eeq
which is defined using the exterior tensor product of vector
bundles. The diagonal map $X\rightarrow X\times X$ induces a natural
transformation $\K^0(X\times X) \rightarrow \K^0(X)$. The 
composition of the two maps thereby leads to the product
\beq
\K^0(X) \times \K^0(X) ~\longrightarrow~ \K^0(X) \ .
\eeq
If $\alg=C(X)$ is the algebra of continuous functions on $X$, then the
diagonal map translates into the product map 
\beq
m\,:\,\alg\otimes\alg~\longrightarrow~\alg
\label{multmap}\eeq
on the algebra $\alg$ given by $m(a\otimes b)=a\,b$ for all
$a,b\in\alg$. Since $\alg$ is commutative, the multiplication
(\ref{multmap}) is an algebra homomorphism and there is an
induced map
\beq
m_*\,:\,\K_0(\alg\otimes\alg) ~\longrightarrow~ \K_0(\alg) \ .
\label{multKthind}\eeq
For a noncommutative $C^*$-algebra the multiplication map is not an
algebra homomorphism and so we cannot expect that in general the map
(\ref{multKthind}) will be defined.

Recall that the suspension of a generic $C^*$-algebra $\alg$ is the 
$C^*$-algebra $\Sigma(\alg):=C_0(\bbr) \otimes\alg$. By definition one has
$\K_p(\alg) = \K_0(\Sigma^p(\alg)) = \K_0(C_0(\bbr^p)
\otimes\alg)$. Bott periodicity ensures that up to isomorphism there
are only two distinct $\K$-theory groups $\K_0(\alg)$ and
$\K_1(\alg)$. One has
\bea
\Sigma^k(\alg_1)\otimes\Sigma^l(\alg_2) & =&
\big(C_0(\bbr^k)\otimes\alg_1\big)\otimes
\big(C_0(\bbr^l)\otimes\alg_2\big) \nonumber \\
& \cong& C_0(\bbr^{k+l}) \otimes (\alg_1\otimes\alg_2) \nonumber \\
& = & \Sigma^{k+l}(\alg_1\otimes\alg_2) \ .
\eea
If we combine this formula with the exterior product
(\ref{Kextprodgen}) defined for $\K_0$-groups then we obtain the
general exterior product
\beq
\K_k(\alg_1) \times \K_l(\alg_2) ~\longrightarrow~
\K_{k+l}(\alg_1\otimes\alg_2) \ .
\eeq
See \cite[\S4.7]{HR} for more details and examples. 

Many important statements in $\K$-theory admit a concise formulation
in terms of the product structure. For example, there exists a
canonical class $\beta\in \K_2(\bbc) = \K_0(C_0(\bbr^2))$, called the
Bott generator, such that the exterior product with $\beta$ defines a
map
\beq
\K_0(\alg) ~\xr{\otimes\beta}~
\K_2(\alg\otimes \bbc) = \K_2(\alg) \ .
\eeq
This provides the isomorphism required by the Bott periodicity theorem
 \cite[\S4.9]{HR}. These observations all find their most natural
generalisation in Kasparov's $\K\K$-theory  \cite{Kas81}, which we now
proceed to describe. (See  \cite[Ch.\ VIII]{Black} for a more detailed
exposition.) 

\subsection{$\K\K$-Theory\label{KK-Theory}}

Let $\balg$ be a $C^*$-algebra. A {\it Hilbert $\balg$-module} $\hil$
is a module over $\balg$ equipped with a $\balg$-valued inner product
\beq
\hil \times \hil~ \longrightarrow~ \balg \ ,
\quad (\zeta,\zeta'\,) ~\longmapsto~ 
\langle \zeta \mid \zeta'\, \rangle \in \balg
\eeq
which satisfies similar properties to those of an inner product 
with values in $\bbc$  \cite{Lance}. 
We denote by $\call (\hil)$ the algebra of linear operators on 
$\hil$ which admit an adjoint with respect to this inner product. 
The closed subalgebra generated by all rank $1$ operators of 
the form $\theta_{\zeta,\zeta'} : \xi \mapsto \zeta\,\langle \zeta'\mid
\xi\rangle $ is denoted $\cK(\hil)$ and called the algebra of {\it
  compact operators} on $\hil$. The algebra $\cK(\hil)$ is a closed
ideal in $\call(\hil)$.

\begin{definition}
Let $\alg$ and $\balg$ be $C^*$-algebras. An {\it odd $\alg$--$\balg$
  Kasparov bimodule} is a triple $(\hil,\rho,F)$ where $\hil$ is a
countably generated Hilbert $\balg$-module, the map $\rho: \alg \rightarrow
\call(\hil)$ is a $*$-homomorphism, and $F\in \call(\hil)$ is a
self-adjoint operator such that for each $a\in\alg$ one has 
\begin{equation}\label{compactness}
\rho(a)\,\left(\Id_\hil-F^2\right) \in \cK(\hil) \qquad \mbox{and}
\qquad F\,\rho(a) - \rho(a)\,F\in \cK(\hil) \ .
\end{equation}
An {\it even $\alg$--$\balg$ Kasparov bimodule} is a triple $(\hil,
  \rho, F)$ where $\hil= \hil^+ \oplus\hil^-$ is $\bbz_2$-graded,
  $\phi$ is an even degree map, $F$ is an odd map, and 
the compactness conditions (\ref{compactness}) are satisfied. In both
cases a triple is called \emph{degenerate} if the operators
$\rho(a)\,(\Id_\hil-F^2)$ and $ F\,\rho(a) - \rho(a)\,F$ are zero for all
$a\in\alg$. 
\end{definition}

We denote by $\cale_0(\alg,\balg)$ and $\cale_1(\alg,\balg)$ the sets
of isomorphism classes of even and odd Kasparov bimodules,
respectively. These two sets are made into semi-groups using the
direct sum of Kasparov bimodules. Two triples $(\hil, \rho, F)$ and
$(\hil, \rho', F'\,)$ are regarded as equivalent if (by adding 
a degenerate triple to both if necessary) $F'$ can be obtained from
$F$ via operator homotopy. Imposing these equivalence relations on
$\cale_0(\alg,\balg)$ and $\cale_1(\alg,\balg)$ yields two abelian groups
$\KK_0(\alg,\balg)$ and $\KK_1(\alg,\balg)$. The functor
$\KK_\bullet(\alg,\balg)$ is homotopy invariant and satisfies excision in both
variables with respect to $\complex$-split exact sequences of
$C^*$-algebras.

The special case where $\balg= \bbc$ is important. A Hilbert
$\bbc$-module $\hil$ is just a Hilbert space, and the algebra $\call (\hil)$
is in this case the $C^*$-algebra of bounded linear operators 
on $\hil$. The compactness conditions (\ref{compactness}) provide an 
abstraction of the essential properties of elliptic
operators \cite{Atiyah}, and a Kasparov bimodule in this
case is just a Fredholm module. Thus we can \emph{define} the
$\K$-homology of the $C^*$-algebra $\alg$ as
\beq
\K^\bullet(\alg) = \KK_\bullet(\alg, \bbc) \ .
\eeq
One can also show that $\KK_\bullet(\bbc, \alg)$ is isomorphic to the
$\K$-theory $\K_\bullet(\alg)$ of the algebra $\alg$. 

The key property of the bivariant functor $\KK_\bullet(\alg,\balg)$ is the
existence of an associative product
\beq
\otimes_\balg:
\KK_i(\alg,\balg) \times \KK_j(\balg,\calg) ~\longrightarrow~
\KK_{i+j}(\alg,\calg)
\label{KKcompprod}\eeq
induced by the composition of bimodules, which is additive in both
variables. This product is called the {\it
  composition} or {\it intersection product} and it is compatible with
algebra homomorphisms in the following sense. There is a functor from
the category of separable $C^*$-algebras to an additive category 
$\underline{\KK}$ whose objects are
separable $C^*$-algebras and whose morphisms 
$\alg\to\balg$ are precisely the elements of $\KK_\bullet(\alg,\balg)$. An
algebra homomorphism  $\phi:\alg\rightarrow\balg$ thus defines an
element $\KK (\phi) \in\KK_0(\alg,\balg)$, and if
$\psi:\balg\rightarrow\calg$ is another homomorphism then
\beq
\KK(\psi\circ \phi) ~=~ \KK(\phi) \otimes_\balg \KK(\psi) \in \KK_0(\alg,\calg)
\ .
\eeq
The intersection product makes $\KK_\bullet(\alg,\alg)$ into a
$\bbz_2$-graded ring whose unit element is $1_\alg:=\KK(\Id_\alg)$, the element
of $\KK_0(\alg,\alg)$ determined by the identity map
$\Id_\alg:\alg\rightarrow\alg$. 

The operation of taking the composition product by a fixed element
$\alpha \in \KK_0(\alg,\balg)$ gives a map
\beq
\KK_i(\bbc, \alg)  ~\longrightarrow~
\KK_{i}(\bbc,\balg) \ ,
\label{KKindexmap}\eeq
i.e., a homomorphism $\alpha_*\colon\K_i(\alg)\rightarrow \K_{i}(\balg)$ in
$\K$-theory, and also a map
\beq
\KK_i(\balg, \bbc) ~\longrightarrow~
\KK_{i}(\alg, \bbc) \ ,
\eeq
i.e., a homomorphism of $\K$-homology groups $\alpha^*\colon\K^i(\balg) \rightarrow
\K^{i}(\alg)$. If $\alpha$ is the class of a bimodule $(\hil,\rho,F)$,
then (\ref{KKindexmap}) is the index map ${\rm
  index}_F:\K_i(\alg)\to\K_{i}(\balg)$. In general, we will say that the
element $\alpha$ is {\it invertible} if there exists $\beta \in
\KK_0(\balg,\alg)$ such that $\alpha\otimes_\balg\beta = 1_\alg \in
\K\K_0(\alg,\alg) $ and $\beta\otimes_\alg\alpha= 1_\balg \in
\KK_0(\balg,\balg)$. We call $\beta$ the {\it inverse}
of $\alpha$ and write $\beta=\alpha^{-1}$. An invertible element of
$\KK_0(\alg,\balg)$ gives an isomorphism $\K_i(\alg) \cong
\K_{i}(\balg)$ of $\K$-theory groups and of $\K$-homology groups
$\K^i(\alg) \cong\K^{i}(\balg)$. This construction will be
generalized in Section~\ref{KKEq}.

The composition product (\ref{KKcompprod}), along with the natural map
$\KK_i(\alg,\balg)\to\KK_i(\alg\otimes\calg,\balg\otimes\calg)$ given by
$\alpha\mapsto\alpha\otimes1_\calg$, imply the existence of a more
general associative product in $\K\K$-theory called the {\it Kasparov
  product}. For any collection of separable $C^*$-algebras $\alg_1,
\balg_1, \alg_2, \balg_2$ and $\dalg$ there is a bilinear map
\begin{equation}\label{product}
\KK_i(\alg_1 , \balg_1\otimes\dalg) \otimes_\dalg\KK_j(\dalg\otimes
\alg_2 ,\balg_2) ~\longrightarrow~ \KK_{i+j} (\alg_1\otimes\alg_2,
\balg_1\otimes\balg_2) \ .
\end{equation}
This product can be thought of as a mixture between the usual cup and
cap products. For $\dalg=\bbc$ it specializes to the exterior product 
(also written $\times$):
\beq
\KK_i(\alg_1,\balg_1) \otimes \KK_j(\alg_2,\balg_2) ~
\longrightarrow~ \KK_{i+j}(\alg_1\otimes\alg_2,\balg_1\otimes\balg_2)
\ . 
\label{KKextprod}\eeq
When $\alg_1=\alg_2 = \bbc$, (\ref{KKextprod}) is just the exterior
product on $\K$-theory that we discussed in Section~\ref{ExtProdsK}
above. On the other hand, when we put $\balg_1 = \bbc$ and $\alg_1
=\bbc$ in (\ref{product}) we recover the original composition product 
(\ref{KKcompprod}) in the form
\beq
\KK_i(\alg_1,\dalg)\otimes_\dalg\KK_{j}(\dalg,\balg_2) ~
\longrightarrow~ \KK_{i+j}(\alg_1,\balg_2) \ .
\label{compprodD}\eeq
Various technical details of the Kasparov product, useful for explicit
computations, are collected in Appendix~A, and a pictorial method for keeping
track of these is given in Appendix~B.

\subsection{Strong Poincar\'e Duality\label{PDKK}}

Poincar\'e duality for $C^*$-algebras was
defined by Connes  \cite{Connes,Connes:CMP} in the context of real
$\KK$-theory as a means of defining noncommutative spin$^c$ manifolds.
It was subsequently extended to more general situations by Kaminker
and Putnam \cite{KP}, Emerson \cite{E1,E2}, amongst others. These
latter works motivate our first definition of the duality in the
context of complex $\KK$-theory.

\begin{definition}[{\bf Strong Poincar\'e Duality}]\label{dual1}
A pair of separable $C^*$-algebras $(\alg, \balg)$ 
is said to be a {\it strong Poincar\'{e} dual pair} ({\it strong PD
  pair} for short) if there exists a class
$\Delta \in \KK_d(\alg\otimes\balg, \bbc)= \K^d(\alg\otimes\balg)$ in
the $\K$-homology of $\alg\otimes\balg$ and a class $\Delta^\vee\in
\KK_{-d}(\bbc,\alg\otimes\balg)=\K_{-d}(\alg\otimes\balg)$ in the
$\K$-theory of $\alg\otimes\balg$ with the properties 
\beq
\Delta^\vee\otimes _\balg\Delta ~=~ 1_\alg \in
\KK_0(\alg,\alg) \quad {\rm and} \quad
\Delta^\vee\otimes _\alg\Delta ~=~ (-1)^d\,1_\balg \in
\KK_0(\balg,\balg) \ .\nonumber\\
\label{StrongPDdef}\eeq
The element $\Delta$ is called a {\it fundamental
  $\K$-homology class} for the pair $(\alg,\balg)$ and
$\Delta^\vee$ is called its {\it
  inverse}. A separable $C^*$-algebra $\alg$ is said to be a {\it strong
  Poincar\'{e} duality algebra} ({\it strong PD algebra} for short) 
 if $(\alg, \alg^\op)$ is a strong PD pair, 
 where $\alg^\op$ denotes the opposite algebra of $\alg$,
i.e., the algebra with the same underlying vector space as $\alg$ but
with the product reversed.
\end{definition}
\begin{remark}
The use of the opposite algebra in this definition is to describe
$\alg$-bimodules as $(\alg\otimes\alg^\op)$-modules. We will see this
explicitly in Section~\ref{NCspin} below.
\end{remark}

Let us indicate how Definition~\ref{dual1} is used to implement
Poincar\'{e} duality. First of all, we note that the tensor product
algebra $\alg\otimes\balg$ is canonically isomorphic to the algebra
$\balg\otimes\alg$ through the ``flip'' map
$\alg\otimes\balg\rightarrow\balg\otimes\alg$ which interchanges the
two factors. Thus $\K^d(\alg\otimes\balg) \cong\K^d(\balg\otimes\alg)$
and $\K_{-d}(\alg\otimes\balg) \cong\K_{-d}(\balg\otimes\alg)$. With this
observation, we can use the Kasparov product (\ref{product}) to induce
a map
\bea
 \otimes_\alg:
\K\K_d(\alg\otimes\balg, \bbc) \otimes \K_i(\alg)\cong &&
\K\K_d(\balg\otimes\alg, \bbc)\otimes \K\K_i(\bbc,\alg)~ \nonumber\\
&& \longrightarrow ~\K\K_{d+i}(\balg, \bbc) = \K^{d+i}(\balg) \ .
\eea
Thus taking the product on the right with the element $\Delta \in
\KK_d(\alg\otimes\balg, \bbc)$ produces a map
\beq
\K_i(\alg) ~\xr{\otimes_\alg \Delta}~ \K^{i+d}(\balg)
\label{betamap}\eeq
from the $\K$-theory $\K_i(\alg)$ of the algebra $\alg$ to the $\K$-homology
$\K^{i+d}(\balg)$ of the algebra $\balg$. Since the element $\Delta$ has
an inverse $\Delta^\vee \in \KK_{-d}(\bbc, \alg\otimes\balg)$, using the
exterior product again we can define a map
\beq
\otimes_\balg:
\KK_{-d}(\bbc,\alg\otimes\balg) \otimes \KK_i(\balg, \bbc) ~
\longrightarrow~ \KK_{-d+i}(\bbc,\alg) = \K_{-d+i}(\alg) \ .
\eeq
Thus multiplying on the left by the element $\Delta^\vee$ establishes
a map
\beq
\K^i(\balg) ~\xr{\Delta^\vee\otimes_\balg}~ \K_{i-d}(\alg)
\label{alphamap}\eeq
from the $\K$-homology of $\balg$ to the $\K$-theory of $\alg$.

Since $\Delta$ and  $\Delta^\vee$ are inverses to each other, for any
$x\in\K_i(\alg)$ one has
\bea
\Delta^\vee\otimes_\balg( x \otimes_\alg \Delta)&=&\left(\Delta^\vee
\otimes_\balg\Delta\right) \otimes_\alg x\nonumber\\ & =& 
1_\alg\otimes_\alg x \nonumber \\ &=& x \ .
\eea
As a consequence the two maps (\ref{betamap}) and (\ref{alphamap})
are inverse to each other, up to the sign given in
(\ref{StrongPDdef}) which results from graded commutativity of the
exterior product (\ref{KKextprod}).  (See \cite[p.~588]{Connes} and
 \cite[\S3]{E1} for further details.) Thus when $(\alg, \balg)$
is a strong PD pair, the elements $\Delta$ and $\Delta^\vee$ establish
isomorphisms
\beq
\K_i(\alg) \cong \K^{i+d}(\balg)\qquad{\rm and}\qquad \K_i(\balg)\cong
\K^{i-d}(\alg) \ ,
\label{fundpropPoincare}\eeq
which is the fundamental property of any form of Poincar\'e duality.

More generally, for any pair of separable $C^*$-algebras $(\cC, \cD)$, 
the maps 
\bea
\quad\Delta\otimes_\alg \,:\, \KK_i(\cC, \alg\otimes\cD) &
\longrightarrow& \KK_{i+d}(\cC\otimes\balg, \cD) \ , \nonumber\\
\Delta^\vee\otimes_\balg \,:\,\KK_i(\cC, \balg\otimes\cD) &
\longrightarrow&\KK_{i-d}(\cC\otimes\alg, \cD) 
\label{eqn:PD}\eea
are also isomorphisms, showing that Poincar\'e duality with arbitrary
coefficients holds in this case (Compare \cite{Connes}). By setting
$\cC,\cD$ equal to various choices from the collection of algebras
$\bbC,\alg,\balg$, we may infer from \eqref{eqn:PD} that the four
maps
\bea
\Delta\otimes_\alg \,:\, \KK_i(\alg, \alg) &
\longrightarrow&\KK_{i+d}(\alg \otimes\balg, \bbC) \ , \nonumber\\
\Delta^\vee\otimes_{\balg} \,:\, \KK_i(\balg, \balg) &
\longrightarrow&\KK_{i-d}(\balg \otimes \alg, \bbC) \ , \nonumber\\
\Delta\otimes_\alg \,:\, \KK_i(\bbC, \alg\otimes\balg) &
\longrightarrow&\KK_{i+d}(\balg, \balg) \ , \nonumber\\
\Delta^\vee\otimes_{\balg} \,:\,\KK_i(\bbC, \balg\otimes\alg) &
\longrightarrow&\KK_{i-d}(\alg, \alg)
\label{eqn:op2}\eea
are all isomorphisms. It follows that if $(\alg,\balg)$ is a strong PD
pair, then a fundamental class for $(\alg,\balg)$ induces isomorphisms
\beq\label{eqn:op}
\K_{i+d}(\alg\otimes\balg) ~\cong~\KK_i(\alg, \alg)  ~\cong ~
\KK_i(\balg, \balg) ~\cong~
\K^{i+d}(\alg\otimes\balg) \ .
\eeq

\begin{proposition}\label{prop:PD-opposite}
Let $\alg$ and $\balg$ be separable $C^*$-algebras. Then:
\begin{enumerate}
\item $\alg$ is a strong PD algebra if and only if $\alg^\op$ is a
  strong PD algebra; and
\item If $\alg$ is Morita equivalent to $\balg$, then $\alg$ is a
  strong PD algebra if and only if $\balg$ is a strong PD algebra.
\end{enumerate}
\end{proposition}
\begin{proof}
(1) follows easily since $\Delta$ is a fundamental class for $\alg$ if
and only if it is a fundamental class for $\alg^\op$, where we
identify the $\KK$-groups of $\alg\otimes\alg^\op$ with those of the
flip $\alg^\op\otimes\alg$. The proof of (2) is in  \cite[\S4 Theorem
7]{Kas80}.
\end{proof}

\begin{example}
Let $X$ be a complete oriented manifold. Then the two pairs of
$C^*$-algebras $(C_0(X)\,,\, C_0(T^*X))$
and $(C_0(X)\,,\, C_0(X, {\sf Cliff}(T^*X)))$ are both strong PD pairs,
where ${\sf Cliff}(T^*X)$ is the Clifford algebra bundle of the
cotangent bundle $T^*X$. If in addition $X$ is spin$^c$ then $C_0({\sf
  Cliff}(T^*X))$ and $C_0(X)$ are Morita equivalent, and so $C_0(X)$
is a strong PD algebra. If moreover $X$ is compact with boundary
$\partial X$ equipped with the induced spin$^c$ structure, then the
$\K$-homology connecting homomorphism takes the fundamental class of
$X$ to the fundamental class of $\partial X$. It follows that
$(C(X)\,,\, C(X, \partial X))$ is a strong PD pair, where $C(X, \partial
X)$ is the $C^*$-algebra of continuous functions on $X$ which
vanish at the boundary $\partial X$.
\label{strongPDEx1}\end{example}

\begin{example}
Let $\Gamma$ be a $\K$-amenable, torsion-free discrete group whose
classifying space $B\Gamma$ is a
smooth oriented manifold. Suppose that $\Gamma$ has the 
  Dirac-dual Dirac property, i.e., for any proper
$\Gamma$--$C^*$-algebra $\alg$, a Dirac element
$\alpha\in\KK_0^\Gamma(\alg,\bbC)$ in the $\Gamma$-equivariant
$\K$-homology of $\alg$ and a dual Dirac element
$\beta\in\KK_0^\Gamma(\bbC,\alg)$ in the $\Gamma$-equivariant
$\K$-theory of $\alg$ satisfy the conditions
\beq
\alpha\otimes_\bbC\beta~=~1_\alg\in\KK_0^\Gamma(\alg,\alg)\quad
\mbox{and}\quad\beta\otimes_\alg\alpha~=~1_\complex\in
\KK_0^\Gamma(\bbC,\bbC) \ .
\label{Diracdualprop}\eeq
The Dirac element $\alpha$ is constructed using a spin$^c$ Dirac
operator. Recall that a
multiplier $\sigma$ on the group $\Gamma$ is a normalized,
$U(1)$-valued group 2-cocycle on $\Gamma$. Its Dixmier-Douady
invariant $\delta_\Gamma(\sigma)\in\H^3(\Gamma,\zed)$ is induced in
the usual way via the short exact sequence of coefficients in
(\ref{gpexactseq}). Given $\sigma$ we consider the reduced twisted
group $C^*$-algebra $C_r^*(\Gamma,\sigma)$. Then $(C_0(T^*B\Gamma)\,,\,
C_r^*(\Gamma, \sigma))$ is a strong PD pair for every multiplier
$\sigma$ on $\Gamma$ with trivial Dixmier-Douady invariant (we have
used the fact that the Baum-Connes conjecture holds for $\Gamma$ in
this case). In particular, this holds whenever $\Gamma$ is:
\begin{itemize}
\item A torsion-free, discrete subgroup of $SO(n,1)$ or of $SU(n,1)$;
  or
\item A torsion-free, amenable group.
\end{itemize}
If moreover $B\Gamma$ is spin$^c$, then $C_0(T^*B\Gamma)$ is a strong
PD algebra and hence $C_r^*(\Gamma, \sigma)$ is a strong PD algebra
for every multiplier $\sigma$ on $\Gamma$ with trivial Dixmier-Douady
invariant. In particular, we conclude that the noncommutative
torus $\mathcal{T}^{2n}_\theta$ is a strong PD algebra, and more generally the
noncommutative higher genus Riemann surface $\mathcal{R}_\theta^g$ is a strong
PD algebra. We will consider these latter examples in more detail in
Section~\ref{noncommutative-surfaces}.
\label{strongPDEx2}\end{example}

\subsection{Duality Groups\label{DualityGroup}}

We will now determine how many fundamental classes a given strong PD
pair admits.
\begin{proposition}\label{kk:fund1}
Let $(\alg,\balg)$ be a strong PD pair, and let $\Delta \in
\K^d(\alg\otimes\balg)$ be a fundamental class with inverse
$\Delta^\vee  \in \K_{-d}(\alg\otimes\balg)$. Let $\ell \in
\KK_0(\alg, \alg)$ be an invertible element. Then
$\ell\otimes_\alg\Delta \in \K^d(\alg\otimes\balg)$ is another
fundamental class, with inverse $\Delta^\vee \otimes_\alg\ell^{-1}
\in \K_{-d}(\alg\otimes\balg)$.
 \end{proposition}
\begin{proof} Using associativity of the Kasparov product along with
  the flip isomorphism $\alg\otimes\balg\to\balg\otimes\alg$, we
  compute
\bea
 \left(\Delta^\vee \otimes_\alg\ell^{-1} \right) \otimes_\alg
 \left(\ell\otimes_\alg\Delta\right)
 & = &\Delta^\vee \otimes_\alg  \left(\ell^{-1}  \otimes_\alg \ell
 \right)\otimes_\alg\Delta \nonumber
 \\ & = &\Delta^\vee \otimes_\alg  (1_\alg \otimes_\alg\Delta) 
\nonumber\\
  & = &\Delta^\vee \otimes_\alg  \Delta \nonumber\\ &=&
(-1)^d\,1_{\balg} \ .
\eea
The calculation in the other direction is similar, but slightly trickier
because of notational quirks in the way the Kasparov product
is written.  The calculation goes as follows:
\bea
\left(\Delta^\vee \otimes_\alg\ell^{-1} \right)\otimes_{\balg} 
  \left(\ell\otimes_\alg\Delta\right) & = &
\ell \otimes_\alg \left(\Delta^\vee \otimes_{\balg} \Delta\right)
\otimes_\alg \ell^{-1} \nonumber\\ &=&\ell \otimes_\alg 1_\alg  \otimes_\alg 
\ell^{-1} \nonumber\\ &=& \ell \otimes_\alg \ell^{-1} \nonumber\\
&=&  1_\alg \ .
\eea
It would appear that we needed to reorder many
of the factors in the product, but in fact,
all we are really using is the associativity of the product, in the
form discussed in Appendix B.  In terms of the diagram calculus
discussed there, we simply need to consider the diagram
depicted in Figure \ref{fig:ellDelta}.
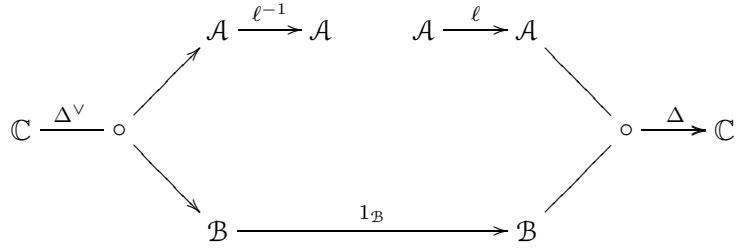
\begin{figure}[ht]
\[
\xymatrix{&&\alg \ar[r]^{\ell^{-1}} &\alg & \alg \ar[r]^\ell &\alg
\ar '[dr] [drr]^\Delta & & \\
\bbC \ar '[r]^{\Delta^\vee} [rru] \ar '[r] [rrd]&\circ&&&&&\circ& \bbC \\
&&\balg \ar[rrr]^{1_\balg}&& &\balg \ar '[ur] [urr]& & }
\]
\caption{Diagram representing the proof of Proposition \ref{kk:fund1}}
\label{fig:ellDelta}
\end{figure}

\end{proof}
\noindent
This result has a converse.
\begin{proposition}\label{kk:fund2}
Let $(\alg,\balg)$ be a strong PD pair, and let
$\Delta_1, \Delta_2 \in \K^d(\alg\otimes\balg)$ be fundamental 
classes with inverses $\Delta_1^\vee, \Delta_2^\vee  \in
\K_{-d}(\alg\otimes\balg)$. Then
$\Delta_1^\vee\otimes_{\balg}\Delta^{\phantom{\vee}}_2$ is an
invertible element in $\KK_0(\alg, \alg)$, with inverse given by
$(-1)^d\,\Delta_2^\vee \otimes_{\balg} \Delta^{\phantom{\vee}}_1 \in
\KK_0(\alg, \alg)$.
\end{proposition}
\begin{proof} 
As above we compute
\bea
  \left(\Delta_1^\vee\otimes_{\balg}\Delta^{\phantom{\vee}}_2
\right) \otimes_{\alg} 
  \left(\Delta_2^\vee \otimes_{\balg} \Delta^{\phantom{\vee}}_1 \right)
& =&\Delta_1^\vee \otimes_{\balg} \left(\Delta_2^\vee\otimes_{\alg}
\Delta^{\phantom{\vee}}_2\right) \otimes_{\balg}
\Delta^{\phantom{\vee}}_1 \nonumber\\
& = &\Delta_1^\vee\otimes_{\balg} (-1)^d\,1_{\balg} \otimes_{\balg}
\Delta^{\phantom{\vee}}_1 \nonumber \\ &=& (-1)^d\,
\Delta_1^\vee\otimes_{\balg}
\Delta^{\phantom{\vee}}_1 \nonumber \\
 & =& (-1)^d\,1_{\alg} \ ,
\eea
and similarly with $\Delta_1$ and $\Delta_2$ interchanged. 
\end{proof}

As an immediate consequence of Propositions \ref{kk:fund1} and
\ref{kk:fund2} above, we deduce the following.
\begin{corollary}
\label{cor:moduliforpair}
Let $(\alg,\balg)$ be a strong PD pair. Then the moduli space of
fundamental classes for $(\alg,\balg)$ is isomorphic to the group of
invertible elements in the ring $\KK_0(\alg, \alg)$.
\end{corollary}
\noindent
We now make some clarifying comments concerning the corollary
above. In the physics literature, $\K$-orientations
on a smooth manifold $X$ 
are generally linked to spin$^c$ structures, which (once an orientation
has been fixed) are an affine space modeled on
the second integral cohomology of $X$, $\H^2(X, \bbZ)$,
i.e., the isomorphism classes of line bundles over $X$.
However, the space of $\K$-orientations of $X$ is in general 
much larger, being a principal homogeneous space for the abelian 
group of units in the ring
$\K^\bullet(X)$, consisting of stable isomorphism classes of virtual vector
bundles over $X$ of virtual rank equal to $1$, and with group operation
given by tensor product. The space of all fundamental classes
of $X$  in $\KK$-theory is in general still larger,  being a principal
homogeneous space for the abelian group of units of $\KK_0(C(X), C(X))$,
which in turn by  \cite{RS87} is an extension of $\Aut \K^\bullet(X)$ by
$\Ext_\bbZ (\K^\bullet(X), \K^{\bullet+1}(X))$.

Recall that in the situation of Corollary \ref{cor:moduliforpair}
there is an isomorphism $\KK_0(\alg,
\alg)\cong\KK_0(\balg,\balg)$. This moduli space is called the {\it
  duality group} of the pair $(\alg,\balg)$ and is denoted
$\KK_0(\alg, \alg)^{-1}$. It can be computed explicitly using
(\ref{eqn:op}) from either the $\K$-theory or the $\K$-homology of the
algebra $\alg\otimes\balg$. We will now describe two illustrative and
broad classes of examples of strong PD algebras.

\subsection{Spectral Triples\label{NCspin}}

There is a natural object that encodes the geometry of a D-brane whose
construction can be motivated by the observation that bounded
Kasparov modules are not the most useful ones in practical
applications of $\K\K$-theory, especially when it comes to defining the
Kasparov product (\ref{product}) \cite{BJ}. A zeroth order elliptic
pseudodifferential operator $F$ on a smooth closed manifold $X$
determines a class in the $\K$-homology of $X$, i.e., a class in $\KK_\bullet
(C(X), \bbc)$. However, the product of two such operators need not be
a pseudodifferential operator. The Kasparov product is handled better
when one uses first order operators instead. But a first order
elliptic pseudodifferential operator $D$ will not in general extend to
a bounded operator on $L^2(X)$ and so will not generally provide a
class in $\KK_\bullet(C(X), \bbc)$. The trick here, due to Kasparov
and reformulated by Baaj and Julg \cite{BJ}, is to replace $D$ by the
operator $D\,(1+D^2)^{-1/2}$ which gives rise to a bounded Fredholm
module and so produces a class in $\KK_\bullet(C(X), \bbc)$. The
Baaj-Julg construction generalises to noncommutative $C^*$-algebras.

\begin{definition}\label{spectraltrip}
A {\it spectral triple} over a unital $C^*$-algebra $\alg$ is  a triple
$(\alg,\hil, D)$, where the algebra $\alg$ is represented faithfully
on a Hilbert space $\hil$ and $D$ is an unbounded self-adjoint
operator on $\hil$ with compact resolvent such that the commutator
$[D,a]$ is bounded for all $a \in\alg$. In the even case we assume
that the Hilbert space $\hil$ is $\bbz_2$-graded and that $D$ is an
odd operator with respect to this grading, i.e., there is an involution
$\gamma$ on $\hil$ which implements the grading and which anticommutes
with~$D$.
\end{definition}
\noindent
One can prove \cite{BJ} that all classes in the $\K$-homology of $\alg$
are obtained from such spectral triples, which in this context are
also called unbounded $\K$-cycles. Together with Connes'
axioms \cite{Connes:CMP} such a $\K$-cycle defines a noncommutative
spin$^c$ manifold. We will not enter into a detailed account of this
latter characterization but refer to \cite{Connes:CMP}
and \cite[Chapter~10]{GVF} for a thorough discussion. In the example
of a flat D-brane $(W,E,\phi)$ in the spacetime $X$, the spectral
triple is $(C(X),L^2(W,{\sf S}_W\otimes E),\Dirac_E)$ as specified in
the proof of Proposition~\ref{propflatDFred}. This definition can be
generalized to  provide unbounded $\alg$--$\balg$ bimodules, which
have the property that every element of $\KK_\bullet(\alg,\balg)$
arises from such an unbounded bimodule. 

Let $(\alg,\hil, D)$ be an unbounded spectral triple. This fixes our
noncommutative spacetime. Let $\alg^\op$ be the opposite algebra of
the algebra $\alg$. The action of the algebra $\alg\otimes\alg^\op$ on
the Hilbert space $\hil$ is described by using commuting actions of the
algebras $\alg$ and $\alg^\op$, making $\hil$ into a bimodule
over the algebra $\alg$. The action of $\alg$ is provided by the
representation of $\alg$ on $\hil$ which is part of the data given by
the spectral triple $(\alg,\hil,D)$. The algebra $\alg^\op$ is
assumed to act by means of operators $b^\op$ for $b\in\alg$. We assume
that the two actions commute, i.e., $[a,b^\op] = 0$ for all $a,
b\in\alg$, and that $[D,b^\op]$ is bounded for all $b\in\alg$.

\begin{definition}
The {\it index class} of the noncommutative spacetime $(\alg,\hil,D)$
is the class $\Delta_D\in\K^d(\alg\otimes\alg^\op)$ given by the data
$(\alg\otimes\alg^\op,\hil, D)$. If the index class is also a
fundamental class for $\alg$, then its inverse $\Delta_D^\vee\in
\K_{-d}(\alg\otimes\alg^\op)$ is called the {\it Bott class} of
$\alg$.
\label{indexclassdef}\end{definition}

One advantage of the spectral triple formulation is that under
suitable circumstances it enables the straightforward construction of
a {\it smooth} subalgebra $\alg^\infty$ of a $C^*$-algebra
$\alg$. This is a dense $*$-subalgebra of $\alg$ which is stable under
holomorphic functional calculus, i.e., if $a\in\alg^\infty$ with
$a=a^*$ and $a>0$, then $f(a)\in \alg^\infty$ for every {holomorphic}
function $f$ on a neighbourhood of the spectrum ${\rm Spec}(a)$. 
By the Karoubi density theorem (see for example  \cite{Connes-Thom}), 
the inclusion homomorphism 
$\iota : \alg^\infty \hookrightarrow \alg$ then induces an
isomorphism in $\K$-theory. Given a spectral triple $(\alg,\hil,D)$,
we assume that the smooth domain
\beq
\hil^\infty=\bigcap_{k\in\nat}\,{\rm Dom}\bigl(D^k\bigr)
\eeq
of the operator $D$ is an $\alg^\infty$-bimodule for some smooth
subalgebra $\alg^\infty\subset\alg$, or equivalently an
$\alg^\infty\otimes(\alg^\infty)^{\rm o}$-module. We further assume
that $\alg^\infty$ is a Fr\'echet algebra for the family of semi-norms
\beq
q_k(a)=\bigl\|\delta^k(a)\bigr\|
\eeq
provided by the derivation $\delta:a\mapsto[\,|D|,a]$ on $\alg$. The
usage of smooth subalgebras, and in particular topological algebras,
will be important later on when we start employing cyclic theory.

\begin{example}
Let $X$ be a compact spin$^c$ manifold of dimension $d$. Let
$\alg^\infty=C^\infty(X)$ be the Fr\'echet algebra of smooth functions on
$X$. It acts by pointwise multiplication on the Hilbert space 
$\hil=L^2(X,{\sf S}_X)$ of square integrable spinors on
$X$. This Hilbert space is $\bbz_2$-graded when $d$ is even, with the
usual grading operator $\gamma$ defining the split ${\sf
  S}^{\phantom{+}}_X={\sf S}_X^+\oplus{\sf S}_X^-$ into irreducible
half-spinor bundles, and ungraded in the odd case. For the operator
$D$ we take the usual spin$^c$ Dirac operator $D=\Dirac$
acting on $\hil$. Then $(\alg^\infty,\hil,\Dirac)$ defines a cycle
$[\Dirac]$ in $\K^d(\alg)$ \cite[Theorem~9.20]{GVF}. This data
determines the index class in
$\K^d(\alg\otimes\alg^\op)$. Because the algebra $\alg$ is
commutative in this case, one has $\alg=\alg^\op$ and the
multiplication map (\ref{multmap}) is an algebra homomorphism. Since
the $\K$-homology functor is contravariant, there is a map
\beq
m^*\,:\,\K^d(\alg) ~\longrightarrow~ \K^d(\alg\otimes\alg)
\eeq
induced by (\ref{multmap}). The image of the class $[\Dirac]$ under
this homomorphism is the index class
$\Delta_\Dirac$ \cite[p.~488]{GVF}.
\label{spincexstrongPD}\end{example}

\subsection{Twisted Group Algebra Completions of Surface
  Groups}\label{noncommutative-surfaces}

The noncommutative two-torus $\mathcal{T}^2_\theta$ provides the original
example of noncommutative Poincar\'{e} duality which was described by
Connes \cite{Connes:CMP} in the context of real spectral triples. It
is the specialization to genus one of the example that we present
here. Let $\Gamma_g$ be the fundamental group of a compact, oriented
Riemann surface $\Sigma_g$ of genus $g\ge 1$. It has the presentation
\beq
\Gamma_g = \Bigl\{\mbox{$U_j, V_j\,,\, j=1, \ldots, g~
\Big|~\prod\limits_{j=1}^g\,[U_j, V_j] = 1$}\Bigr\} \ ,
\eeq
and $B\Gamma_g=\Sigma_g$ is a smooth spin manifold. Since
$\H^2(\G_g, { U}(1)) \cong
\real/\zed$, for each $\theta\in [0, 1)$
we can identify a unique multiplier $\sigma_\theta$ on $\G_g$ up to
isomorphism. Let $\complex(\Gamma_g,\sigma_\theta)$ be the
$\sigma_\theta$-twisted convolution algebra of finitely supported maps
$\G_g\to\complex$, which is spanned over $\complex$ by a set of formal
letters $\delta_\gamma$, $\gamma\in\G_g$ satisfying
$\delta_\gamma\,\delta_\mu=\sigma_\theta(\gamma,\mu)~\delta_{\gamma\,\mu}$.
Let $\|f\|$ denote the operator norm of the operator on
$\ell^2(\Gamma_g)$ given by left convolution with $f\in\G_g$. Then the
completion of $\complex(\Gamma_g,\sigma_\theta)$ with respect to this
norm is the reduced twisted group $C^*$-algebra $\twga$. It can also
be viewed as the $C^*$-algebra generated by unitaries $U_j$ and $V_j$
satisfying the commutation relation
\beq
\prod_{j=1}^g \,[U_j, V_j] = \exp (2\pi \ii\theta) \ .
\eeq
On $\twga$ there is a canonical trace $\tau$ defined by evaluation at
the identity element of $\Gamma_g$.

Let $D$ be the operator defined by
\beq
D \delta_\gamma = \ell(\gamma)\,\delta_\gamma
\eeq
where $\ell(\gamma)\in[0,\infty)$ is the word length of $\gamma\in\G_g$.
Let $\delta = {\rm ad}(D)$ denote the commutator $[D,-]$. Then
$\delta$ is an unbounded closed derivation on the reduced twisted 
group $C^*$-algebra $\twga$. Consider the smooth subalgebra
\beq
\cR^\infty(\G_g, \sigma_\theta) := \bigcap_{k\in \mathbb N}\,{\rm
  Dom}\bigl(\delta^k\bigr) \ .
\eeq
Since $\cR^\infty(\G_g, \sigma_\theta)$ contains
$\delta_\gamma\ \ \forall \gamma \in  \Gamma_g$, it contains
${\mathbb C}(\Gamma_g, \sigma_\theta)$. Hence it is dense in
$\twga$. Since $\cR^\infty(\G_g, \sigma_\theta)$ is defined as a domain of
derivations, it is closed under holomorphic functional calculus.
Because $\Gamma_g$ is a surface group, it
follows from a variant of a result by Jolissaint  \cite{Jol} that there exists
$k\in \mathbb N$ and a positive constant $C'$ such that for all $f\in
\mathbb C(\Gamma_g,  \sigma_\theta)$ one has the {\em Haagerup
  inequality}
\begin{equation}\label{Haagerup}
\|f\| \le C'\, \nu_k(f) \ ,
\end{equation}
where
\beq
\nu_k(f) = \Bigl(\,\sum\limits_{\gamma\in \Gamma_g}\,\bigl
(1+\ell(\gamma)\bigr)^{2k}\,\bigl|f(\gamma)\bigr|^2\,\Bigr)^{1/2}\ .
\label{seminormg}\eeq
Using this, it is routine to show that $\cR^\infty(\G_g,
\sigma_\theta)$ is a Fr\'echet algebra, complete in the semi-norms
(\ref{seminormg}) induced by $\delta^k$.

To define the fundamental class of the noncommutative Riemann surface
$\cR_\theta^g:=\cR^\infty(\G_g, \sigma_\theta)$, we recall from \S 8 \cite{CHMM},
the $\K$-theory of $\twga$. For any $\theta$, the
multiplier $\sigma_\theta$ has trivial Dixmier-Douady invariant, as
$\delta_{\Gamma_g}(\sigma_\theta)\in\H^3(\Sigma_g,\zed)=0$, and so
(via the Baum-Connes assembly map)
$\K_0(\twga) \cong\K^0(\Sigma_g)= \mathbb{Z}^2$. For irrational $\theta$,
the algebras $\twga$ are distinguished for different values of
$\theta$ by the image of the trace map induced on $\K$-theory by the
trace $\tau$. In a basis $e_0$, $e_1$ of $\K_0(\twga)$ the trace is
given by
\beq
\tau(n\,e_0 + m\,e_1) = n + m\,\theta \ .
\eeq
We choose $e_0 = [1]$, the class of the identity element, and $e_1$
such that $\tau(e_1) = \theta$. Another result in \S8 \cite{CHMM} is
$\K_1(\twga) \cong \K^1(\Sigma_g)=\mathbb{Z}^{2g}$. Moreover, the
unitaries $U_j$ and $V_j$ form a basis for
$\K_1(\twga)$, \S6 \cite{CM}. Then the inverse fundamental class of $\twga$
is given by 
\beq
\Delta^\vee= e^{\phantom{\op}}_0\otimes e^\op_1 - e^{\phantom{\op}}_1
\otimes e^\op_0 + \sum_{j=1}^g\,\left(U_j^{\phantom{\op}}
\otimes V^\op_j - V_j^{\phantom{\op}}\otimes U^\op_j\right) \ .
\label{BottRiemanng}\eeq
The trace $\tau$ also leads to an inner product on $\cR_\theta^g$
defined by $(a,b)=\tau(b^*\,a)$ for $a,b\in\cR_\theta^g$. Let
$L^2(\cR_\theta^g)$ denote the completion of $\cR_\theta^g$ with respect
to this inner product, and define $\hil:=L^2(\cR_\theta^g)\oplus
L^2(\cR_\theta^g)$. Then the element (\ref{BottRiemanng}) is the Bott
class of the spectral triple $(\cR_\theta^g,\hil,D)$, with
$\cR_\theta^g$ acting diagonally on $\hil$ by left multiplication
and $D$ odd with respect to the canonical $\zed_2$-grading $\gamma$ on
$\hil$.

\subsection{Other Notions of Poincar\'e Duality\label{PoincareOther}}

We now return to the general theory and introduce some alternative
weaker forms of the duality described in Section~\ref{PDKK} above, all
of which imply the fundamental property (\ref{fundpropPoincare}). We
start with a ``pointwise'' version of Definition~\ref{dual1}.

\begin{definition}[{\bf Weak Poincar\'e Duality}]\label{dual2}
A pair of separable $C^*$-algebras $(\alg, \balg)$ is said to be 
a {\em weak Poincar\'e duality pair} ({\it weak PD pair} for short) if
there exists a class
$\Delta \in \KK_d(\alg\otimes\balg, \bbc)= \K^d(\alg\otimes\balg)$ in
the $\K$-homology of $\alg\otimes\balg$ and a class $\Delta^\vee\in
\KK_{-d}(\bbc,\alg\otimes\balg)=\K_{-d}(\alg\otimes\balg)$ in the
$\K$-theory of $\alg\otimes\balg$ with the properties 
\beq
\left(\Delta^\vee\otimes _\balg\Delta
\right) \otimes_\alg x= x \qquad \forall \,
x \in \KK_0(\bbc,\alg)
\eeq
and
\beq
\left(\Delta^\vee\otimes_\alg\Delta\right) \otimes_\balg y= (-1)^d\,y
 \qquad \forall \,y \in \KK_0(\bbc,\balg) \ .
\eeq
A separable $C^*$-algebra $\alg$ is said to be a {\it weak
  Poincar\'{e} duality algebra} ({\it weak PD algebra} for short) 
if $(\alg, \alg^\op)$ is a weak PD pair.
\end{definition}

\begin{example}
Let $\Gamma$ be a torsion-free, discrete group having the
Dirac-dual Dirac property such that $B\Gamma$ is a smooth oriented
manifold. Then $(C_0(T^*B\Gamma)\,,\, C_r^*(\Gamma, \sigma))$ is a weak PD
pair for every multiplier $\sigma$ on $\Gamma$ with trivial
Dixmier-Douady invariant. If moreover $B\Gamma$ is spin$^c$, then
$C_0(T^*B\Gamma)$ is a weak PD algebra and so $C_r^*(\Gamma,
\sigma)$ is a weak PD algebra for every multiplier $\sigma$ on
$\Gamma$ with trivial Dixmier-Douady invariant. In particular, this
holds whenever $\Gamma$ is:
\begin{itemize}
\item A torsion-free, word hyperbolic group;
\item A torsion-free, cocompact lattice in a product of a finite
  number of groups among Lie or $p$-adic groups of rank one, or in
  $SL_3(\mathbb F)$ with $\mathbb F$ a local field, $\mathbb H$ or
  $E_{6(-26)}$; or
\item A torsion-free lattice in a reductive Lie group or in reductive
  groups over non-archimedean local fields.
\end{itemize}
\end{example}

The duality of Definition~\ref{dual2} can also be weakened to hold
only modulo torsion elements of the $\K_0$-groups of the algebras
involved.

\begin{definition}[{\bf Rational Poincar\'e Duality}]\label{dual3}
A pair of separable $C^*$-algebras $(\alg, \balg)$ is said to be 
a {\em rational Poincar\'e duality pair} ({\it $\rat$--PD pair} for
short) if there exists a class
$\Delta \in \KK_d(\alg\otimes\balg, \bbc)= \K^d(\alg\otimes\balg)$ in
the $\K$-homology of $\alg\otimes\balg$ and a class $\Delta^\vee\in
\KK_{-d}(\bbc,\alg\otimes\balg)=\K_{-d}(\alg\otimes\balg)$ in the
$\K$-theory of $\alg\otimes\balg$ with the properties 
\beq
\left(\Delta^\vee\otimes _\balg\Delta\right)
\otimes_\alg x= x \qquad \forall \,
x \in \KK_0(\bbc,\alg)\otimes \rat
\eeq
and
\beq
\left(\Delta^\vee\otimes _\alg\Delta\right) \otimes_\balg y= (-1)^d\,y
 \qquad \forall \, y \in \KK_0(\bbc,\balg)\otimes \rat \ .
\eeq
A separable $C^*$-algebra $\alg$ is said to be a {\it rational
  Poincar\'{e} duality algebra} ({\it $\rat$--PD algebra} for short) 
if $(\alg, \alg^\op)$ is a $\rat$--PD pair.
\end{definition}

\begin{example}
Let $X$ be an oriented rational homology manifold, such as the
quotient of a manifold by an orientation-preserving action of a finite
group. Then $C_0(X)$ is a $\rat$--PD algebra.
\end{example}

\begin{example}
Let $\Gamma$ be a discrete group with the Dirac-dual Dirac property and
with a torsion-free subgroup $\Gamma_0$ of finite index such that
$B\Gamma_0$ is a smooth oriented manifold. Then
$(C_0(T^*B\Gamma)\,,\, C_r^*(\Gamma, \sigma))$ is a $\rat$--PD pair for
every multiplier $\sigma$ on $\Gamma$. (Note that the Dixmier-Douady
invariant in this case is always torsion.) If moreover $B\Gamma_0$ is
spin$^c$, then $C_0(T^*B\Gamma)$ 
is a $\rat$--PD algebra and hence $C_r^*(\Gamma,
\sigma)$ is a $\rat$--PD algebra for every multiplier $\sigma$ on
$\Gamma$. In particular, this holds whenever $\Gamma$ is:
\begin{itemize}
\item A word hyperbolic group; or
\item A cocompact lattice in a product of a finite number of
groups among Lie or $p$-adic groups of rank one, or in $SL_3(\mathbb
F)$ with $\mathbb F$ a local field, $\mathbb H$ or $E_{6(-26)}$.
\end{itemize}
\end{example}

Finally, we can take the fundamental property (\ref{fundpropPoincare})
itself as the weakest form of the duality.
\begin{definition}[{\bf Poincar\'e Duality}]\label{dual4}
A pair of separable $C^*$-algebras $(\alg, \balg)$ is said to be 
a {\em Poincar\'e duality pair} ({\it PD pair} for short) 
if there exist isomorphisms
\beq
\K_i(\alg) \cong \K^{i+d}(\balg)\qquad{\rm and}\qquad \K_i(\balg)\cong
\K^{i-d}(\alg) \ .
\eeq
A separable $C^*$-algebra $\alg$ is said to be a {\it 
Poincar\'e duality algebra} ({\it PD algebra} for short) 
if $(\alg, \alg^\op)$ is a PD pair.
\end{definition}

\begin{remark}
\label{rem:Kequivop}
For any $C^*$-algebra $\alg$, the $\K$-theory groups of $\alg$ and
$\alg^\op$ are isomorphic. This follows easily from the fact that if $p$ is
a projection in $\alg$ and $p^\op$ is the corresponding element of
$\alg^\op$, then $p^\op$ is also a projection. Similarly for a 
unitary $u\in \alg$, the corresponding element $u^\op$ of $\alg^\op$ is
also unitary.

In fact, there is a more general statement, which we will need later.  
The additive category
$\underline{\KK}$ with separable $C^*$-algebras as objects and with
$\KK_\bullet(\alg,\balg)$ as the morphisms from $\alg$ to $\balg$ may be
viewed as a certain  completion of the stable homotopy category of
separable $C^*$-algebras  \cite[\S 22]{Black}. As such, it has an
involution $^\op$ induced by the involution $f\mapsto f^\op$ sending
$f:\alg \to \balg$ to the $*$-homomorphism $f^\op:\alg^\op \to
\balg^\op$ which sends $a^\op$ to $(f(a))^\op$. The above isomorphism
from $\K_\bullet(\alg)$ to $\K_\bullet(\alg^\op)$ is simply this
involution $\KK_\bullet(\bbC, \alg) \cong \KK_\bullet(\bbC^\op=\bbC,
\alg^\op)$. 
\end{remark} 

\newsection{$\K\K$-Equivalence\label{Sect:KKEq}}

In this section we introduce the notion of $\KK$-equivalence and
describe its intimate connection to Poincar\'e duality and Morita
equivalence for $C^*$-algebras. As in the previous section, we will
describe various criteria for the equivalence and describe several
natural inequivalent versions of it, giving illustrative commutative
and noncommutative examples.

\subsection{Strong $\K\K$-Equivalence\label{KKEq}}

Our first notion of equivalence in $\K\K$-theory is a generalization
of the standard definition that was essentially already described in
Section~\ref{KK-Theory}.

\begin{definition}[{\bf Strong $\KK$-Equivalence}]\label{kk-equiv1}
A pair of separable $C^*$-algebras $(\alg, \balg)$
are said to be {\em strongly $\KK$-equivalent} if there
are elements 
\beq
\alpha \in \KK_n(\alg, \balg) \qquad\mbox{and}\qquad
\beta \in \KK_{-n}(\balg, \alg)
\eeq
such that 
\beq
\alpha \otimes_\balg \beta ~=~ 1_\alg \in \KK_0(\alg, \alg)
\qquad\mbox{and} \qquad \beta \otimes_\alg \alpha ~=~ 1_\balg \in
\KK_0(\balg, \balg) \ .
\eeq
\end{definition}
\noindent
The significance of this definition stems from the following results.

\begin{lemma}\label{lemma:kk-conseq}
Suppose that the pair of separable $C^*$-algebras $(\alg, \balg)$
are strongly $\KK$-equivalent. Then the maps
\bea\label{eqn:KK1}
\alpha \otimes_\balg \,:\, \K^i(\balg) ~\longrightarrow~ \K^{i+n}(\alg)
\quad&,& \quad \beta \otimes_\alg \,:\,  \K^i(\alg)
~\longrightarrow~\K^{i-n}(\balg) \ , \\
\label{eqn:KK2}
\otimes_\balg \alpha  \,:\, \K_i(\balg) ~\longrightarrow~\K_{i+n}(\alg)
\quad &,& \quad \otimes_\alg \beta \,:\,  \K_i(\alg) ~\longrightarrow~
\K_{i-n}(\balg)
\eeq
are all isomorphisms.
\label{KKisolemma}\end{lemma}

\begin{proof}
By the associativity property of the Kasparov product, the maps in
\eqref{eqn:KK1} satisfy
\bea
\beta \otimes_\alg (\alpha \otimes_\balg x)  &=& (\beta \otimes_\alg
\alpha) \otimes_\balg x\nonumber\\ &=& 1_\balg  \otimes_\balg x
\nonumber\\ & =& x \qquad \forall\, x\in \K^i(\balg) \ , \nonumber\\[4pt]
\alpha \otimes_\balg (\beta \otimes_\alg y) & =& 
(\alpha \otimes_\balg \beta) \otimes_\alg y\nonumber\\
& =& 1_\alg \otimes_\alg y\nonumber\\ & =& y \qquad \forall\, y\in
\K^i(\alg) \ ,
\eea
and are therefore all isomorphisms. Again by the associativity
property of the Kasparov product, the maps in \eqref{eqn:KK2} satisfy
\bea
(z \otimes_\balg \alpha) \otimes_\alg \beta &  =&
  z \otimes_\balg (\alpha \otimes_\alg \beta) \nonumber\\ &  = &
z \otimes_\balg 1_\balg \nonumber\\ & =& z \qquad \forall\, z\in
\K_i(\balg) \ , \nonumber\\[4pt]
(w \otimes_\alg \beta) \otimes_\balg \alpha &=&
 w \otimes_\alg (\beta \otimes_\balg \alpha)\nonumber\\ & =&
 w \otimes_\alg 1_\alg\nonumber\\ & =& w \qquad \forall\, w\in
 \K_i(\alg) \ ,
\eea
and thus are also isomorphisms. 
\end{proof}

\begin{remark}
As in (\ref{eqn:PD}), one can generalize the isomorphisms of
Lemma~\ref{KKisolemma} above to arbitrary coefficients. For any pair
of separable $C^*$-algebras $(\cC,\cD)$, the maps
\bea
\alpha \otimes_\balg \,:\,\KK_i(\cC\otimes\balg,\cD) &
\longrightarrow&\KK_{i+n}(\cC\otimes\alg,\cD) \ , \nonumber\\
\beta \otimes_\alg \,:\,  \KK_i(\cC\otimes\alg,\cD)
&\longrightarrow&\KK_{i-n}(\cC\otimes\balg,\cD) \ , \nonumber\\
\otimes_\balg \alpha  \,:\,  \KK_i(\cC,\balg\otimes\cD) &
\longrightarrow& \KK_{i+n}(\cC,\alg\otimes\cD)
\ , \nonumber\\ \otimes_\alg \beta \,:\,  \KK_i(\cC,\alg
\otimes\cD) &\longrightarrow&
\KK_{i-n}(\cC,\balg\otimes\cD)
\label{eqn:KKgencoeff}\eea
are all isomorphisms.
\end{remark}

\begin{lemma}\label{lemma:kk1}
Suppose that the pair of separable $C^*$-algebras $(\alg, \balg)$
are strongly $\KK$-equivalent. Then $\alg$ is a PD algebra
\textup{(}resp., strong PD algebra, weak PD algebra\textup{)} if and only if 
$\balg$ is a PD algebra \textup{(}resp., strong PD algebra, weak PD
algebra\textup{)}.  
\end{lemma}
\begin{proof}
If $\alg$ is a PD algebra, then there are isomorphisms $\K_i(\alg)
\cong \K^{i+d}(\alg)$. Combining this with the isomorphisms given in
Lemma~\ref{lemma:kk-conseq}, we deduce that there are isomorphisms
$\K_i(\balg) \cong \K^{i+d}(\balg)$, showing that $\balg$ is also a PD
algebra. The argument is symmetric, proving the result.
\end{proof}

We will now investigate some circumstances under which
$\KK$-equivalence holds. Let $\alg$ be a $C^*$-algebra, and let $\hil$
be a Hilbert $\alg$-module. Recall from Section~\ref{KK-Theory}
that the norm closure of the linear span of the set $\left\{\langle
  \zeta\mid \zeta'\,\rangle~|~ \zeta,\zeta' \in \hil \right\}$ is
the algebra $\cK(\hil)$ of compact operators on $\hil$. The module
$\hil$ is said to be {\em full} if $\cK(\hil)$ is equal to $\alg$.

\begin{definition}[{\bf Strong Morita equivalence}]\label{Morita-equiv}
Two $C^*$-algebras $\alg$ and $\balg$ are said to be {\it strongly
  Morita equivalent} if there is a full Hilbert $\alg$-module $\hil$
such that $\cK(\hil) \cong \balg$.
\end{definition}
\noindent
Upon identifying $\balg$ with $\cK(\hil)$, we define a Hilbert $(\alg,
\balg)$-bimodule $\hil^\vee$ as follows. As sets
(or real vector spaces), one has $\hil^\vee 
= \hil$. Let $\zeta\mapsto \zeta^\vee$ denote the identity map on
$\hil\to\hil^\vee$. Since $\lambda\,\zeta^\vee=
(\,\overline{\lambda}\,\zeta)^\vee$ for $\lambda\in\complex$ and
$\zeta\in\hil$, it follows that the identity map is 
conjugate linear. For $\zeta_1, \zeta_2 \in \hil^\vee$, $a \in \alg$ and
$b\in \balg$ we set
\beq
a \,\zeta_1^\vee = (\zeta^{\phantom{\vee}}_1a^*)^\vee \ , \quad 
\zeta_1^\vee\,b = (b^*\,\zeta^{\phantom{\vee}}_1\!)^\vee \ , \quad
\mbox{and}\quad
\langle \zeta_1^\vee, \zeta_2^\vee\,\rangle=
\zeta^{\phantom{\vee}}_1\langle \zeta^{\phantom{\vee}}_2,
-\rangle \ .
\eeq
Then $\hil^\vee$ is a Hilbert $\balg$-module which is full by
definition. Moreover, the map $\zeta_1^\vee\,\langle \zeta_2^\vee,
-\rangle$ $ \mapsto \langle \zeta^{\phantom{\vee}}_1,
\zeta^{\phantom{\vee}}_2\rangle$ identifies $\cK(\hil^\vee\,)$ with
$\alg$. From the point of view of the present paper, the importance of
this notion stems from the fact that Morita equivalent algebras encode
the same physics. The following well-known lemma relates strong Morita
equivalence to strong $\KK$-equivalence.

\begin{lemma}\label{lemma:kk-examples}
Let $\alg$ and $\balg$ be separable $C^*$-algebras. 
If $\alg$ is strongly Morita equivalent to $\balg$,
then $\alg$ is strongly $\KK$-equivalent to $\balg$. 
\end{lemma}

\begin{proof}
Let $\hil$ be a full Hilbert $\alg$-module implementing the Morita
equivalence between $\alg$ and $\balg$. Define elements $\beta \in
\KK_0(\balg, \alg)$ by the equivalence class of $(\hil, i, 0)$ where
$i : \balg \to \cK(\hil)$ is the identity, and $\alpha \in \KK_0(\alg,
\balg)$ by the equivalence class of $(\hil^\vee, i^\vee, 0)$ in the
notation above. Then the map $\zeta^{\phantom{\vee}}_1\!\otimes
\zeta_2^\vee \mapsto \zeta^{\phantom{\vee}}_1\langle
\zeta^{\phantom{\vee}}_2,-\rangle$ identifies the $\balg$-bimodule
$\hil\otimes_\alg \hil^\vee$ with $\balg$, and hence
$\beta \otimes_\alg \alpha = 1_\balg$. Similarly, the map 
$\zeta_1^\vee \otimes \zeta^{\phantom{\vee}}_2 \mapsto
\zeta_1^\vee\,\langle \zeta_2^\vee,-\rangle$ identifies the
$\alg$-bimodule $\hil^\vee \otimes_\balg \hil$ with $\alg$. Therefore
$\alpha \otimes_\balg \beta= 1_\alg$, proving that $(\alg, \balg)$ is
a strongly $\KK$-equivalent pair.
\end{proof}

There are many examples of strongly $\KK$-equivalent algebras that 
are {\em not} strongly Morita equivalent. For example, by \cite{RS87}
any two type I separable $C^*$-algebras with the same
$\K_0$ and $\K_1$ groups are automatically strongly $\KK$-equivalent.
Another famous example concerns the two-dimensional noncommutative
tori $\mathcal{T}^2_\theta$. We recall \cite{Rieffel,P-V}
that $\mathcal{T}^2_\theta$ is Morita equivalent to
$\mathcal{T}^2_{\theta'}$ if and only if $\theta$ and $\theta'$ belong
to the same ${GL}_2(\bbZ)$ orbit. On the other hand, the algebras
$\mathcal{T}^2_\theta$ and $C(\bbT^2)$ are strongly $\KK$-equivalent
for all $\theta$ \cite{P-V}.

The following lemma from \cite{Kas80} gives us more examples of
strongly $\KK$-equivalent algebras.

\begin{lemma}\label{lemma:kk-Thom}
The Thom isomorphism for an oriented real vector bundle $E\to X$ gives
a natural strong $\KK$-equivalence between
the algebra $C_0(E)$ of continuous functions on $E$ vanishing at
infinity and the algebra $C_0(X, {\sf Cliff}(E))$ of continuous
sections, vanishing at infinity, of the  Clifford algebra bundle $
{\sf Cliff}(E)$ of $E$.
\end{lemma}

\begin{remark}
If $E\to X$ is a spin$^c$ vector bundle then $\delta_X({\sf
  Cliff}(E))=0$, and the $C^*$-algebras $C_0({\sf
  Cliff}(E))$ and $C_0(X)$ are strongly Morita equivalent. In this
case the space of sections $C_0(X, {\sf Cliff}(E))$ can be replaced
by $C_0(X)$ in Lemma~\ref{lemma:kk-Thom}.
\end{remark}

\subsection{Other Notions of $\K\K$-Equivalence\label{KKEqOther}}

We now introduce variants of the concept of strong $\KK$-equivalence.

\begin{definition}[{\bf Weak $\KK$-Equivalence}]\label{kk-equiv2}
A pair of separable $C^*$-algebras $(\alg, \balg)$
are said to be {\em weakly $\KK$-equivalent} if there
are elements 
\beq
\alpha \in \KK_n(\alg, \balg) \qquad \mbox{and} \qquad
\beta \in \KK_{-n}(\balg, \alg)
\eeq
such that 
\beq
 (\alpha \otimes_\balg \beta) \otimes_\alg y  = y \quad \forall\, y\in
 \K^i(\alg), \quad  z \otimes_\balg (\alpha \otimes_\alg
 \beta)   =  z \quad \forall\, z\in \K_i(\balg)
\eeq
and
\beq
(\beta \otimes_\alg \alpha) \otimes_\balg x =  x \quad \forall\, x\in
\K^i(\balg), \quad w \otimes_\alg (\beta \otimes_\balg
\alpha) = w \quad \forall\, w\in \K_i(\alg) \ .
\eeq
\end{definition}

\begin{definition}[{\bf Rational $\KK$-Equivalence}]\label{kk-equiv3}
A pair of separable $C^*$-algebras $(\alg, \balg)$
are said to be {\em rationally $\KK$-equivalent} if there
are elements 
\beq
\alpha \in \KK_n(\alg, \balg) \qquad \mbox{and} \qquad
\beta \in \KK_{-n}(\balg, \alg)
\eeq
such that 
\beq
 (\alpha \otimes_\balg \beta) \otimes_\alg y  = y \quad \forall\, y\in
 \K^i(\alg)\otimes \rat, \quad  z \otimes_\balg (\alpha
 \otimes_\alg \beta)   =  z \quad \forall\, z\in
 \K_i(\balg)\otimes\rat \nonumber\\
\eeq
and 
\beq
(\beta \otimes_\alg \alpha) \otimes_\balg x =  x \quad \forall\, x\in
\K^i(\balg)\otimes\rat, \ w \otimes_\alg (\beta
\otimes_\balg \alpha) = w \quad \forall\, w\in \K_i(\alg)\otimes\rat \
. \nonumber\\
\eeq
\end{definition}

\begin{definition}[{\bf $\K$-Equivalence}]\label{kk-equiv4}
A pair of separable $C^*$-algebras $(\alg, \balg)$
are said to be {\em $\K$-equivalent} if there
are isomorphisms
\beq
\K_i(\alg) \cong \K_{i-n}(\balg)\qquad \mbox{and} \qquad
\K^i(\alg) \cong \K^{i-n}(\balg) \ .
\eeq
\end{definition}

With a proof along the lines of Lemma~\ref{lemma:kk1}, one can prove
the following.

\begin{lemma}\label{lemma:kk4}
Suppose that the pair of separable $C^*$-algebras $(\alg, \balg)$
are weakly $\KK$-equivalent \textup{(}resp., rationally $\KK$-equivalent,
$\K$-equivalent\textup{)}. Then $\alg$ is a weak PD algebra
\textup{(}resp., $\rat$--PD 
algebra, PD algebra\textup{)} if and only if $\balg$ is a weak PD algebra
\textup{(}resp., $\rat$--PD algebra, PD algebra\textup{)}.
\end{lemma}
\noindent
In the remainder of this section we will give some classes of examples
illustrating the various notions of $\KK$-equivalence introduced
above.

\subsection{Universal Coefficient Theorem}

To understand the relation between weak and\linebreak strong $\KK$-equivalence,
we appeal to the universal coefficient theorem of Rosenberg and
Schochet \cite{RS87}. It holds precisely for the class
$\underline{\mathfrak{N}}$ of $C^*$-algebras which are
$\KK$-equivalent to commutative $C^*$-algebras. For every pair of
$C^*$-algebras $(\alg, \balg)$  with $\alg
\in\underline{\mathfrak{N}}$ and $\balg$ separable, there is a split
short exact sequence of abelian groups given by
\beq
0 ~\to~{\Ext_\bbZ}\bigl(\K_{\bullet+1}(\alg) ,
\K_\bullet(\balg)\bigr) ~\to~ \KK_\bullet\bigl(\alg,
\balg\bigr)~\to ~ {\Hom_\bbZ}\bigl(\K_\bullet(\alg),
\K_\bullet(\balg) \bigr)~\to~ 0\nonumber\\
\label{UCTKK}\eeq 
Since there are many examples of $C^*$-algebras which are not in
$\underline{\mathfrak{N}}$ \cite{Ska88}, the notion of
$\K$-equivalence may strictly contain that of strong
$\KK$-equivalence. More precisely, suppose that $\alg$ is not in
$\underline{\mathfrak{N}}$. If $\alg$ satisfies the universal
coefficient theorem for one-variable $\K$-homology but not for
$\KK$-theory, then $\alg$ is $\K$-equivalent to a commutative
$C^*$-algebra but not strongly $\KK$-equivalent to such an algebra. (We
do not know if such algebras exist, but this is a possibility.)

\begin{remark}
\label{rem:KKequivop}
{}From Remark~\ref{rem:Kequivop} and the universal coefficient theorem
(\ref{UCTKK}) it follows that if $\alg$ lies in the category
$\underline{\mathfrak{N}}$ of $C^*$-algebras discussed above, then
$\alg$ and $\alg^\op$ are strongly $\KK$-equivalent. (It is easy to
construct examples, however, where they are not Morita equivalent.) We
are not sure if $\alg$ and $\alg^\op$ are always strongly
$\KK$-equivalent, without any hypotheses on $\alg$.
\end{remark}

\subsection{Deformations}

Let $\alg$ and $\balg$ be $C^*$-algebras. A {\it deformation} of
$\alg$ into $\balg$ is a continuous field of $C^*$-algebras over a
half-open interval $[0, \varepsilon)$, locally trivial over the open
interval $(0, \varepsilon)$, whose fibre over $0$ is $\alg$ and 
whose fibres over $\hbar\in(0,\varepsilon)$ are all isomorphic to
$\balg$.  A deformation gives rise to an extension of $C^*$-algebras
\beq
\label{eq:deformation}
0 ~\longrightarrow ~C_0\bigl((0, \varepsilon)\,,\, \balg\bigr) ~
\longrightarrow~ \calg ~\longrightarrow~ \alg~ 
\longrightarrow~ 0 \ .
\eeq
Connes and Higson observed that any deformation from $\alg$ into
$\balg$  defines a morphism $\K_\bullet(\alg) \to\K_\bullet(\balg)$,
which is simply the connecting homomorphism $\partial$ in the six-term exact
$\K$-theory sequence associated to \eqref{eq:deformation}. Moreover,
when $\alg$ is nuclear, the extension \eqref{eq:deformation} has a
completely positive cross-section and thus defines an element in
$\KK_0(\alg,\balg)$ which induces the map on $\K$-theory groups.

\subsection{Homotopy Equivalence}

Let $\alg$ and $\balg$ be $C^*$-algebras. Two algebra homomorphisms
$\phi_{0}, \phi_1 : \alg \rightarrow \balg$ are {\em homotopic} if
there is a path $ \gamma_t,t\in[0,1]$ of homomorphisms $\gamma_t:
\alg\rightarrow \balg$ such that $t\mapsto \gamma_t(a)$ is a norm
continuous path in $\balg$ for every $a\in \alg$ and such that
$\gamma_0 = \phi_0$, $ \gamma_1 =\phi_1$. The algebras $\alg$ and
$\balg$ are said to be {\em homotopy equivalent} if there exist
morphisms $\phi: \alg \rightarrow \balg $ and $\eta: \balg\rightarrow
\alg$ whose compositions  $\eta\circ \phi$ and $\phi\circ\eta$  are
homotopic to the identity maps on $\alg$ and $\balg$,
respectively. The algebra $\alg$ is called {\em contractible} if it is
homotopy equivalent to the zero algebra.

\begin{lemma}
If $\alg$ and $\balg$ are homotopy equivalent $C^*$-algebras, then the
pair $(\alg, \balg)$ are strongly $\KK$-equivalent.
\end{lemma}
\begin{proof}
Suppose that $\phi: \alg \rightarrow \balg $ and $\eta: \balg\rightarrow
\alg$ are $*$-homomorphisms which are homotopy inverses to one
another. Then they define classes $\KK(\phi)\in \KK_0(\alg, \balg )$
and $\KK(\eta)\in \KK_0(\balg, \alg )$ whose Kasparov products are
simply $\KK(\eta\circ \phi)\in \KK_0(\alg, \alg )$ and  $\KK(\phi\circ \eta)\in
\KK_0(\balg, \balg )$. Since $\eta\circ \phi$ is homotopic to $\Id_\alg$,
one has $\KK(\eta\circ \phi) = 1_\alg$ by homotopy invariance of
$\KK$-theory, and similarly $\KK(\phi\circ \eta) = 1_\balg$.
\end{proof}

\newsection{Cyclic Theory\label{Cyclic}}

As was crucial in the definition of D-brane charge given in
Section~\ref{FlatD}, the topological $\K$-theory of a spacetime
$X$ is connected to its cohomology through the rational isomorphism
provided by the Chern character
$\ch:\K^\bullet(X)\otimes\rat\to\H^\bullet(X,\rat)$. While this
works well in the case of flat D-branes, in the more general settings described
before we need a more general cohomological framework in which to
express the D-brane charge, particularly when the branes are described
by a noncommutative algebra $\alg$. The appropriate receptacle for the Chern
character in analytic $\K$-theory is a suitable version of the cyclic
cohomology of
the given algebra $\alg$. In this section we will present an overview
of the general aspects of cyclic homology and cohomology. As we
will see later on, this general formulation provides a nice
intrinsic definition of the D-brane charge even in the flat
commutative case, which moreover has a suitable extension to the
noncommutative situations.

\subsection{Formal Properties of Cyclic Homology Theories\label{CyclicHom}}

Cyclic cohomology of a complex algebra, from its introduction by
Connes, was developed in parallel with $\K$-theory as a  
noncommutative analogue of the de Rham cohomology of a differentiable
manifold. One of the main  
features of that theory, which made it a very useful tool to aid in
computations of $\K$-theory, is the  
existence of the Chern character from $\K$-theory of the algebra to
the  cyclic cohomology of the  
algebra. Compared to $\K$-theory, cyclic type homology theories exhibit
a major weakness: they are  
all defined using a suitably chosen deformation of the tensor
algebra. In the case of a topological algebra $\alg$,  there are many
ways to topologize the tensor algebra $T\alg$: \emph{this makes cyclic
type homology theories very sensitive to the choice of topology}. 

In this section we introduce the properties that we shall require of
the cohomology theory to make  
it suitable for our purposes. We shall then briefly outline the main
points in the construction of cyclic type  
homology theories that will satisfy those properties. 

Let $\alg$ and $\balg$ be topological algebras, whose topology will be
specified shortly. We shall denote  
by $\HE_i(\alg,\balg)$, $i=0,1$, a bivariant cyclic theory
associated with $\alg$ and $\balg$ that has the following  
formal properties: 
\begin{enumerate}
\item $\HE_i(\alg,\balg)$ is covariant in the second variable and
  contravariant in the first variable;  
\item For any three algebras $\alg$, $\balg$ and $\calg$ there is a
  natural composition product 
$$
\otimes_\balg:
\HE_i(\alg,\balg)\times \HE_j(\balg,\calg) \rightarrow
\HE_{i+j}(\alg,\calg) \ ;
$$
\item The functor $\HE_i(-,-)$ is split exact and satisfies excision
  in each variable;
\item $\HE_i(-,-)$  is homotopy invariant; 
\item\label{property:intprod}
For any algebras $\alg_1$, $\alg_2$, $\balg_1$, and $\balg_2$, there is an
exterior product
\[
\HE_i(\alg_1,\balg_1)\times \HE_j(\alg_2,\balg_2) \rightarrow 
\HE_{i+j}(\alg_1 \otimes \alg_2,\balg_1 \otimes \balg_2)
\]
compatible with the composition product; and
\item \label{property:Chern}
When $\alg$ and $\balg$ are
$C^*$-algebras, there is a natural transformation of functors, the
Chern character,  
$$
\ch: \KK_i(\alg,\balg) \rightarrow \HE_i(\alg,\balg)
$$
which is compatible with the product on $\KK$ and the composition
product on $\HE$.  
\end{enumerate}
Often we suppress the subscript $i$ when $i=0$. Axiom (2) ensures that
$\HE_\bullet(\alg,\alg)$ is a $\bbZ_2$-graded ring.
There are now many definitions of bivariant cyclic
theories of this kind, each suited to a specific category of 
algebras. (See  \cite{PuschDoc} for a survey of these theories,
as well as the relationships among them.)
With every choice of a class of algebras we need to specify the notions
of  homotopy and  
stability which are suitable for the given category. In many cases, 
for example when $\alg$ and $\balg$ are multiplicatively convex ($m$-convex) 
algebras, $\KK$ in 
property \eqref{property:Chern} needs to be replaced by a different
form of bivariant 
$\K$-theory, for example, Cuntz's $kk$  \cite{Cuntz:Documenta},
which is defined on a class 
of $m$-convex algebras. Cuntz's $kk$ is much easier to define than $\KK$,
but it is harder to compute, and 
at present, the precise relation between $kk$ and
$\KK$ is unclear. This is why
we are led to consider Puschnigg's \emph{local bivariant cyclic
  cohomology}, developed primarily in  \cite{PuschDoc}, which 
we shall denote $\HE_i(-,-)$.\footnote{Puschnigg calls it
${\sf HC}^{\text{loc}}$ or
${\sf HE}^{\text{loc}}$, but as this is a bit cumbersome,
 we have chosen to simplify the notation.}
This theory, which can be defined on a class of $C^*$-algebras which is
suitable for our purposes, 
is closest  to Kasparov's $\KK$-theory. 

Furthermore we need to point out that
the correct notion of tensor product in property
\eqref{property:intprod} 
depends on the theory.  When working with nuclear $C^*$-algebras, the usual
$C^*$-tensor product is appropriate, but when working with Fr\'echet algebras,
one might need the projective tensor product.
When we do not assume that $\alg$ and $\balg$ are topological
algebras, then the natural cyclic type homology theory to consider is the  
bivariant periodic cyclic homology $\HP_i(\alg,\balg)$ of Cuntz and
Quillen  \cite{CQ1}. This theory is closest  
to the cyclic homology and cohomology defined by Connes. We have that
$\HP_i(\alg,\complex)$ is  
the same as the periodic cyclic cohomology of $\alg$, $\HP^i(\alg)$,
while $\HP_i(\complex, \alg)$  
coincides with the periodic cyclic homology, $\HP_i(\alg)$. 

When the algebra $\alg$ is equipped with a topology, Meyer's work
indicates  \cite{Meyer} that in  
the construction of a reasonable cyclic type theory one should
consider \emph{bounded} rather than  
\emph{continuous} maps. More precisely, this means the following. As
is well known, a  map of topological vector spaces
$f: E\rightarrow F$ is bounded if and only if
it sends bounded sets in $E$ to bounded sets in $F$. If $E$ and 
$F$ are locally convex,  then a reasonable definition of a bounded set 
states that a subset of $E$ is bounded if and only if it is absorbed
by every open neighbourhood of  
zero. Since the choice of open neighbourhoods in this definition is
dictated by topology, the class of  
bounded sets in $E$ is fixed by the choice of topology. The class of
bounded sets in a topological space is called a \emph{bornology} $\fB$; 
the bornology associated with the topology of a space is called the 
\emph{von Neumann} bornology. A space equipped with a chosen family of
bounded sets is called a \emph{bornological space}. A bornology
on a vector space $E$ is a class $\fB$ of subsets of $E$, which have the 
properties that $\{e\}\in \fB$ for all $e\in E$;
if $S\in \fB$ and $T\subset S$, then $T\in \fB$ (a subset of a 
bounded set is bounded); if $S,T\in \fB$, then $S\cup T\in
\fB$ (the union of two bounded sets is bounded); 
$c\cdot S\in \fB$ if $S\in \fB$ and $c\in
\bbC$ (a scalar multiple of a bounded set is bounded); and
if $S,T\in \fB$ then $S+T\in \fB$ (vector addition in $E$ is a bounded map). 
A vector space equipped with a bornology is called a \emph{bornological 
vector space}.
A map $f: E\rightarrow F$ of bornological 
spaces is called \emph{bounded} if and only
if it sends elements of the bornology in $E$ (i.e., 
the \lq bounded' sets in $E$) to elements of the bornology in $F$. 

In the study of  bornological spaces it became clear that it is useful
to consider the choice of bornology 
to be independent from the choice of topology. This observation lies
at the basis of Meyer's construction 
of his analytic cyclic homology $\HA_i(-,-)$, which is a bivariant
functor defined on a class  of  
bornological algebras. A bornological vector space $\alg$ is a
bornological algebra if and only if  
it is equipped with a multiplication $m: \alg\times \alg\rightarrow
\alg$ which is bounded in the bornological sense. Meyer's analytic
theory is very flexible and can be used in a variety of contexts. For
example,  
it contains the Cuntz-Quillen bivariant cyclic theory $\HP_i(\alg,\balg)$ of
 \cite{CQ1}. Moreover, it can be defined 
for Fr\'{e}chet algebras and, in particular, for Banach
algebras. Meyer showed in his thesis 
that for a suitable choice of bornology on a locally convex algebra
$\alg$ his analytic cyclic cohomology  
$\HA_i(\alg, \bbc) = \HA^i(\alg)$ is isomorphic to Connes' entire
cyclic cohomology  
${\sf HE}^i(\alg)$  \cite[Thm.~3.47]{Meyer}. A very important example
of an entire cyclic cohomology class  
is given by the JLO cocycle, which provides the character of a
$\theta$-summable Fredholm module  \cite{JLO1}. 

\subsection{Local Cyclic Theory}
\label{BCCoh}
We shall now outline, in broad terms, the construction and main
properties of Puschnigg's local cyclic  
theory. For any algebra $\alg$, unital or not, the (non-unital) tensor
algebra $T\alg$ of $\alg$ is defined by 
$$
T\alg = \alg \oplus (\alg\otimes \alg)\oplus (\alg\otimes \alg\otimes
\alg)\oplus \dots 
$$
In applications, $\alg$ will be assumed to be complete with respect to
some additional structure. For example, $\alg$ may be a Fr\'{e}chet or
Banach algebra, or in the bornological case, $\alg$ will be assumed to
be a complete bornological algebra  \cite[2.2]{Meyer}. In each case the
definition of $T\alg$ will require  
a choice of a completed tensor product which is relevant to the given
situation. For example, for  
Banach algebras a reasonable choice is the completed projective tensor
product $\hat{\otimes}_\pi$  
while for bornological algebras we need to take the completed
bornological tensor product  
  \cite[2.2.3]{Meyer}. The tensor algebra is closely related to the algebra of 
\emph{noncommutative  differential forms} $\Omega \alg$, which 
is defined (as a vector space) by 
\[\Omega ^n\alg =  
\alg^{\otimes n+1} \oplus \alg^{\otimes 
 n}, \quad n>0.\]
We put $\Omega^0 \alg= \alg$.\footnote{Caution: In  \cite{PuschDoc},
Puschnigg forgets to mention this, i.e., to mention that the 
definition of $\Omega^n\alg$ is different when $n=0$.}
One defines a differential $\dd$ on $\Omega \alg$ of degree
$+1$, which for $n\geq 1$ is given by the two-by-two matrix
\[
\dd = \left(
\begin{array}{cc}
0 & 0\\
1 & 0
\end{array}
\right)\; ,
\]
while in degree zero we put $\dd= (0,1) : \alg \rightarrow \alg^{\otimes 2}
\oplus \alg$. One can then show that,
for $n > 0$, $\Omega^n\alg \cong \text{Span} 
\{\tilde{a}_0\,\dd a_1\,\dots \,\dd a_n\} $,
where $\tilde{a}_0 $ is an element of the unitization $\tilde{\alg}$
of $\alg$, and $a_i\in \alg$.  The multiplication on $\Omega\alg$
is uniquely determined by the requirements that $\dd$ be a derivation
(satisfying the Leibnitz rule) and that
\[
\begin{aligned}
(\dd a_1\,\dots \,\dd a_n)\cdot(\dd a_{n+1}\,\dots \,\dd a_{n+m})
&=\dd a_1\,\dots \,\dd a_{n+m}\,,\\
(\tilde a_0\,\dd a_1\,\dots \,\dd a_n)\cdot(\dd a_{n+1}\,\dots \,\dd a_{n+m})
&=\tilde a_0\,\dd a_1\,\dots \,\dd a_{n+m}\,.
\end{aligned}
\]

A key point in the construction of any cyclic type homology theory is
 the choice of a suitable completion (depending on whether $\alg$ is
 considered to be a topological or a bornological algebra)   
 of $\Omega \alg$. 
 To retain the important universal property of the tensor algebra,
 this completion is also usefully described as a deformation of the
 tensor algebra denoted $X(T\alg)$. This is a $\bbz_2$ graded
 complex defined very simply as follows:  
 $$
 \Omega^1(T\alg) / [\Omega^1 (T\alg), \Omega^1(T\alg)]
 \xrightarrow{\b}\Omega^0(T\alg) 
 $$
 where $[\Omega^1 (T\alg), \Omega^1(T\alg)]$, the commutator space of
 $\Omega^1(T\alg)$, is  
 spanned by the set of all commutators $[\omega, \eta]$ with $\omega,
 \eta\in \Omega^1(T\alg)$. The map 
 $\b$ is given by $\omega_0\,\dd\omega_1\mapsto [\omega_0, \omega_1]$ 
for any two 
 $\omega_0, \omega_1 \in T\alg$. There is a differential going the
 other way, which is  
 the composition of the differential $\dd: T\alg \rightarrow
 \Omega^1(T\alg)$ with the quotient map  
 $\Omega^1(T\alg) \rightarrow 
  \Omega^1(T\alg) / [\Omega^1 (T\alg), \Omega^1(T\alg)] 
  $. 
  
Let $\alg$ and $\balg$ be two complete bornological algebras and let
$X(T\alg)^c$ be the Puschnigg  
completion of the $X(T\alg)$ (see  \cite[\S5]{Puschnigg},
 \cite[\S23]{CuntzSurvey}).  
There is a $\bbz_2$-graded complex of bounded maps 
$\Hom_\bbc(X(T\alg)^c, X(T\balg)^c)$. We define the bivariant local
cyclic homology by 
$$
\HE_i(\alg,\balg) = \H_i(\Hom_\bbc(X(T\alg)^c, X(T\balg)^c))
$$
where $i=0,1$  \cite{CuntzSurvey}.  
This homology theory coincides with other theories discussed 
there under suitable conditions. For 
example, when $\balg$ is a Fr\'{e}chet algebra whose bornology is 
specified by the family of pre-compact 
sets (or is nuclear) then $\HA_\bullet(\balg) = \HE_\bullet(\balg)$
and there is a natural map  
$\HA_\bullet(\alg,\balg) \rightarrow \HE_\bullet(\alg,\balg)$. 

We recall the notion of a smooth subalgebra of a complete bornological
algebra.
\begin{definition}  \cite[23.3]{CuntzSurvey}. 
Let $\alg$ be a complete bornological algebra with bornology $\fB(\alg)$, 
which is 
a dense subalgebra of a Banach algebra $\balg$ with closed unit ball  $U$. 
Then $\alg$ is a \emph{smooth 
subalgebra} of $\balg$ if and only if for every element $S\in \fB (\alg)$ 
such that $S\subset r\,U$, for some $r< 1$, the set 
$S^\infty = \bigcup _n\, S^n $ is an element of $\fB(\alg)$. 
\end{definition}
\noindent
Smooth subalgebras of Banach algebras are closed under the 
holomorphic functional calculus. The following result will be
important to us.
\begin{theorem}  \cite[23.4]{CuntzSurvey}. 
\label{thm:approxprop}
Let $\balg$ be a Banach algebra with the metric approximation 
property. Let $\alg$ be a smooth 
subalgebra of $\balg$. Then $\alg$ and $\balg$ are $\HE$-equivalent, that
is the inclusion map $\alg 
\rightarrow \balg$ induces an invertible element of $\HE_0(\alg,\balg)$. 
\end{theorem}
\noindent
Note by  \cite{ChoiE} that all nuclear $C^*$-algebras have the metric
approximation property.  Some, but not all, non-nuclear $C^*$-algebras have 
it as well.

\begin{example}
Let $X$ be a compact manifold. Then the Fr\'{e}chet algebra $C^\infty(X)$ is 
a smooth subalgebra of the algebra $C(X)$ of continuous functions
on $X$. Furthermore, the inclusion $C^\infty(X) \hookrightarrow
C(X)$ is an invertible element in $\HE(C^\infty(X),C(X))$
by Theorem \ref{thm:approxprop} above. In particular,
both the local homology and cohomology of these two algebras are
isomorphic. Puschnigg also proves that $\HE_\bullet(C^\infty(X))\cong
\HP_\bullet(C^\infty(X))$, and so, in this case, Puschnigg's local 
cyclic theory coincides with the
standard periodic cyclic homology. The following fundamental result of 
Connes makes it possible
to establish contact between Puschnigg's local cyclic theory of $C(X)$ and the 
de Rham cohomology of $X$.
\end{example}
\begin{theorem}
For $X$ a compact manifold, the periodic cyclic homology 
$\HP_\bullet(C^\infty(X))$ is canonically isomorphic to the periodic
de~Rham cohomology:
\beq
\HP_0\bigl(C^\infty(X)\bigr)\cong\H_{\rm dR}^{\rm
even}\bigl(X\bigr) \qquad \mbox{and} \qquad
\HP_1\bigl(C^\infty(X)\bigr)\cong
\H_{\rm dR}^{\rm odd}\bigl(X\bigr) \ .
\label{HPHdR}\eeq
\label{HKRthm}\end{theorem}
\noindent
It is in the sense of this theorem that we may regard cyclic homology
as a generalization of de~Rham cohomology to other (possibly
noncommutative) settings. 

The local cyclic theory $\HE$ admits a Chern character with the
required properties.  
\begin{theorem}  \cite[23.5]{CuntzSurvey} Let $\alg$ and $\balg$ be
  separable $C^*$-algebras. Then  
there exists a natural bivariant Chern character 
$$
\ch: \KK_\bullet(\alg,\balg) \rightarrow \HE_\bullet(\alg,\balg)
$$
which has the following properties: 
\begin{enumerate}
\item $\ch$ is multiplicative, i.e., if
$\alpha\in \K\K_i(\alg,\balg)$ and $\beta\in \K\K_j(\balg,\calg)$ then
\beq
\ch(\alpha\otimes_\balg\beta) = \ch (\alpha) \otimes_\balg \ch(\beta) \ ; 
\eeq
\item $\ch$ is compatible with the exterior product; and
\item $\ch(\KK(\phi))=\HE(\phi)$ for any algebra homomorphism
  $\phi:\alg\to\balg$.
\end{enumerate}
The last property implies that the Chern character sends invertible
elements of $\K\K$-theory to invertible elements of bivariant local
cyclic cohomology. 

Moreover, 
if $\alg$ and $\balg$ are in the class $\underline{\mathfrak N}$
of $C^*$-algebras 
for which the Universal Coefficient Theorem 
holds in $\KK$-theory, then 
$$
\HE_\bullet(\alg,\balg)
\cong\Hom_\bbC(\K_\bullet(\alg)\otimes_{\bbz}\complex ,
\K_\bullet(\balg)\otimes_\bbz\complex).  
$$
If $\K_\bullet(\alg)$ is finitely generated, this is also equal to
$\KK_\bullet(\alg,\balg) \otimes_\bbz\complex$.
\end{theorem}

\newsection{Duality in Bivariant Cyclic
  Cohomology\label{PDinPCC}}

In this section we shall formulate and analyse Poincar\'e
duality in the context of bivariant cyclic cohomology of generic
noncommutative algebras. Our analysis of Poincar\'{e} duality in
$\KK$-theory from Section~\ref{Pairings} and of $\KK$-equivalence in
Section~\ref{Sect:KKEq} indicates that it is possible to 
define analogous notions in any bivariant theory that has the same
formal properties as $\KK$-theory. An important example of such a
situation is provided by the bivariant local cyclic theory as
introduced in Section~\ref{BCCoh}, where we have
the additional tool of the bivariant Chern
character from $\KK$. Rather than repeating all the details, we shall
simply state the main 
points. Duality in cyclic homology and periodic cyclic homology has
also been considered by Gorokhovsky \cite[\S5.2]{Gor99}.

\subsection{Poincar\'{e} Duality\label{PDHP}}

We will now develop the periodic cyclic theory analogues of the
versions of Poincar\'e duality introduced in Section~\ref{Pairings}.
Because we want to link everything with $\KK$ and not with $kk$
or its variants, we will work throughout with $\HE$ and not with
$\HP$, even though the latter is probably more familiar to most readers.

\begin{definition}\label{cyclicdual}
Two complete bornological algebras $\alg, \balg$ are a
{strong cyclic Poincar\'e dual pair} (strong C-PD pair for
short) if there exists a class $\Xi\in \HE_d(\alg\otimes\balg, \bbc)
= \HE_d(\alg\otimes\balg)$ in the local cyclic cohomology of
$\alg\otimes\balg$ and a class $\Xi^\vee\in
\HE_d(\bbc,\alg\otimes\balg)=\HE_d(\alg\otimes\balg)$ in the
local cyclic homology of $\alg\otimes\balg$ with the properties
$$
\label{cyclPDalgcond}
\qquad \Xi^\vee\otimes _\balg\Xi~=~
1_\alg\in \HE_0(\alg,\alg) \quad \mbox{and}\quad
\Xi^\vee\otimes _\alg\Xi~=~(-1)^d\,
1_\balg\in \HE_0(\balg,\balg) \ .
$$
The class $\Xi$ is called a {\it fundamental
  cyclic cohomology class} for the pair $(\alg,\balg)$ and $\Xi^\vee$
is called its {\it inverse}. A complete bornological algebra
$\alg$ is a {\it strong cyclic Poincar\'{e} duality algebra} ({\it
  strong C-PD algebra} for short) if $(\alg,\alg^\op)$ is a strong
C-PD pair.
\end{definition}
\noindent
As in the case of $\KK$-theory, these hypotheses establish an
isomorphism between the periodic cyclic homology and cohomology of the
algebras $\alg$ and $\balg$ as
\beq
\HE_\bullet(\alg)\cong\HE^{\bullet+d}(\balg) \qquad \mbox{and} \qquad
\HE_\bullet(\balg)\cong\HE^{\bullet+d}(\alg) \ .
\eeq
One also has the isomorphisms
\beq\label{eqn:opHP}
\HE_{\bullet+d}(\alg\otimes\balg) ~\cong~\HE_\bullet
(\alg, \alg)  ~\cong ~
\HE_\bullet(\balg, \balg) ~\cong~
\HE^{\bullet+d}(\alg\otimes\balg) \ .
\eeq
The moduli space of fundamental cyclic cohomology classes for the pair
$(\alg,\balg)$ is the {\it cyclic duality group}
$\HE_0(\alg,\alg)^{-1}$ of invertible elements of the ring
$\HE_0(\alg,\alg)\cong\HE_0(\balg,\balg)$. Similarly to
Section~\ref{PoincareOther}, one has the alternative notions of weak
C-PD pairs and of cyclic Poincar\'e duality.

\begin{example}
Let $\alg= C^\infty(X)$ be the algebra of smooth functions on a
compact oriented manifold $X$ of dimension $d$. Then the image of the
class $[\varphi^{~}_X]$ of the cyclic $d$-cocycle 
\beq
\varphi^{~}_X(f^0,f^1, \dots, f^d) = \frac1{d!}\,
\int\limits_X f^0~\dd f^1\wedge\dots\wedge \dd f^d
\label{varphiX}\eeq
for $f^i \in \alg$,
under the homomorphism
$m^*:\HP^\bullet(\alg)\cong
\HE^\bullet(C(X))\to\HE^\bullet(C(X)\otimes C(X))$ induced by the
product map (\ref{multmap}), is the fundamental class
$\Xi\in\HE^d(C(X)\otimes C(X))$ of $X$ in cyclic cohomology.
Thus in this case $\Xi$ corresponds to the orientation cycle $[X]$ and
our notion of Poincar\'{e} duality agrees 
with the classical one. More generally, if $X$ is non-orientable,
i.e., ${\sf w}_1(X)\neq0$, we choose the local coefficient system ${\sf
  C}_X\to X$ associated to ${\sf w}_1(X)$ whose fibres are each
(non-canonically) isomorphic to $\zed$. Then
$(C(X)\,,\,C(X,{\sf C}_X\otimes\complex))$ is a strong
C-PD pair.
\label{CPDcommex}\end{example}

Thanks to the existence of a universal, multiplicative Chern character
which maps the bivariant $\K\K$-theory to bivariant local cyclic
cohomology, we can show that Definitions~\ref{StrongPDdef}
and~\ref{cyclicdual} are compatible. Let $(\alg,\balg)$ be a
strong PD pair of algebras in $\KK$-theory with fundamental class
$\Delta\in\K^d(\alg\otimes\balg)$ and inverse
$\Delta^\vee\in\K_{-d}(\alg\otimes\balg)$. Then there is a 
commutative diagram
\beq
\xymatrix{
\K_\bullet(\alg)\ar[d]_{\ch}\ar[rr]^{\Delta \otimes_\alg}&&
 \K^{\bullet+d}(\balg)\ar[d]^{\ch} \\
\HE_\bullet(\alg)
\ar[rr]_(.45){\ch(\Delta)\otimes_{\alg}}&&
\HE^{\bullet+d}(\balg) \ . }
\label{ChernPDdiag}\eeq
Since the Chern character is a unital homomorphism
the cocycle $\ch(\Delta)$ is an invertible class in
$\HE^d(\alg\otimes\balg)$ with inverse
$\ch(\Delta^\vee)\in\HE_d(\alg\otimes\balg)$, and so it
establishes Poincar\'{e} duality in local cyclic cohomology,
i.e., Poincar\'e duality in $\KK$-theory implies Poincar\'e duality in
cyclic theory. However, the converse is not true, since the cyclic
theories constructed in Section~\ref{Cyclic} give vector spaces over
$\complex$ and are thus insensitive to torsion. A simple example is
provided by any compact oriented manifold $X$ for which ${\sf
  W}_3(X)\neq0$. Then the algebra $\alg=C(X)$ is a
strong C-PD algebra but not a PD algebra. In the cases
where the Chern characters $\ch_\alg$ and $\ch_\balg$ are both
isomorphisms after tensoring the $\K$-groups with $\complex$,
Poincar\'e duality in cyclic theory implies rational Poincar\'e
duality in $\K$-theory.

The commutative diagram (\ref{ChernPDdiag}) allows us to transport
the structure of Poincar\'e duality in $\KK$-theory to local cyclic
cohomology. In particular, all examples that we presented
in Section~\ref{Pairings} in the context of $\KK$-theory also apply to
local cyclic cohomology. Note, however, that if a strong PD pair
of algebras $(\alg,\balg)$ are equipped with their own fundamental cyclic
cohomology class $\Xi\in\HE^d(\alg\otimes\balg)$,
then generically $\ch(\Delta)\neq\Xi$. We will see an example
of this in Section~\ref{NCSpinHP} below. In fact, this will be the
crux of our construction of D-brane charge cycles in
Section~\ref{CyclicDbrane}. The choice $\ch(\Delta)=\Xi$ has
certain special properties which will be discussed in
Section~\ref{sect:Todd}.

\subsection{$\HE$-Equivalence\label{HPEq}}

Exactly as we did above, it is possible to define the analogous notion
of $\KK$-equivalence from Section~\ref{Sect:KKEq} in the bivariant
local cyclic cohomology. We now briefly discuss how this works.

\begin{definition}\label{hp-equiv1}
Two complete bornological algebras $\alg$ and $\balg$
are said to be {\em strongly $\HE$-equivalent} if there
are elements 
\beq
\xi \in \HE_n(\alg, \balg) \qquad\mbox{and}\qquad
\eta \in \HE_n(\balg, \alg)
\eeq
such that 
\beq
\xi \otimes_\balg \eta ~=~ 1_\alg \in \HE_0(\alg, \alg)
\qquad\mbox{and} \qquad \eta \otimes_\alg \xi ~=~ 1_\balg \in
\HE_0(\balg, \balg) \ .
\eeq
\end{definition}
\noindent
As in the case of $\KK$-theory, these hypotheses induce isomorphisms
between the local cyclic homology and cohomology groups of the
algebras $\alg$ and $\balg$ as
\beq
\HE^\bullet(\alg)\cong\HE^{\bullet+n}(\balg) \qquad \mbox{and} \qquad
\HE_\bullet(\alg)\cong\HE_{\bullet+n}(\balg) \ .
\label{HPisos}\eeq
One similarly has the notions of weak equivalence and of
$\HE$-equivalence. The Chern character again allows us to transport
results of Section~\ref{Sect:KKEq} to cyclic theory. If $(\alg,\balg)$
is a pair of strongly $\KK$-equivalent algebras, with the equivalence
implemented by classes $\alpha\in\KK_n(\alg,\balg)$ and
$\beta\in\KK_{-n}(\balg,\alg)$, then there are commutative diagrams
\beq
\xymatrix{
\K^\bullet(\balg) \ar[d]_{\ch}
\ar@<0.5ex>[rr]^{\alpha \otimes_\balg} &&
\ar@<0.5ex>[ll]^{\beta\otimes_\alg}
\K^{\bullet+n}(\alg)\ar[d]^{\ch} \\
\HE^\bullet(\balg)
\ar@<0.5ex>[rr]^{\ch(\alpha)\otimes_{\balg}} &&
\ar@<0.5ex>[ll]^{\ch(\beta)\otimes_{\alg}}
\HE^{\bullet+n}(\alg) }
\label{ChernKKdiag1}\eeq
and
\beq
\xymatrix{
\K_\bullet(\balg) \ar[d]_{\ch}
\ar@<0.5ex>[rr]^{\otimes_\balg\alpha } &&
\ar@<0.5ex>[ll]^{\otimes_\alg\beta}
\K_{\bullet+n}(\alg)\ar[d]^{\ch} \\
\HE_\bullet(\balg)
\ar@<0.5ex>[rr]^{\otimes_{\balg}\ch(\alpha)} &&
\ar@<0.5ex>[ll]^{\otimes_{\alg}\ch(\beta)}
\HE_{\bullet+n}(\alg) \ . }
\label{ChernKKdiag2}\eeq
Thus $\KK$-equivalence implies $\HE$-equivalence, but not conversely.

\begin{remark}
In the various versions of cyclic theory,
one needs different notions of Morita
equivalence. For example, one has an isomorphism
$\HP_\bullet(\alg)\cong\HP_\bullet(\alg\widehat\otimes\mathcal{L}^1)$
in periodic cyclic homology, where $\mathcal{L}^1$ is the algebra of
trace-class operators on a separable Hilbert space.
Fortunately, since $\HE$ is well behaved for $C^*$-algebras, which are
our main examples of interest, we will usually not have to worry about
this point.
\end{remark}

\subsection{Spectral Triples\label{NCSpinHP}}

Let us model a D-brane by an even spectral triple $(\alg,\hil,D)$ as
prescribed in Section~\ref{NCspin}. Assume that the resolvent
of the operator $D$ is of order $p$, i.e., its eigenvalues $\mu_k$
decay as $k^{-1/p}$. This is the situation, for example, for the case
when $D$ is the canonical Dirac operator over a finite-dimensional
spin$^c$ manifold. We further make the following regularity assumption
on the spectral triple. Let us assume that both $\alg$ and $[D,\alg]$
are contained in $\bigcap_{k>0}\,{\rm Dom}(\delta^k)$, where
$\delta={\rm ad}(|D|)$. Let
$\Sigma\subset\complex$ denote the set of all singularities of the
spectral zeta-functions $\zeta_P(z)=\Tr^{~}_{\hil}(P\,|D|^{-z})$, where
$z\in\complex$ and $P$ is an element of the algebra generated by
$\delta^k(\alg)$ and $\delta^k([D,\alg])$. The set $\Sigma$ is
called the {\it dimension spectrum} of the spectral triple
$(\alg,\hil,D)$. We will assume that $(\alg,\hil,D)$ has
discrete and simple dimension spectrum $\Sigma$, i.e., that $\zeta_P$,
$\forall P$ can be extended as a meromorphic function to
$\complex\setminus\Sigma$ with simple poles in $\Sigma$. Such a
spectral triple is said to be {\it regular}.

Under these circumstances, the residue formula
\beq
\ncint\,P=\Res^{~}_{z=0}\Bigl(\Tr^{~}_{\hil}\left(P\,|D|^{-2z}\right)\Bigr)
\label{ncintPdef}\eeq
defines a trace on the algebra generated by $\alg$, $[D,\alg]$ and
$|D|^z$, $z\in\complex$. Using it we may define the {\it
  Connes-Moscovici cocycle} $\varphi^{\rm CM}=(\varphi^{\rm CM}_{2n})_{n\geq0}$
in the $(\b,\B)$ bi-complex of the algebra $\alg$ \cite{CM2}. For
this, we denote by $a^{[k]}$ the $k$-th iterated commutator of
$a\in\alg$ with the operator $D^2$,
\beq
a^{[k]}:=\underbrace{\big[D^2\,,\,\big[D^2\,,\,\cdots
\big[D^2}\limits_{k~{\rm times}}\,,\,a\big]~\big]\cdots\big] \ .
\label{kthiterdef}\eeq
For $n=0$ and $a_0\in\alg$ we set
\beq
\varphi^{\rm CM}_0(a_0)=\Tr^{~}_{\hil}\left(\gamma\,a_0\,\Pi_{\ker(D)}\right)+
{\rm Res}^{~}_{z=0}\Bigl(\mbox{$\frac1z$}\,\Tr^{~}_\hil\left(\gamma\,
a_0\,|D|^{-2z}\right)\Bigr) \ ,
\label{varphiCM0}\eeq
where $\Pi_{\ker(D)}$ is the orthogonal projection onto the kernel of
the operator $D$ on $\hil$. For $n>0$ and $a_i\in\alg$, we define
\bea
&& \!\!\! \varphi^{\rm CM}_{2n}(a_0,a_1,\dots,a_{2n})\\
&& =\sum_{\vec k}\,
\frac{(-1)^{|\vec k|}\,\bigl(|\vec k|+n-1\bigr)!}{2\,\vec k!\,
\prod\limits_{j=1}^{2n}\,(k_1+\dots+k_j+j)}~
\ncint\,\gamma\,a_0~\Bigl(\,\mbox{$
\prod\limits_{j=1}^{2n}$}\,[D,a_j]^{[k_j]}\Bigr)~|D|^{-2|\vec k|-2n
} \ , \nonumber
\label{varphiCM2n}\eea
where the sum runs through all multi-indices $\vec k=(k_1,\dots,k_{2n})$
with $|\vec k|:=k_1+\dots+k_{2n}$ and $\vec k!:=k_1!\cdots
k_{2n}!$. It can be shown that this formula has only a finite number
of non-zero terms. The class $\ch(D)=[\varphi^{\rm
  CM}_{2n}]\in\H\E^0(\alg)$ is called the (even)
{\it cyclic cohomology Chern character} and it may be regarded as a
map
\beq
\ch\,:\,\K^0(\alg)~\longrightarrow~\H\E^0(\alg) \ .
\label{chbulcyclic}\eeq

It is instructive again to look at the case where $X$ is a compact,
smooth spin manifold of even dimension $d$. For the spectral triple we
then take $\alg=C^\infty(X)$, $\hil^\pm=L^2(X,{\sf S}_X^\pm)$, and
$D=\Dirac:C^\infty(X,{\sf S}_X^+)\to C^\infty(X,{\sf S}_X^-)$ the usual
(untwisted) Dirac operator. Then the dimension spectrum $\Sigma$ consists of
relative integers $<d$, and is simple. (Multiplicities would arise in
the case that the spacetime $X$ is a singular orbifold, for
example.) We can thereby apply the Connes-Moscovici cocycle
construction to this situation. One finds that the contributions to
(\ref{varphiCM2n}) vanish unless $\vec k=\vec0$, and hence its
components are given explicitly by \cite{Ponge1}
\beq
\varphi^{\rm CM}_{2n}(f^0,f^1,\dots,f^{2n})=\frac1{(2n)!}\,\int\limits_X
f^0~\dd f^1\wedge\cdots\wedge\dd f^{2n}\wedge\widehat{A}(X)
\label{varphiCM2nX}\eeq
with $f^i\in C^\infty(X)$. In this case, the entire cyclic cohomology
${\sf HE}^\bullet(\alg)$
is naturally isomorphic to the local cyclic cohomology
$\HE^\bullet(\alg)$ and to the periodic cyclic cohomology
$\H\P^\bullet(\alg)$ \cite{Puschnigg}; the resulting entire
cocycle is cohomologous to an explicit periodic cyclic cocycle given 
in terms of the spectral triple.

This result implies an important characterization that will be crucial
to the construction of brane charges as cyclic classes.

\begin{theorem}
Let $X$ be a compact, smooth spin manifold of even dimension. Then the
cyclic cohomology Chern character of the spectral triple
$(C^\infty(X),L^2(X,{\sf S}_X),\Dirac)$  coincides with the
Atiyah-Hirzebruch class of $X$ in even de~Rham homology,
\beq
\ch(\Dirac)={\sf Pd}_X^{-1}\bigl(\widehat{A}(X)\bigr) \ .
\label{cyclchhatA}\eeq
\label{cyclchX}\end{theorem}

\begin{remark}
In some cases the $p$-summability requirements are not met, most
notably when the spectral triple is infinite-dimensional. In such
instances we can still compute the cyclic cohomology Chern character
if the spectral triple $(\alg,\hil,D)$ is {\it $\theta$-summable},
i.e., $[D,a]$ is bounded for all $a\in\alg$, and the eigenvalues
$\mu_k$ of the resolvent of $D$ grow no faster than $\log(k)$. This
implies that the corresponding heat kernel is trace-class,
$\Tr^{~}_\hil(\e^{-t\,D^2})<\infty\quad\forall t>0$.
Within this framework, we can then represent the Chern character of
the even spectral triple $(\alg,\hil,D)$ in the entire cyclic
cohomology of $\alg$ by using the {\it JLO cocycle} $\varphi^{\rm
JLO}=(\varphi^{\rm
  JLO}_{2n})_{n\geq0}$ \cite{JLO1}. With $a_0,a_1\dots,a_{2n}\in\alg$, it is
defined by the formula
\bea
&& \varphi^{\rm JLO}_{2n}(a_0,a_1\dots,a_{2n})\\
&&=\int\limits_{\triangle_{2n}}\,
\dd t_0~\dd t_1\cdots\dd t_{2n}~\Tr^{~}_\hil\biggl(\gamma\,a_0~
\e^{-t_0\,D^2}\,\Bigl(\,\mbox{$
\prod\limits_{j=1}^{2n}$}\,[D,a_j]~\e^{-t_j\,D^2}\Bigr)\biggr)
\ , \nonumber
\label{JLOdef}\eea
where $\triangle_n=\{(t_0,t_1,\dots,t_n)\,|\,t_i\geq0\,,~\sum_i\,t_i=1\}$
denotes the standard $n$-simplex in $\real^{n+1}$. This entire cyclic
cocycle is cohomologous to the Chern character. Once again, consider
the example of the canonical triple $(C^\infty(X),L^2(X,{\sf
  S}_X),\Dirac)$ over a spin manifold $X$, and replace $\Dirac$
everywhere in the formula (\ref{JLOdef}) by $s\,\Dirac$ with $s>0$. By
using asymptotic symbol calculus, one can then
show \cite{CM1,Quillen1} that the character (\ref{JLOdef})
retracts as $s\to0$ to the Connes-Moscovici
cocycle~(\ref{varphiCM2nX}).
\end{remark}

\newsection{T-Duality}\label{sect:T-duality}

In this section we will show that there is a strong link between
$\KK$-equivalence for crossed product algebras and T-duality in string
theory. This will lead to a putative axiomatic characterization of
T-duality for $C^*$-algebras. We also describe an analogous
characterization in local cyclic cohomology.

\subsection{Duality for Crossed Products}

We begin with some general results regarding $\KK$-equivalence and
Poincar\'e duality for crossed product algebras, and then use them to
give some more examples of PD algebras. Let $G$ be a locally compact,
connected Lie group. Recall \cite{CCJJV} that $G$ is
said to satisfy the {\em Haagerup property} if it has a metrically
proper isometric action on some Hilbert space. Examples are $SO(n,
1)$, $SU(n, 1)$ and locally compact, connected, amenable Lie
groups. An {\it amenable} Lie group is one that has an invariant mean,
examples of which include abelian Lie groups, nilpotent Lie groups and
solvable Lie groups.

Let $K$ be a maximal compact subgroup of $G$. Let
$V$ denote the cotangent space to the symmetric space $G/K$ at the
point $(K)$. Let $\Cl(V)$ be the Clifford algebra of $V$ with respect
to some positive definite inner product on $V$. We start by recalling
a theorem of Higson-Kasparov \cite{HK} and Tu \cite{Tu}, generalizing
a theorem of Kasparov \cite[\S6~Theorem~2]{Kas80}.

\begin{theorem}\label{thm:crossed1}
Let $\alg$ be a $G$-$C^*$-algebra where $G$ is a locally compact,
connected Lie group satisfying the Haagerup property. Then in the
notation above, the pair of crossed product $C^*$-algebras
$(\alg\rtimes G\,,\,(\alg\otimes\Cl(V)) \rtimes K)$ are strongly
$\KK$-equivalent. If in addition the coadjoint action of $K$ on $V$
is spin then the pair $(\alg\rtimes G\,,\,  (\alg\rtimes K) \otimes
C_0(\bbR^d)) $ are strongly $\KK$-equivalent, where $d=\dim(G/K)$.
\end{theorem}
\noindent
A special case of Theorem~\ref{thm:crossed1}, which was proved
earlier by Fack and Skandalis \cite{FS} generalising an argument of
Connes \cite{Connes-Thom}, is as follows.

\begin{corollary}\label{cor:crossed3}
Let $\alg$ be a $G$-$C^*$-algebra where $G$ is a simply connected,
locally compact, solvable Lie group of dimension $k$. Then the pair of
$C^*$-algebras $(\alg \otimes C_0(\bbR^k),$ $\alg\rtimes G)$ are
strongly $\KK$-equivalent.
\end{corollary}

As an immediate consequence of Theorem~\ref{thm:crossed1} and
Lemma~\ref{lemma:kk1} we obtain the following.

\begin{corollary}\label{cor:crossed1}
Let $\alg$ be a $G$-$C^*$-algebra where $G$ is a locally compact,
connected Lie group satisfying the Haagerup property. Then in the
notation of Theorem~\ref{thm:crossed1}, $(\alg\otimes\Cl(V)) \rtimes
K$ is a \textup{(}strong, weak\textup{)} PD algebra if and only if
$\alg\rtimes G$ is a 
\textup{(}strong, weak\textup{)} PD algebra. If in addition the
  coadjoint action of $K$ 
on $V$ is spin, then $\alg\rtimes K$ is a  \textup{(}strong,
weak\textup{)} PD algebra if  
and only if $\alg\rtimes G$ is a  \textup{(}strong, weak\textup{)} PD
algebra. 
\end{corollary}
\noindent
In addition, an immediate consequence of Corollary~\ref{cor:crossed3}
and Lemma~\ref{lemma:kk1} is as follows.

\begin{corollary}\label{cor:crossed2}
Let $\alg$ be a $G$-$C^*$-algebra where $G$ is a simply connected,
locally compact, solvable Lie group. Then $\alg$ is a
\textup{(}strong, weak\textup{)} 
PD algebra if and only if $\alg\rtimes G$ is a  \textup{(}strong,
weak\textup{)} PD algebra. 
\end{corollary}

\begin{example}
Let $\Gamma$ be a torsion-free, discrete subgroup of a connected
semisimple Lie group $G$ with finite center. Let $P$ be a minimal parabolic
subgroup of $G$ and $K$ a maximal compact subgroup of $G$. Then $G/P$
is the Furstenberg boundary (at infinity) of the symmetric space
$G/K$. By Green's theorem, $C(G/P) \rtimes \Gamma$ is strongly Morita
equivalent to $C_0(\Gamma\backslash G) \rtimes P$. By
Lemmas~\ref{lemma:kk1} and~\ref{lemma:kk-examples} it follows that
$C(G/P)\rtimes \Gamma$ is a strong PD algebra if and only if
$C_0(\Gamma\backslash G) \rtimes P$ is a strong PD algebra. By
Corollary~\ref{cor:crossed1} above, $C_0(\Gamma\backslash G) \rtimes
P$ is a strong PD algebra if and only if $C_0(\Gamma\backslash G)
\rtimes K$ is a strong PD algebra, i.e., if and only
if $\Gamma\backslash G/ K$ is a spin$^c$
manifold. We conclude that $C(G/P) \rtimes \Gamma$ is a strong PD
algebra if and only if $\Gamma\backslash G/K$ is a spin$^c$ manifold. In
particular, $C(\bbS^1)\rtimes \Gamma_g$ is a strong PD algebra whenever
$\Gamma_g$ is the fundamental group of a compact, oriented Riemann
surface of genus $g\geq1$. There is a deep variant of this example,
analysed in detail by Emerson \cite{E1}, dealing with the crossed
product $C^*$-algebra $C(\partial\Gamma)\rtimes\Gamma$ for a
hyperbolic group $\Gamma$ with Gromov boundary $\partial\Gamma$.
\end{example}

Another important property of such crossed products is {\it Takai
  duality}. If $G$ is a locally compact,
abelian group we denote by $\tilde G$ its Pontrjagin dual, i.e., the
set of characters of $G$, which is also a locally compact, abelian
group. Pontrjagin duality $\tilde{\tilde G}\cong G$ follows by Fourier
transformation. For example, $\tilde \bbR^n =\bbR^n$, $\tilde\bbT^n=
\zed^n$, and $\tilde \zed^n = \bbT^n$. If $\alg$ is a
$G$-$C^*$-algebra, then the crossed product $\alg\rtimes G$ carries a
$\tilde G$-action.

\begin{theorem}[{\bf Takai Duality}]
Let $\alg$ be a $G$-$C^*$-algebra where $G$ is a locally compact,
abelian Lie group. Then there is an isomorphism of $C^*$-algebras
\beq
(\alg \rtimes G) \rtimes \tilde G\cong\alg\otimes\mathcal{K}\bigl(L^2(G)
\bigr) \ .
\eeq
\end{theorem}
\noindent
In other words, the algebras $\alg$ and $(\alg \rtimes G) \rtimes \tilde
G$ are strongly Morita equivalent. If we interpret the crossed product
$\alg\rtimes G$ as the noncommutative analogue of an abelian orbifold
spacetime $X/G$, then Takai duality asserts that ``orbifolding twice''
gives back a spacetime which is physically equivalent to the original
spacetime $X$. The essential physical phenomenon is that the states
which were projected out by $G$ are restored by $\tilde G$.

\subsection{T-Duality and $\KK$-Equivalence}

We next explain how $\KK$-equivalence of crossed products is related
to T-duality. Throughout $X$ will be assumed to be a locally compact,
finite-dimensional, homotopically finite space. Consider first the
simplest case of flat D-branes in
Type~II superstring theory on a spacetime $X=M\times \bbT^n$ which is
compactified on an $n$-torus $\bbT^n=\mathbb{V}_n/\Lambda_n$, where
$\Lambda_n$ is a lattice of maximal rank in an $n$-dimensional, real
vector space $\mathbb{V}_n$. As shown in \cite{Hor}, T-duality in
this instance is explained by using the correspondence
\begin{equation} \label{eqAc}
\xymatrix @=4pc { &  M\times \bbT^n \times \widehat{\bbT}{}^n \ar[dl]^{p} 
\ar[dr]_{\widehat{p}} & \\
M\times \bbT^n   &   &    M\times \widehat{\bbT}{}^n }
\end{equation}
where $ \widehat{\bbT}{}^n = ({\mathbb{V}}{}_n)^\vee/(\Lambda_n)^\vee$
denotes the dual 
torus, with $(\Lambda_n)^\vee$ the dual lattice in the dual vector
space $(\mathbb{V}_n)^\vee$. This gives rise to an isomorphism of
$\K$-theory groups
\begin{equation} \label{eqAd} \begin{CD}
  T_! \,:\, \K^\bullet\big(M\times \bbT^n\big)~ 
@>\approx>> ~\K^{\bullet+n}\big(M\times \widehat{\bbT}{}^n\big) \end{CD}
\end{equation}
given by
\begin{equation}
  T_!(-) = \widehat{p}{\,}_!\ \big( p^!(- )\, \otimes \, \mathcal P
  \big) \ ,
\label{eqAe}\end{equation}
where $\mathcal P$ is the Poincar\'e line bundle over the torus
$\bbT^n \times \widehat{\bbT}{}^n$ pulled back to $M\times \bbT^n
\times \widehat{\bbT}{}^n$ via the projection map ${\rm pr}_1 :
M\times \bbT^n \times \widehat{\bbT}{}^n \to \bbT^n \times
\widehat{\bbT}{}^n$. 
Thus T-duality can be viewed in this case 
as a smooth analog of the Fourier-Mukai transform.
If $G$ is the metric of the torus
$\bbT^n$ inherited from the non-degenerate bilinear form of the
lattice $\Lambda_n$, then the dual torus $\widehat{\bbT}{}^n$ has
metric $G^{-1}$ inherited from the dual lattice $(\Lambda_n)^\vee$.

As argued by \cite{MM}, detailed in  \cite{Horava,FW,OS99,MW,W}, 
and discussed in 
Section~\ref{ChargeDef}, RR-fields are classified by $\K^1$-groups and
RR-charges by $\K^0$-groups of the spacetime $X$ in Type~IIB string
theory, whereas RR-fields are classified by $\K^0$-groups and
RR-charges by $\K^1$-groups in Type~IIA string theory. Thus if
spacetime $X=M\times \bbT^n$ is compactified on a torus of rank $n$,
then the isomorphism (\ref{eqAd}) is consistent with the fact that
T-duality is an equivalence between the Type~IIA and Type~IIB string
theories if $n$ is odd, while if $n$ is even it is a self-duality for
both string theories. Given this compelling fact, we will take this
isomorphism to mean the equivalence itself here (although in string
theory the duality is much more complicated  and involves many more
ingredients).

As was observed in \cite{MR}, all of this can be reformulated in terms
of the $C^*$-algebra $C_0(M \times \bbT^n)$. The locally compact,
abelian vector Lie group $\mathbb{V}_n\cong\real^n$ acts on $ \bbT^n=
\mathbb{V}_n/\Lambda_n$ via left translations,
and consider the crossed product algebra $C_0(M \times
\bbT^n) \rtimes \mathbb{V}_n$ with $\mathbb{V}_n$ acting trivially on
$M$. By Rieffel's imprimitivity
theorem  \cite{Rie}, there is a strong Morita equivalence
\beq
C_0(M \times \bbT^n) \rtimes \mathbb{V}_n \sim C_0(M) \rtimes \Lambda_n
\eeq
where the discrete group $\Lambda_n$ acts trivially on $C_0(M)$. One
therefore has
\beq
C_0(M \times \bbT^n) \rtimes \mathbb{V}_n \sim C_0(M) \otimes C^*(\Lambda_n)
\ .
\eeq
By Fourier transformation the group $C^*$-algebra of $\Lambda_n$ can be
identified as $C^*(\Lambda_n) \cong C(\,\widehat{\bbT}{}^n)$, and as a
consequence there is a strong Morita equivalence
\beq
C_0\big(M \times \bbT^n\big) \rtimes \mathbb{V}_n ~\sim~
C_0\big(M\big) \otimes C\big(\,\widehat{\bbT}{}^n\big) \cong
C_0\big(M \times \widehat{\bbT}{}^n\big) \ .
\eeq

By Lemma~\ref{lemma:kk-examples} and Corollary~\ref{cor:crossed3}, the
pair of $C^*$-algebras $(C_0(M \times \bbT^n)\,,\, C_0(M \times
\bbT^n) \rtimes \mathbb{V}_n \sim C_0(M \times \widehat{\bbT}{}^n))$ are
strongly $\KK$-equivalent (with a degree shift of $n$ mod $2$). Thus
by Lemma~\ref{lemma:kk-conseq} there 
are isomorphisms
\bea
 T_! \,:\, \K_\bullet\big(C_0(M \times \bbT^n)\big) &
\stackrel{\approx}{\longrightarrow} &
  \K_{\bullet+n}\big(C_0(M \times \widehat{\bbT}{}^n)\big) \ ,
\nonumber\\
   T^! \,:\, \K^\bullet\big(C_0(M \times \bbT^n)\big) &
\stackrel{\approx}{\longrightarrow} &
  \K^{\bullet+n}\big(C_0(M \times \widehat{\bbT}{}^n)\big) \ .
\label{TKKisos}\eea
The upshot of this analysis is that the Fourier-Mukai transform, 
or equivalently T-duality for flat D-branes in
Type II string theory, on a spacetime $X$ that is compactified on a
torus $\bbT^n$, can be interpreted as taking a crossed product with
the natural action of $\mathbb{V}_n\cong\real^n$ on the $C^*$-algebra
$C_0(X)$.  

This point of view was generalized in a series of papers
 \cite{BEM,BEMb,MR,MR2,BHM1,BHM2} to twisted D-branes in Type II
superstring theory in a $B$-field $(X,H)$ and
for a spacetime $X$ which is a possibly non-trivial principal torus
bundle $\pi: X\to M$ of rank $n$. As described in
Section~\ref{TwistedD}, in this case the type~I, separable
$C^*$-algebra in question is the algebra $C_0(X,\mathcal{E}_H)$ of
continuous sections vanishing at infinity of a locally trivial
$C^*$-algebra bundle $\mathcal{E}_H\to X$ with fibre
$\mathcal{K}(\hil)$ and Dixmier-Douady invariant
$\delta_X(\mathcal{E}_H)=H \in {\rm
  Br}^\infty(X)\cong\H^3(X,\bbZ)$. This is a stable, continuous trace
algebra with spectrum $X$. It is a fundamental theorem of Dixmier and
Douady \cite{DD} that $H$ is trivial in cohomology if and only if
$C_0(X,\mathcal{E}_H)$ is strongly Morita equivalent to $C_0(X)$ (in
fact $C_0(X,\mathcal{E}_0)\cong C_0(X,\mathcal{K}(\hil))$), consistent
with the above discussion. If in addition $X$ is a Calabi-Yau
threefold, then T-duality in these instances coincides with mirror
symmetry.

As above, the abelian Lie group $\mathbb{V}_n$ acts on $X$ via left
translations of the fibres $\bbT^n$. In \cite{MR2} the following
fundamental technical theorem was proven.

\begin{theorem}\label{thm:technical}
In the notation above, the natural $\mathbb{V}_n$-action on $X$ lifts to a
$\mathbb{V}_n$-action on the total space $\mathcal{E}_H$, and hence to a
$\mathbb{V}_n$-action on $C_0(X,\mathcal{E}_H)$, if and only if the
restriction of $H$ to the fibres of $X$ is trivial in cohomology.
\end{theorem}
\noindent
This is a non-trivial obstruction if and only if the fibres of $X$ are
of rank $n\ge 3$. The T-dual is then defined as the crossed product
algebra $C_0(X,\mathcal{E}_H)\rtimes \mathbb{V}_n$. This algebra is a continuous
trace algebra if and only if $\pi_*(H)=0$ in $\H^1(M, \H^2(\bbT^n,
\bbZ))$. In the general case, the crossed product $C_0(X,\mathcal{E}_H)
\rtimes \mathbb{V}_n$ is not of type~I but is rather a continuous field of
(stabilized) rank $n$ noncommutative tori fibred over
$M$. The fibre over the point $m\in M$ is isomorphic to
$\mathcal{T}^n_{f(m)}\otimes\mathcal{K}(\hil)$, where $\pi_*(H)=[f]$
is the Mackey obstruction class with $f:M \to\H^2(\Lambda_n,
U(1))\cong(\real/\zed)^k, \, k = \binom{n}{2}$ a continuous map. This
obstruction is due to the presence of {\it discrete torsion} in the
fibres of the string background, represented by multipliers $f(m)$ on
the discrete group $\Lambda_n$, which is essentially due to the
presence of a non-trivial global $B$-fields along the fibres of $X$.

\begin{corollary}
In the notation of Theorem~\ref{thm:technical}, suppose that the
restriction of $H$ to the fibres of $X$ is trivial in cohomology. Then
the T-dual $C_0(X,\mathcal{E}_H)\rtimes \mathbb{V}_n$ is a strong PD algebra
if and only if $X$ is a spin$^c$ manifold. 
\end{corollary}
\begin{proof} By a theorem of Parker \cite{Parker92}, the continuous
  trace algebra $C_0(X,\mathcal{E}_H)$ is a strong PD algebra if and only if
  $X$ is a spin$^c$ manifold. The result now follows from
  Theorem~\ref{thm:technical} and Corollary~\ref{cor:crossed2}.
\end{proof}

The Takai duality theorem in these examples implies that
$(C_0(X,\mathcal{E}_H)\rtimes\mathbb{V}_n)\rtimes\mathbb{V}_n$ is
strongly Morita equivalent to $C_0(X,\mathcal{E}_H)$, i.e., T-duality
applied twice returns the original string theory. We can now combine
all of these observations to formulate a generically noncommutative
version of T-duality for $C^*$-algebras in very general settings.

\begin{definition}[{\bf $\K$-Theoretic T-Duality}]
Let $\underline{\mathfrak{T}}$ be a suitable category of separable
$C^*$-algebras, possibly equipped with some extra structure (such as the
$\real^n$-action above). Elements of $\underline{\mathfrak{T}}$ are called
{\it T-dualizable algebras}, with the following properties:
\begin{enumerate}
\item There is a covariant functor ${\sf
T}:\underline{\mathfrak{T}}\to\underline{\mathfrak{T}}$ which sends an
algebra $\alg$ to an algebra ${\sf T}(\alg)$ called its {\it T-dual};
\item There is a functorial map
$\alg\mapsto\gamma_\alg\in\KK_n(\alg\,,\,{\sf T}(\alg))$ such that
$\gamma_\alg$ is a $\KK$-equivalence; and
\item The pair $(\alg\,,\, \sT(\sT(\alg)))$
  are Morita equivalent, and the Kasparov product 
$\gamma_\alg\otimes_{\sT(\alg)} \gamma_{\sT(\alg)}$ is
the $\KK$-equivalence associated to this Morita equivalence.
\end{enumerate}
\label{TdualKdef}\end{definition}

\subsection{T-Duality and $\HE$-Equivalence}

As we have mentioned, the isomorphisms (\ref{TKKisos}) are only part
of the story behind T-duality, as they only dictate how topological
charges behave under the duality. In particular, the isomorphism $T_!$
on $\K$-theory bijectively relates the RR-fields in T-dual spacetimes,
while the bijection $T^!$ relates the RR-charges themselves. As
explained in Sections~\ref{FlatD} and~\ref{RRFields} in the case of
flat D-branes, the RR-fields are represented by closed differential forms
on the spacetime $X$ while the branes themselves are associated to
non-trivial (worldvolume) cycles of $X$. It is therefore natural to
attempt to realize our characterization of T-duality above in the
language of cyclic theory, in order to provide the bijections
between the analogues of these (and other) geometric structures.

We begin with the following observation.

\begin{theorem}
Let $\alg$ be separable $C^*$-algebra, and suppose that $\alg$ admits
an action by a locally compact, real, abelian vector Lie group
$\mathbb{V}_n$ of dimension $n$. Then there is a commutative diagram
\beq
\xymatrix{
 \K_\bullet(\alg)\ar[d]_{\ch}\ar[r]^(.4){T_*} &
 \K_{\bullet+n}(\alg\rtimes\mathbb{V}_n)\ar[d]^{\ch} \\
\HE_\bullet(\alg)\ar[r]_(.4){T_*} &
\HE_{\bullet+n}(\alg\rtimes\mathbb{V}_n) }
\label{KKHPTdiag}\eeq
whose horizontal arrows are isomorphisms.
\label{ENNTthm}\end{theorem}
\begin{proof}
The isomorphism
$\K_\bullet(\alg)\cong\K_{\bullet+n}(\alg\rtimes\real^n)$ in the top
row is the Connes-Thom isomorphism \cite{Connes-Thom} (cf.\
Corollary~\ref{cor:crossed3} above), while the isomorphism
in the bottom row comes from transporting this isomorphism
to local cyclic homology, as in  \cite{Shir}. (See also
the review of  \cite{Shir} in MathSciNet, MR2117221 (2005j:46041), for
more of an explanation.)
\end{proof}
\noindent
The isomorphism in the bottom row of (\ref{KKHPTdiag}) is the local
cyclic homology version of T-duality. Theorem~\ref{ENNTthm} shows
that the T-duality isomorphisms in $\K$-theory descend to
isomorphisms of cyclic cohomology, giving the
mappings at the level of RR-field representatives in
$\HE_\bullet(\alg)$. This motivates the local cyclic
cohomology version of the axioms spelled out in
Definition~\ref{TdualKdef}.

\begin{definition}[{\bf Cohomological T-Duality}]
Let $\alg$ be a complete bornological algebra. A {\it cyclic
  T-dual} of $\alg$ is a complete bornological algebra ${\sf
  T}^{\HE}(\alg)$ which satisfies the following three axioms:
\begin{enumerate}
\item The map $\alg \mapsto {\sf T}^{\HE}(\alg)$ is a covariant functor 
on an appropriate category of algebras;
\item The pair $(\alg\,,\, {\sf T}^{\HE}(\alg))$ are $\HE$-equivalent; and
\item The pair $(\alg\,,\, {\sf T}^{\HE}({\sf T}^{\HE}(\alg)))$
  are topologically Morita equivalent.
\end{enumerate}
As in Definition \ref{TdualKdef}, there should be an explicit functorial
$\HE$-equivalence in (2) compatible with the Morita equivalence in (3).
\label{TdualHPdef}\end{definition}
\noindent
{}From these definitions it follows that $\K$-theoretic T-duality for a
separable $C^*$-algebra $\alg$ implies cohomological T-duality for 
the same algebra, but the converse need not necessarily be true
(because of torsion in $\K$-theory, for example).

\begin{remark}
There are competing points of view concerning T-duality in the nonclassical
case, that is, in the case when the T-dual of a spacetime $X$, which
is a principal 
torus bundle with nontrivial $H$-flux, is not another principal torus bundle. 
Unlike the approach discussed in this section, where the T-dual is a globally
defined but possibly noncommutative algebra, the T-dual in the competing points
of view is not globally defined.  For example,
in  \cite{Hull}, Hitchin's generalized complex geometry is used
to construct a T-dual which is a purely {\em local}  object, that does not 
patch together to give a global object, and is referred to as a
T-fold. See also  \cite{STW} for a related point of view. 
\end{remark}

\newsection{Todd Classes and Gysin Maps\label{sect:gysin}}

In this section we apply the concept of Poincar\'e duality in
$\KK$-theory and bivariant  cyclic cohomology to define the
notion of a Todd class for a very general class of $C^*$-algebras. As
follows from the discussion of Section~\ref{FlatD}, this will be one
of the main building blocks of our definitions of generalized D-brane
charges. Another crucial ingredient in these definitions is the
application of Poincar\'e duality to the construction of
Gysin maps (or ``wrong way'' maps) in $\KK$-theory and in cyclic
theory, which also came up in our discussion of T-duality above. These
general constructions combine together to yield a generalization of
the Grothendieck-Riemann-Roch theorem for an appropriate class of
$C^*$-algebras, which in turn yields another perspective on the
concept of a T-dual $C^*$-algebra that was introduced in the previous
section.

\subsection{The Todd Class\label{sect:Todd}}

Our general definition of Todd classes is motivated by
Theorem~\ref{cyclchX} and Proposition~\ref{kk:fund2}. We begin with
the following observation.

\begin{lemma}
Let $\alg,\balg_1,\balg_2$ be separable $C^*$-algebras such that
$(\alg,\balg_1)$ and $(\alg,\balg_2)$ are both strong PD pairs. Then
the pair of algebras $(\balg_1,\balg_2)$ are strongly
$\KK$-equivalent.
\end{lemma}
\begin{proof}
Let $\Delta_1\in\K^d(\alg\otimes\balg_1)$ and
$\Delta_2\in\K^d(\alg\otimes\balg_2)$ be the respective fundamental
classes. Then with a proof along the lines of
Proposition~\ref{kk:fund2}, one shows that the classes
$\alpha:=\Delta_1^\vee\otimes_\alg\Delta^{\phantom{\vee}}_2\in
\KK_0(\balg_1,\balg_2)$ and
$\beta:=(-1)^d\,\Delta_2^\vee\otimes_\alg\Delta^{\phantom{\vee}}_1\in
\KK_0(\balg_2,\balg_1)$ implement the required equivalence.
\end{proof}

Let $\underline{\mathfrak{D}}$ denote the class of all separable
$C^*$-algebras $\alg$ for which there exists another separable
$C^*$-algebra $\balg$ such that $(\alg,\balg)$ is a strong PD
pair. For any such $\alg$, we fix a representative of the
$\KK$-equivalence class of $\balg$ and denote it by $\tilde\alg$. In
general there is no canonical choice for $\tilde\alg$. If
$\alg$ is a strong PD algebra, the canonical choice
$\tilde\alg:=\alg^\op$ will always be made.
\begin{definition}\label{defn:Todd}
Let $\alg\in\underline{\mathfrak{D}}$, let
$\Delta\in\K^d(\alg\otimes\tilde\alg)$ be a fundamental $\K$-homology
class for the pair $(\alg,\tilde\alg)$ and let
$\Xi\in\HE^d(\alg\otimes\tilde\alg)$ be a
fundamental cyclic cohomology class. Then the
{\it Todd class} of $\alg$ is defined to be the class
\beq
\Todd\big(\alg\big)=\Todd_{\Delta,\Xi}\big(\alg\,,\,\tilde\alg\big)
:=\Xi^\vee\otimes_{\tilde\alg}\ch\big(\Delta\big)
\label{Toddgendef}
\eeq
in the ring $\HE_{0}(\alg,\alg)$.
\end{definition}
\noindent
Recall that the map $\Xi^\vee\otimes_{\tilde\alg}(-)$
implements an isomorphism 
$\HE^d(\alg\otimes\tilde\alg) \cong \HE_{0}(\alg,\alg)$. The element
(\ref{Toddgendef}) is invertible with inverse given by
\beq
\Todd\big(\alg\big)^{-1}=(-1)^d~\ch\big(\Delta^\vee\,\big)
\otimes_{\tilde\alg}\Xi \ .
\label{Toddinverse}\eeq

\begin{remark}
Observe that the Todd class of an algebra
$\alg\in\underline{\mathfrak{D}}$ is trivial,
i.e., $\Todd(\alg)=1_{\alg}$ in $\HE_{0}(\alg,\alg)$, if and only if
$\ch(\Delta)=\Xi$ in
$\HE^d(\alg\otimes\tilde\alg)$.
\label{proptrivialTodd}\end{remark}

The Todd class depends on a number of choices, but this dependence can
be described by ``covariant'' actions on the classes.

\begin{theorem}
In the notation above, the Todd class of an algebra
$\alg\in\underline{\mathfrak{D}}$ has the following properties:
\begin{enumerate}
\item Suppose $\Delta$ and $\Xi$ are fundamental classes for the
strong PD pair $(\alg,\tilde\alg)$, with inverse fundamental classes 
$\Delta^\vee$ and $\Xi^\vee$. If there are $\KK$-equivalences 
  $\alpha\in\KK_0(\alg$, $\alg_1)$ and 
  $\beta\in\KK_0(\tilde\alg,\tilde\alg_1)$, then
$(\alg_1,\tilde\alg_1)$ is a strong PD pair, with fundamental
classes 
\[
\Delta_1 = (\alpha^{-1}\times \beta^{-1})\otimes_{\alg\otimes
\tilde \alg} \Delta,\qquad 
\Xi_1 = (\ch(\alpha)^{-1}\times \ch(\beta)^{-1})
\otimes_{\alg\otimes \tilde \alg} \Xi
\]
(where $\times$ denotes the exterior product) and inverse fundamental classes
\[
\Delta_1^\vee = \Delta^\vee 
\otimes_{\alg\otimes
\tilde \alg} (\alpha\times \beta),\qquad 
\Xi_1^\vee = \Xi^\vee \otimes_{\alg\otimes \tilde \alg}
(\ch(\alpha)\times \ch(\beta)) .
\]
Furthermore,
\eq
\label{KKeqchange}
\Todd_{\Delta_1, \Xi_1}(\alg_1,\tilde\alg_1) = \ch(\alpha)^{-1}\otimes_\alg 
\Todd_{\Delta, \Xi}(\alg,\tilde\alg) \otimes_\alg \ch(\alpha).
\end{equation}
\item If $(\ell,\ell_\HE)$ is an element of the duality group
  $\KK_0(\alg,\alg)^{-1}\times\HE_{0}(\alg,\alg)^{-1}$, then
\beq
\Todd_{\ell\otimes_\alg\Delta\,,\,\ell_\HE\otimes_{\alg}\Xi}
\big(\alg\,,\,\tilde\alg\big)~=~\ch\big(\ell\big)\otimes_{\alg}
\Todd_{\Delta,\Xi}\big(\alg\,,\,\tilde\alg\big)\otimes_{\alg}
\ell^{-1}_\HE \ .
\label{fundclasschange}\eeq
\end{enumerate}
\label{Toddcovtransfrules}\end{theorem}

\begin{proof}
The fact that $(\alg_1,\tilde\alg_1)$ is a strong PD pair
with fundamental classes $\Delta_1$ and $\Xi_1$ is routine and
quite similar to the calculation in Proposition \ref{kk:fund1}.
We proceed to compute the Todd class.
In terms of the diagram calculus of Appendix B, the picture is:
\[
\xymatrix{&&\alg_1 & &\alg_1
\ar '[dr] [drr]^{\ch(\Delta_1)} & & \\
\bbC \ar '[r]^{\Xi_1^\vee} [rru] \ar '[r] [rrd]&\circ&&&&\circ& \bbC \\
&&\tilde\alg_1 \ar[rr]^{1}&&\tilde\alg_1 \ar '[ur] [urr]& & }
\]
\begin{equation}
\xymatrix{&&&\alg \ar[r]^{\ch(\alpha)}&\alg_1 & & \alg_1 \ar[r]^{\ch(\alpha^{-1})}&\alg
\ar '[dr] [drr]^{\ch(\Delta)} & & \\
=&
\bbC \ar '[r]^{\Xi^\vee} [rru] \ar '[r] [rrd]
&\circ&&&&&&\circ& \bbC \\
&&&\tilde\alg  \ar[r]^{\ch(\beta)}&
\tilde\alg_1\ar[rr]^{1}&&\tilde\alg_1 \ar[r]^{\ch(\beta)^{-1}}
&\tilde\alg \ar '[ur] [urr]& & }
\end{equation}
This yields formula \eqref{KKeqchange} via associativity
in the formulation of Appendix B. 

The proof of (2)
is done with a very similar diagram.
\end{proof}
\begin{corollary}
\label{cor:Toddexample}
Suppose $\alg$ is a $C^*$-algebra that is strongly $\KK$-equivalent to $C(X)$,
where $X$ is an even-dimensional compact spin$^c$ manifold. Let
$\Todd(X) $ be the usual Todd class of $X$, but viewed as 
a bivariant cyclic homology class as above.
If $\alpha \in \KK(\alg, C(X))$ and  $\beta \in \KK(C(X), \alg)$ are
explicit $\KK$-equivalences inverse to one another, then
\beq
\Todd(\alg) = \ch(\alpha)\otimes_{C(X)}\Todd(X) \otimes_{C(X)}
\ch(\beta)\,.
\eeq
\end{corollary}
\begin{proof}
Immediate from  Theorem \ref{Toddcovtransfrules}, since $\Dirac_{X\times X}$
is a $\KK$ fundamental class for $C(X)$ and the usual homology
fundamental class provides another fundamental class in cyclic homology.
\end{proof}

\begin{remark}
There is more subtle real version of Kasparov's $\KK$-theory defined
for complex $C^*$-algebras with
involution \cite[Definition~9.18]{GVF}. For any separable
$C^*$-al\-ge\-bra $\alg$, the algebra $\alg\otimes\alg^\op$ may be
equipped with a canonical involution $\tau$ defined by
\beq
\tau(a\otimes b^\op)=b^*\otimes(a^*)^\op
\eeq
for all $a,b\in\alg$. The corresponding real $\K$-homology groups are
denoted $\KR^\bullet(\alg\otimes\alg^\op)$. There is a forgetful map
\beq
\frf\,:\,\KR^\bullet(\alg\otimes\alg^\op)~\longrightarrow~
\K^\bullet(\alg\otimes\alg^\op)
\label{forgetful}\eeq
from the real to the complex $\K$-homology of the algebra
$\alg\otimes\alg^\op$. Suppose that $\alg$ is a strong PD algebra
which admits a fundamental $\KR$-homology class
$\Delta_\real\in\KR^d(\alg\otimes\alg^\op)$. Let
$\Xi\in\HE^d(\alg\otimes \alg^\op)$ be a fundamental
cyclic cohomology class for $\alg$. The image of $\Delta_\real$ under the
homomorphism (\ref{forgetful}) is a fundamental $\K$-homology class
for $\alg$ and the corresponding Todd class (\ref{Toddgendef}) in 
$\HE_{0}(\alg,\alg)$ is called the {\it Atiyah-Hirzebruch
  class} of $\alg$, denoted
\beq
\widehat{A}\big(\alg\big):=\Xi^\vee\otimes_{\alg^\op}
\ch\big(\frf(\Delta_\real)\big) \ .
\label{Aroofgenusalg}\eeq
It satisfies the same basic properties as the Todd class above, and
may be related to (\ref{Toddgendef}) for any other fundamental
$\K$-homology class $\Delta$ by using the action of the duality group
$\KK_0(\alg,\alg)^{-1}$ in (\ref{fundclasschange}).
\label{AroofPDrem}\end{remark}

\subsection{Gysin Homomorphisms}

Let $f : \alg \rightarrow \balg$ be a morphism of separable
$C^*$-algebras. It induces morphisms in $\K$-theory,
\beq
f_* \,:\, \K_\bullet (\alg)  ~\longrightarrow~ \K_\bullet(\balg) \ ,
\eeq
and morphisms in $\K$-homology,
\beq
f^* \,:\, \K^\bullet (\balg)  ~\longrightarrow~ \K^\bullet(\alg) \ .
\eeq
We will now describe how to construct Gysin maps (or ``wrong way''
maps) on these groups. If both $\alg$ and $\balg$ are PD algebras,
then they are easily constructed as analogues of the classical
``Umkehrhomomorphismus''. In this case, there are isomorphisms
\beq
{\sf Pd_\alg} \,:\, \K_\bullet (\alg)
~\stackrel{\approx}{\longrightarrow}~ \K^{\bullet-d_\alg} (\alg^\op)\quad
\mbox{and} \quad
{\sf Pd_\balg} \,:\, \K_\bullet (\balg)
~\stackrel{\approx}{\longrightarrow}~ \K^{\bullet-d_\balg} (\balg^\op) \
.
\eeq
We can then define the Gysin map in $\K$-theory,
\beq
f_! \,:\, \K_\bullet (\balg)  ~\longrightarrow~
\K_{\bullet+d}(\alg) \ ,
\eeq
(where $d = d_\alg-d_\balg$) as the composition
\beq
f_!\,:\,\K_\bullet (\balg) ~\xr{{\sf Pd}^{\phantom{1}}_\balg}~
\K^{\bullet -d_\balg}(\balg^\op)~\xr{(f^\op)^*}~
\K^{\bullet -d_\balg}(\alg^\op)~\xr{{\sf Pd}_\alg^{-1}}~
\K_{\bullet+d}(\alg) \ .
\eeq
Under the same hypotheses, we can similarly define the 
Gysin map in $\K$-homology,
\beq
f^! \,:\, \K^\bullet (\alg)  ~\longrightarrow ~
\K^{\bullet+d}(\balg) \ ,
\eeq
as the composition
\beq
f^!\,:\,\K^\bullet (\alg)~\xr{{\sf Pd}_\alg^{-1}}~
\K_{\bullet +d_\alg}(\alg^\op)~\xr{f^\op_*}~
\K_{\bullet +d_\alg}(\balg^\op)~\xr{{\sf Pd}^{\phantom{1}}_\balg}~
\K^{\bullet+d}(\balg) \ .
\eeq
However, this relies on the fact that $\alg$ and $\balg$ are PD
algebras, which is in general too stringent of a requirement. We will
therefore proceed to some more general constructions.

\subsection{Strongly $\K$-oriented Maps}

We will consider a  subcategory $\underline\fC$  of the category of
separable $C^*$-algebras and morphisms of $C^*$-algebras, 
consisting of {\em strongly $\K$-oriented morphisms}. $\underline\fC$ comes
equipped with a contravariant functor $!:\underline\fC \to
\underline\KK$, sending  
$$\underline\fC \ni (\alg\stackrel{f}\longrightarrow\balg)
\longrightarrow f! \in \KK_d(\balg, \alg)$$ 
and having the following properties:

\begin{enumerate}
\item For any $C^*$-algebra $\alg$, the identity morphism $\Id_\alg:
  \alg \to \alg$ is strongly $\K$-oriented 
with $(\Id_\alg)!=1_\alg$, and the $0$-morphism $0_\alg:\alg\to 0$ is
  strongly $\K$-oriented 
with $(0_\alg)!=0\in \KK(0,\alg)$;

\item If $(\alg\stackrel{f}\longrightarrow\balg) \in \underline\fC$, then
  $(\alg^\op\stackrel{f^\op}\longrightarrow\balg^\op) \in \underline\fC$ and
  moreover $(f!)^\op = (f^\op)! $; 

\item If $\alg$ and $\balg$ are strong PD algebras, then any morphism 
$(\alg\stackrel{f}\longrightarrow\balg)\in \underline\fC$, and $f!$ is
  determined as follows: 
$$f! = (-1)^{d_\alg}
\Delta_\alg^\vee \otimes_{\alg^\op} [f^\op] \otimes_{\balg^\op}
  \Delta_\balg, $$ 
where for the rest of the paper, $[f] = \KK(f)$ denotes the class in
  $\KK(\alg, \balg)$ of the morphism  
$(\alg\stackrel{f}\longrightarrow\balg)$ and $[f^\op]$ is defined similarly.

As Kasparov products like this are rather hard to visualize when
written this way, it is useful 
to use the diagram calculus developed in Appendix B.  In these terms,
$f!$ is represented by the picture depicted by Figure \ref{fig:fshriek}.
\begin{figure}[ht]
\[
\xymatrix{&&\alg & &\balg
\ar '[dr] [drr]^{\Delta_\balg} & & \\
\bbC \ar '[r]^{\Delta^\vee_\alg} [rru] \ar '[r] [rrd]&\circ&&&&\circ& \bbC \\
&&\alg^\op \ar[rr]^{f^\op}&&\balg^\op \ar '[ur] [urr]& & }
\]
\caption{Diagram representing the construction of $f!$.  The 
``free ends'' are on the top line and concatenation is done
on the bottom line.}
\label{fig:fshriek}
\end{figure}
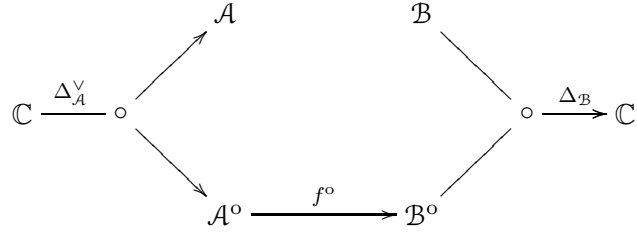
\end{enumerate}

Actually, it is not immediately obvious that property (3) above is
compatible with the required 
functoriality. However, consistency of the definition follows from the
following: 

\begin{lemma}[{\bf Functoriality of the Gysin map}]\label{lem:functofshriek}
If $\alg$, $\balg$ and $\calg$ are strong PD algebras, and if
$f:\alg\to\balg$, $g:\balg\to\calg$ are morphisms of $C^*$-algebras, then
\[
\begin{aligned}
&\bigl( (-1)^{d_\alg} \Delta^\vee_\alg \otimes_{\alg^\op} [f^\op]
\otimes_{\balg^\op} \Delta_\balg \bigr)
\otimes_\balg
\bigl( (-1)^{d_\balg}
\Delta^\vee_\balg \otimes_{\balg^\op} [g^\op] \otimes_{\calg^\op}
\Delta_\calg \bigr) \\
&=\bigl( (-1)^{d_\alg} \Delta^\vee_\alg \otimes_{\alg^\op} [(g\circ f)^\op]
\otimes_{\calg^\op} \Delta_\calg \bigr).
\end{aligned}
\]
\end{lemma}
\begin{proof}
Note that, by associativity of the Kasparov product,
\[
\begin{aligned}
&\bigl(\Delta^\vee_\alg \otimes_{\alg^\op} [f^\op]
\otimes_{\balg^\op} \Delta_\balg)
\otimes_\balg
\bigl(\Delta^\vee_\balg \otimes_{\balg^\op} [g^\op]
\otimes_{\calg^\op} \Delta_\calg \bigr)\\
&=\Delta^\vee_\alg \otimes_{\alg^\op} \bigl( [f^\op]
\otimes_{\balg^\op} \Delta_\balg
\otimes_\balg
\Delta^\vee_\balg \otimes_{\balg^\op} [g^\op] \bigr)
\otimes_{\calg^\op} \Delta_\calg .
\end{aligned}
\]
But
\[
\begin{aligned}\,
[f^\op]\otimes_{\balg^\op} \Delta_\balg
\otimes_\balg \Delta^\vee_\balg \otimes_{\balg^\op} [g^\op]
&= [f^\op] \otimes_{\balg^\op} (\Delta_\balg
\otimes_\balg \Delta^\vee_\balg ) \otimes_{\balg^\op} [g^\op]\\
&= [f^\op] \otimes_{\balg^\op} (-1)^{d_\balg}
1_{\balg ^\op}\otimes_{\balg^\op}[g^\op]\\
&= (-1)^{d_\balg} [f^\op] \otimes_{\balg^\op} [g^\op] = (-1)^{d_\balg}
[(g\circ f)^\op]
\end{aligned}
\]
and the result follows.
\end{proof}

We now exhibit more examples of elements in this category $\underline\fC$. 
In the following example, the $C^*$-algebras are not strong PD algebras,
but yet we can get an element in this category $\underline\fC$.
Suppose that we 
are given oriented manifolds $X$ and $Y$, and classes $H_X \in \H^3(X,
\bbZ)$ and   
$H_Y \in \H^3(Y, \bbZ)$. A smooth map $f\colon X \to Y$ defines a
morphism $f^*\colon 
C_0(Y, \cE_{H_Y}) \longrightarrow C_0(X, \cE_{H_X})$ if $f^*H_Y = H_X$. Since 
$X$ and $Y$ are oriented, then by Example 2.5, the pair $(C_0(X),
C_0(X, \Cl(TX)))$ is a strong PD pair, that is, there is a fundamental class $\Delta_X \in 
\KK(C_0(X) \otimes C_0(X, \Cl(TX), \bbC).$  Since $\cE_{H_X} \otimes \cE_{H_X}^\op$
is stably isomorphic to the trivial bundle $X \times \cK$, it follows that 
$C_0(X) \otimes C_0(X, \Cl(TX)$ is stably isomorphic to $C_0(X, \cE_{H_X}) \otimes 
C_0(X, \cE_{H_X}^\op \otimes  \Cl(TX).$
Therefore $$\KK(C_0(X) \otimes C_0(X, \Cl(TX), \bbC) \cong \KK( C_0(X, \cE_{H_X}) \otimes 
C_0(X, \cE_{H_X}^\op \otimes  \Cl(TX), \bbC),$$ giving rise to a
fundamental class in $\KK( C_0(X, \cE_{H_X}) \otimes C_0(X, \cE_{H_X}^\op \otimes  \Cl(TX), \bbC)$,
and showing that $(C_0(X, \cE_{H_X}),  C_0(X, \cE_{H_X}^\op \otimes  \Cl(TX)))$ is a
strong PD pair. The analogous statement is true for
$Y$. Finally, if $f^*\W_3(Y) 
 = \W_3(X)$, where  
$\W_3(X) \in \H^3(X, \bbZ)$ is the third integral Stiefel-Whitney
class of $X$, then we get the commutative diagram,
\beq
\xymatrix{
\K_\bullet(C_0(Y, \cE_{H_Y}^\op \otimes  \Cl(TY)))\ar[d]_{{\sf Pd}_Y}\ar[r]^{f^*}&
\K_\bullet(C_0(X, \cE_{H_X}^\op \otimes  \Cl(TX)))\ar[d]^{{\sf Pd}_X} \\
\K_\bullet(C_0(Y, \cE_{H_Y}))
\ar[r]_{f^*}&
\K_\bullet(C_0(X, \cE_{H_X})) \ , }
\label{CT-PD}\eeq
where the vertical arrows are isomorphisms.
Then 
\[
(f^*)! \in \KK(C_0(X, \cE_{H_X}),C_0(Y, \cE_{H_Y}))
\]
is defined as  
the Kasparov product
$(-1)^{\dim Y}\Delta^\vee_Y \otimes [f^*] \otimes \Delta_X$.  

This is a special case of the more general situation given as
follows. Let $(\alg_i, \balg_i), \, i=1, 2$  
be strong PD pairs with fundamental classes $\Delta_i, \, i=1,2$ respectively, 
and let
$f: \alg_1 \to \alg_2$ be a morphism. Then $f! \in \KK(\balg_2, \balg_1)$
is defined using the diagram calculus in Appendix B as 
$(-1)^{d_1}\Delta_1^\vee \otimes_{\alg_1} [f] \otimes_{\alg_2} \Delta_2$.

There are also many interesting examples of strongly $\K$-oriented
maps between noncommutative foliation $C^*$-algebras constructed
in  \cite{HilSk}.

\subsection{Weakly $\K$-oriented Maps}

We will consider a  subcategory $\underline\fC_w$  of the
category of separable $C^*$-algebras and morphisms of $C^*$-algebras, 
consisting of ``weakly $\K$-oriented morphisms''. $ \underline\fC_w$ comes
equipped with a contravariant functor ${}_!: \underline\fC_w \to
\underline{\sf Ab}$ ($ \underline{\sf Ab}$ denotes the  
category of $\bbZ_2$-graded abelian groups) 
sending $$\underline\fC_w \ni (\alg\stackrel{f}\longrightarrow\balg)
\longrightarrow f_! \in \Hom_\bbZ(\K_\bullet(\balg),  
\K_{\bullet+d}(\alg))$$
and having the following properties:

\begin{enumerate}
\item For any $C^*$-algebra $\alg$, the identity morphism $\Id_\alg:
  \alg \to \alg$ is weakly $\K$-oriented;

\item If $(\alg\stackrel{f}\longrightarrow\balg) \in \underline\fC_w$,
  then $(\alg^\op\stackrel{f^\op}\longrightarrow\balg^\op) \in
  \underline\fC_w$ and moreover $(f_!)^\op = (f^\op)_! $; 

\item If $\alg$ and $\balg$ are weak PD algebras, then any morphism
  $(\alg\stackrel{f}\longrightarrow\balg)\in \underline\fC_w$, and
  $f_!$ is determined as follows: 
$$f_! ={\PD}_\alg^{-1} \circ (f^\op)^* \circ {\PD}_\balg, $$
where $(f^\op)^*$ denotes the morphism in
$\Hom_\bbZ(\K^{\bullet}(\alg^\op),\K^\bullet(\balg^\op))$.  

\end{enumerate}

This definition generalizes the one in the previous subsection in the
following sense: 
\begin{proposition}
\label{prop:strongtoweakshriek}
$\underline\fC$ can be taken to be a subcategory of $\underline\fC_w$.
In other words, if 
$f:\alg \to \balg$ is a morphism in $\underline\fC$, then $f_!=
(f!)_*:\K_\bullet(\balg) \to
\K_{\bullet+d}(\alg)$ satisfies the above requirements.
\end{proposition}

\begin{proof}
Functoriality is obvious since we are merely composing the functor
from $\underline\fC$ to $\underline\KK$ with the functor from
$\underline\KK$ to $\underline{\sf Ab}$ that sends $\alg \mapsto
\K_\bullet(\alg)$, $\KK(\alg,\balg)\ni x
\mapsto x_* \in \Hom_\bbZ(\K_\bullet(\alg), \K_\bullet(\balg))$.

We need to check property (3).  In other words, suppose $\alg$ and $\balg$
are
strong PD algebras and $f!\in \KK_d(\balg, \alg)$ is defined to be
$(-1)^{d_\alg}\Delta^\vee_\alg \otimes_{\alg^\op} [f^\op]
\otimes_{\balg^\op} \Delta_\balg$. We want to show that the induced map on
$\K_\bullet$ is $\PD_\alg^{-1} \circ (f^\op)^* \circ \PD_\balg$. However, this
is obvious, since the Kasparov product with
$\Delta_\balg \in \KK_{d_\balg}(\balg\otimes\balg^\op,\bbC)$ is
$\PD_\balg: 
\K_\bullet(\balg) \to \K^{\bullet - d_\balg}(\balg^\op)$ and the Kasparov
product with 
$(-1)^{d_\alg}\Delta^\vee_\alg \in \KK_{d_\alg}(\bbC,\alg\otimes\alg^\op)$
is $\PD_\alg^{-1}:
\K^\bullet(\alg) \to \K_{\bullet + d_\alg}(\alg^\op)$.
\end{proof}

\begin{remark} The Gysin maps in \K-homology 
$f^!\in{\rm Hom}_\zed(\K^\bullet(\alg)\,,\K^{\bullet+d}(\balg))$
can also be defined with completely 
analogous properties.  There are also the obvious $\HE$-theory
analogues, used in the next subsection.
\end{remark}

\subsection{Grothendieck-Riemann-Roch Formulas: the Strong Case}

The Grothendieck-Rie\-mann-Roch formula compares 
the two bivariant cyclic classes ${\ch}(f!)$ and $f^\HE! $.

\begin{theorem}\label{thm:GRR-strong}
Suppose $\alg$ and $\balg$ are strong PD algebras with given
$\HE$ fundamental classes.
Then one has the Grothendieck-Rie\-mann-Roch formula,
\beq
{\ch}(f!) = (-1)^{d_\balg}~
{\sf Todd}(\balg) \otimes_{\balg} f^\HE!  \otimes_{\alg}
{\sf Todd}(\alg)^{-1}. 
\label{eqn:GRR-strong}\eeq
\end{theorem}

\begin{proof} We will write out the right-hand side of
\eqref{eqn:GRR-strong} and simplify. 
In the notation of Definition \ref{defn:Todd}, the {Todd class} of
$\balg$ is the class 
\beq
\Todd\big(\balg\big)=
 \Xi_\balg^\vee\otimes_{\tilde\balg}\ch\big(\Delta_\balg\big) \in  
\HE_{0}(\balg, \balg)
\eeq
and the inverse of  the {Todd class} of $\alg$ is the class
\beq
\Todd\big(\alg\big)^{-1}=(-1)^{d_\alg}~\ch\big(\Delta_\alg^\vee\,\big)
\otimes_{\tilde\alg}\Xi_\alg \in \HE_{0}(\alg, \alg)\ .
\eeq
Since $\alg$ and $\balg$ are strong PD algebras, then $f^\HE!$ is
determined as follows: 
$$f^\HE! = (-1)^{d_\alg}~\Xi_\alg^\vee \otimes_{\tilde\alg} [(f^{\HE})^\op]
\otimes_{\tilde\balg} \Xi_\balg, $$  
where $[f^\HE] = \HE(f)$ denotes the class in $\HE(\alg, \balg)$ of
the morphism  
$(\alg\stackrel{f}\longrightarrow\balg)$ and $[(f^{\HE})^\op]$ is
defined similarly. 

Therefore the right hand side of \eqref{eqn:GRR-strong} is equal to
$$
(-1)^{d_\balg}~
\Bigl(\Xi_\balg^\vee\otimes_{\tilde\balg}\ch\big(\Delta_\balg\big)\Bigr)
\otimes_{\balg}  \bigl(\Xi_\alg^\vee \otimes_{\tilde\alg} [(f^{\HE})^\op]
\otimes_{\tilde\balg} \Xi_\balg \bigr) 
\otimes_{\alg} \Bigl(\ch\big(\Delta_\alg^\vee\,\big)
\otimes_{\tilde\alg}\Xi_\alg\Bigr)\ ,
$$
which by the associativity of the intersection product, or
equivalently by the diagram calculus 
of Appendix B (there it is worked out for $\KK$, but it works the same way
for $\HE$), is equal to
$$
(-1)^{d_\balg}~\Bigl(\Xi_\alg^\vee \otimes_{\alg}
\bigl( \ch\big(\Delta_\alg^\vee\,\big) \otimes_{\tilde\alg} \Xi_\alg 
 \bigr) \Bigr)
 \otimes_{\tilde\alg} [(f^{\HE})^\op] \otimes_{\tilde\balg}
 \Bigl(\bigl(\Xi_\balg^\vee\otimes_{\tilde\balg}\ch\big(\Delta_\balg\big)
 \bigr) \otimes_{\balg} \Xi_\balg \Bigr)\ . 
$$

On the other hand, 
$$f! = (-1)^{d_\alg}~\Delta_\alg^\vee \otimes_{\tilde\alg} [f^\op]
\otimes_{\tilde\balg} \Delta_\balg.$$ 
Therefore the left hand side of \eqref{eqn:GRR-strong} is equal to
$$
(-1)^{d_\alg}~
{\sf ch}(\Delta_\alg^\vee) \otimes_{\tilde\alg} {\sf ch}[f^\op]
\otimes_{\tilde\balg} {\sf ch}(\Delta_\balg). 
$$
By the functorial properties of the bivariant Chern character, one has
\beq
{\sf ch}[f^\op]  = [(f^{\HE})^\op] .
\eeq
In order to prove the theorem, it therefore  suffices to prove that 
\beq
\bigl( \Xi_\balg^\vee\otimes_{\tilde\balg}\ch\big(\Delta_\balg\big)\bigr)
\otimes_{\balg} \Xi_\balg 
= (-1)^{d_\balg}~{\sf ch}(\Delta_\balg),
\eeq
and 
\beq
\Xi_\alg^\vee \otimes_{\alg}  \bigl(\ch\big(\Delta_\alg^\vee\,\big)
\otimes_{\tilde\alg}\Xi_\alg\bigr)
= (-1)^{d_\alg}~{\sf ch}(\Delta^\vee_\alg).
\eeq
But both of these equalities also follow easily from the diagram
calculus:
\beq
\begin{aligned}
(\Xi_\balg^\vee\otimes_{\tilde\balg}\ch\big(\Delta_\balg\big))
  \otimes_{\balg} \Xi_\balg 
&= (\Xi_\balg^\vee \otimes_{\balg}
  \Xi_\balg)\otimes_{\tilde\balg}\ch\big(\Delta_\balg\big)\\ 
& = (-1)^{d_\balg}
1_{\tilde\balg} \otimes_{\tilde\balg}\ch\big(\Delta_\balg\big)\\
& = (-1)^{d_\balg} \ch\big(\Delta_\balg\big)
\end{aligned}
\eeq
and
\beq
\begin{aligned}
\Xi_\alg^\vee \otimes_{\alg}  \bigl(\ch\big(\Delta_\alg^\vee\,\big)
\otimes_{\tilde\alg}\Xi_\alg\bigr)
&= \ch\big(\Delta_\alg^\vee\,\big) \otimes_{\tilde\alg}
\bigl (\Xi_\alg^\vee \otimes_{\alg}
  \Xi_\alg \bigr)\\ 
& = \ch\big(\Delta_\alg^\vee\,\big) \otimes_{\tilde\alg}
(-1)^{d_\alg} 1_{\tilde\alg} \\
& = (-1)^{d_\alg} \ch\big(\Delta_\alg^\vee\,\big) \ .
\end{aligned}
\eeq
\end{proof}

\subsection{Grothendieck-Riemann-Roch Formulas: the Weak Case}

For $(\alg\stackrel{f}\longrightarrow\balg) $, weakly $\K$-oriented, the
Grothendieck-Riemann-Roch formula repairs the noncommutativity of the 
diagram, 
\beq
\xymatrix{
\K_\bullet(\balg)\ar[d]_{{\sf ch}}\ar[r]^{f_!}&
\K_{\bullet+d}(\alg)\ar[d]^{{\sf ch}} \\
\HE_\bullet(\balg)
\ar[r]_(.42){f^\HE_!}&
 \HE_{\bullet+d}(\alg) \  . }
\label{GRR-false}\eeq
The following can be proved in an analogous way to the strong case of the 
Grothendieck-Riemann-Roch formula, Theorem \ref{thm:GRR-strong},
so we will omit the proof.

\begin{theorem}\label{thm:GRR-weak}
If $f\colon\alg \to  \balg$ is a morphism in $\underline\fC_w$,  and
$\xi\in \K_\bullet(\balg)$, 
then one has the Grothendieck-Riemann-Roch formula,
\beq
{\sf ch}(f_!\xi) \otimes_{\alg}  {\sf Todd}(\alg)= 
(-1)^{d_\balg}~f^\HE_! \Bigl({\sf ch}(\xi) \otimes_{\balg} 
{\sf Todd}(\balg) \Bigr)\,. 
\label{eqn:GRR-weak}\eeq
\end{theorem}

\begin{remark}
\label{rem:ASfromGRR}
Let $\alg$ be a unital PD algebra having an even degree fundamental class
in $\K$-theory. Then there
is a canonical morphism, $\lambda : \bbc \to \alg$,
given by $\bbc \ni z \mapsto z\cdot 1 \in \alg$, where $1$ denotes the 
unit in $\alg$.  Observe that
$\lambda$ is always weakly $\K$-oriented, since $\bbc$ is 
a PD algebra, and the Gysin map
$\lambda_! : \K_0(\alg) \to \bbz$
is the analog of the topological index morphism (for
compact manifolds). Theorem 
\ref{thm:GRR-weak} above applied to this situation says that,
$$
\lambda_!(\xi) = \lambda^\HE_!\bigl(\ch(\xi) \otimes_\alg
\Todd(\alg)  \bigr),
$$
where $\lambda^\HE_! : \HE_0(\alg) \to \bbc$
is the associated Gysin morphism in cyclic theory.
In the case where $\alg=C(X)$, $X$ a compact spin$^c$ manifold,
this is just the usual Atiyah-Singer index theorem. Indeed,
$\lambda_!(\xi) = \Ind \PD_X(\xi) = \Ind(\Dirac_\xi)$, while
 the other side of the index formula is
$\lambda^\HE_!\bigl(\ch(\xi) \otimes_\alg
\Todd(\alg)  \bigr) = 
\bigl( \Todd(X)\cup \ch(\xi)\bigr) [X]$. Note
in particular that when $\xi$ is the canonical rank one free module
over $\alg$, then we obtain a numerical invariant which we call the
\emph{Todd genus} of $\alg$, a characteristic number of the algebra.
\end{remark}

\newsection{Noncommutative D-Brane Charges\label{CyclicDbrane}}

In this final section we will come to the main motivation for the
present work, the  ``D-brane charge formula'' for very general noncommutative
spacetimes. The crux of the definition of D-brane charge in
Section~\ref{ChargeDef} relied upon the introduction of natural
pairings in $\K$-theory and singular cohomology, which in turn arose  
as a consequence of Poincar\'e duality. We are now ready to describe 
the analogs of the natural pairings in appropriate noncommutative
cases. The key point is that the
multiplication map $m:\alg\otimes\alg\to\alg$ is an algebra
homomorphism only in the commutative case and one needs to replace its
role with some new construct. This is where the formalism of $\KK$-theory
plays a crucial role. Mathematically, the problem is concerned with taking
the square root of the Todd class of a noncommutative spacetime,
under mild hypotheses. This then enables one to ``correct'' the
Chern character so that the index pairing in $\K$-theory and the
given pairing in $\HE$-theory agree.

\subsection{Poincar\'e Pairings\label{PPairings}}

In the notation of Section 7.1, let $\alg\in\underline{\mathfrak{D}}$ and
$\alpha\in\K_i(\alg)$, $\beta\in\K_{-d-i}(\tilde\alg)$. Then there
is a pairing
\beq
(\alpha,\beta)~\longmapsto~
\langle\alpha,\beta\rangle~=~(\alpha\times\beta)
\otimes_{\alg\otimes\tilde\alg}\Delta\in\KK_0(\bbC, \bbC)=\zed \ .
\label{pairing}\eeq
In the case where $\alg=\tilde\alg=C(X)$ is the algebra of continuous functions
on a spin$^c$ manifold and the fundamental class $\Delta$ comes from
the Dirac operator, this is the same as the pairing \eqref{Kthpairing}
introduced earlier, and is the $\K$-theory analogue of the
cup-product pairing \eqref{cohpairing}. Indeed, in this case,
\[
\langle\alpha,\beta\rangle = \PD_X(\alpha) \otimes_{C(X)} \beta
= \Dirac_\alpha \otimes_{C(X)} \beta = {\rm index}(\Dirac_{\alpha\otimes\beta})\,.
\]
If $\alg$ and $\tilde\alg$ have
finitely generated $\K$-theory and satisfy the Universal Coefficient
Theorem (UCT), then
the pairing (\ref{pairing}) is nondegenerate modulo torsion. 

In the case of
a strong PD algebra, since we have $\tilde\alg = \alg^\op$,
whose $\K$-theory is canonically isomorphic to that of $\alg$ itself
(by Remark \ref{rem:Kequivop}),
the pairing \eqref{pairing} can be
viewed as a pairing of $\K_\bullet(\alg)$ with itself. Then we are led
to consider the following additional condition.
\begin{definition}
\label{def:symm}
A fundamental class $\Delta$ of a strong PD algebra $\alg$ is said to be
{\em symmetric} if $\sigma(\Delta)^\op = \Delta \in \K^d(\alg\otimes 
\alg^\op)$ where
\beq\label{sigma}
\sigma :
\alg\otimes\alg^\op \longrightarrow \alg^\op \otimes \alg
\eeq
is the involution $x\otimes y^\op \mapsto y^\op \otimes x$
and $\sigma$ also denotes the induced map on $\K$-homology.
In terms of the diagram calculus of Appendix B, $\Delta$ being symmetric
implies that
\[
\xymatrix{\alg \ar[r]^{x}&\alg \ar '[rd] [drr]^\Delta
& & &&\alg \ar[r]^{y}&\alg \ar '[rd] [rrd] &&\\
&&\circ&\bbC&= & &&\circ&\bbC\\
\alg^\op \ar[r]^{y^\op}& \alg^\op \ar '[ur] [urr]& & 
&&\alg^\op \ar [r]^{x^\op}& \alg^\op\ar '[ur] [urr]^{\Delta} &&}
\]
for all $x$ and $y$.
\end{definition}
Symmetry is a natural condition to consider, since
the intersection pairing
on an even-dimensional manifold is symmetric. 

\begin{proposition}\label{ppairing}
For any strong Poincar\'{e} duality algebra $\alg$ there exists a
bilinear pairing on $\K$-theory:
\[
\langle-,-\rangle\,:\,\K_i(\alg) \times \K_{d-i}(\alg) ~
\longrightarrow~ \bbz 
\]
defined by 
\beq
\label{eq:k-pairing}
\langle\alpha,\beta\rangle ~=~(\alpha\times\beta^\op)
\otimes_{\alg\otimes\alg^\op} \Delta\in\KK_0(\bbC, \bbC)=\zed \ .
\eeq
Moreover, if the fundamental class $\Delta$ is symmetric,
then the bilinear pairing \eqref{eq:k-pairing} on $\K$-theory is  
symmetric.  If $\alg$ satisfies the UCT in $\K$-theory and has
finitely generated $\K$-theory, then the pairing \eqref{eq:k-pairing}
is nondegenerate modulo torsion.
\end{proposition}
\begin{proof}
Immediate from the remarks above.
\end{proof}

If $\alg$ is a strong C-PD algebra, then the local cyclic homology and
cohomology of $\alg$ are isomorphic. This is equivalent to saying that the
canonical pairing
\beq
(-,-)\,:\,\HE_i(\alg)\otimes_\complex\HE_{d-i}(\alg)~\longrightarrow~
\complex
\label{HPpairingcan}\eeq
on cyclic homology, given by
\beq
(x,y)=(x\times y^\op)\otimes_{\alg\otimes\alg^\op}\Xi
\label{HPpairing2}\eeq
for $x\in\HE_i(\alg)$ and $y\in\HE_{d-i}(\alg)$, is non-degenerate, since
the pairing between $\HE_\bullet(\alg)$ and $\HE^\bullet(\alg)$ is always
non-degenerate for any algebra, at least if the universal
coefficient theorem holds. In the
commutative case, this pairing coincides with the intersection form
(\ref{cohpairing}).

If $\alg$ is a strong PD algebra, then one can also define a bilinear form
on cyclic homology determined by the class $\ch(\Delta)$ as
\beq
(-,-)_\frh\,:\,\HE_i(\alg)\otimes_\complex\HE_{d-i}(\alg)~
\longrightarrow~\complex
\label{HPpairing}\
\eeq
by setting
\beq
(x,y)_\frh=(x\times y^\op)\otimes_{\alg\otimes\alg^\op}\ch(\Delta)
\label{HPpairing1}\eeq
for $x\in\HE_i(\alg)$ and $y\in\HE_{d-i}(\alg)$.

A fundamental class in $\HE$-theory is said to be {\em symmetric} if
$\sigma(\Xi)^\op = \Xi \in \HE^d(\alg\otimes\alg^\op)$, where $\sigma$
is the involution defined earlier in \eqref{sigma}
and $\sigma$ also denotes the induced map on $\HE$-theory.

\subsection{D-Brane Charge Formula for Noncommutative Spacetimes\label 
{GenConstr}}

If $\alg$ is a strong PD algebra, then we have defined in the previous
subsection
two pairings, one given by the formula~\eqref{HPpairing1},
and the other by the formula \eqref{HPpairing2}.
These two pairings will {\it a priori} be different. Comparing them is
the crux of our definition of D-brane charge. Let us begin with the
following observation.

\begin{proposition}
\label{prop:isometry}
If $\alg$ is a strong PD algebra, then the 
Chern character $\ch:\K_\bullet(\alg) \rightarrow \HE_{\bullet}(\alg)$
is an isometry with respect to the inner products given in equations
\textup{\eqref{eq:k-pairing}} and \textup{\eqref{HPpairing1}},
\beq
\big\langle p\,,\,q\big\rangle=\big(\ch(p)\,,\,\ch(q)\big)_\frh \ .
\label{chNCisometry}\eeq
\end{proposition}
\begin{proof}
Using multiplicativity of the Chern character, one has
\beq
\label{eq:chisom2}
\big(\ch(p)\,,\,\ch(q)\big)_\frh=\ch\bigl( (p\times q^\op)
\otimes_{\alg\otimes\alg^\op}\Delta \bigr) \ .
\eeq
Now use the fact that $\ch$ is a unital homomorphism (this is essentially
the index theorem, i.e., the statement that the index pairing
(\ref{eq:k-pairing}) coincides with the canonical pairing between the
corresponding Chern characters in local cyclic homology).
\end{proof}
\noindent
 From this Proposition it follows that the bilinear form
(\ref{HPpairing1}) is the analogue of the twisted inner product
defined in (\ref{hpairing}). Finding the appropriate modified Chern
character which maps (\ref{HPpairing2}) onto (\ref{HPpairing1}) will
thereby yield the formula for D-brane charge that we are looking for.

The technical problem that we are
faced with is to take the square root of the Todd class ${\sf Todd} 
(\alg)$
in $\HE(\alg, \alg)$. To do this, we will assume that the Universal  
Coefficient
Theorem holds for the noncommutative spacetime $\alg$. Then
$\HE(\alg, \alg) = \End(\HE_\bullet(\alg))$. In addition, we  
will assume that $\dim_{\bbc}\HE_\bullet(\alg)$ is finite,
say equal to $n$.  Then since ${\sf Todd}(\alg)$
is in $GL(\HE_\bullet(\alg))\cong GL_n(\bbC)$ 
and every matrix in $GL_n(\bbC)$ has
a square root (use the Jordan canonical form to prove this!), 
we can take a square root, $\sqrt{{\sf Todd}(\alg)}$.
Using the UCT, $\sqrt{{\sf Todd}(\alg)}$ can again be considered as  
an element
in $\HE(\alg, \alg)$.  The square root is not unique, but we fix a 
choice.  In some cases, the Todd class may be self-adjoint and positive
with respect to a suitable inner product on $\HE_\bullet(\alg)$,
which might help to pin down a more canonical choice.
In any event, we have the following theorem.

\begin{theorem}[{\bf Isometric pairing formula}]
\label{thm:chNCisometry1}
Suppose that  the noncommutative spacetime $\alg$ satisfies 
the UCT for local cyclic homology, and that $\HE_\bullet(\alg)$ is a finite
dimensional vector space.  
If  $\alg$ has symmetric \textup{(}even-dimensional\textup{)}
fundamental classes both in $\K$-theory and in cyclic theory, then
the modified Chern character
\beq
\ch \otimes_\alg \sqrt{{\sf Todd}(\alg)}:\K_\bullet(\alg) \rightarrow 
\HE_ {\bullet}(\alg)
\eeq
is an isometry with respect to the inner products \eqref{pairing} and
\eqref{HPpairing2},
\beq
\big\langle p\,,\,q\big\rangle=\big(\ch(p) \otimes_\alg
\sqrt{{\sf Todd}(\alg)} \,,\,\ch(q)\otimes_\alg \sqrt{{\sf
    Todd}(\alg)}~\big) \ .
\label{chNCisometry1}\eeq
\end{theorem}

\begin{proof}
To prove the theorem, we use Proposition \ref{prop:isometry} 
and observe that it's
enough to show that the right-hand sides of equations \eqref{chNCisometry1}
and \eqref{eq:chisom2} agree.  For this we
use the diagram calculus of appendix B,
\[
 \xymatrix@C-.5pc{\alg \ar[r]^{\sf Todd}&\alg \ar '[rd] [drr]^{\Xi}
& & &&\alg\ar[r]^{\sqrt{\sf Todd}}
&\alg\ar[r]^{\sqrt{\sf Todd}}&\alg\ar '[rd] [rrd] &&\\
&&\circ&\bbC&= &&&&\circ&\bbC\\
& \alg^\op\ar '[ur] [urr]& & 
&& &\alg^\op \ar [r]^{1_{\alg^\op}}& \alg^\op \ar '[ur] [urr]^{\Xi} &&}
\]

\[
\xymatrix@C-.5pc{
& \alg  \ar[r]^{\sqrt{\sf Todd}}
&\alg \ar[r]^{1_\alg}&\alg \ar '[rd] [rrd] & &
&\alg \ar[r]^{\sqrt{\sf Todd}}&\alg \ar '[rd] [rrd]
& & \\
=&&&&\circ& \bbC
\quad  =&&&\circ& \bbC \\
&&\alg^\op\ar[r]^{\sqrt{\sf Todd}^\op}&\alg^\op \ar '[ur] [urr]^(.4){\Xi}& &  
&\alg^\op\ar[r]^{\sqrt{\sf Todd}^\op}&\alg^\op \ar '[ur] [urr]^{\Xi}& & .} 
\]
Note that the symmetry of $\Xi$ is used here in a crucial way. This
computation shows that
\[ 
(\ch(p)\otimes_\alg \sqrt{\sf Todd}(\alg), \ch(p)\otimes_\alg
\sqrt{\sf Todd}(\alg)) =
(\ch(p)\otimes \ch(q))\otimes_{\alg\otimes\alg^o}({\sf Todd}(\alg)\otimes_\alg \Xi)
\]
Since, by definition, ${\sf Todd}(\alg)= \Xi^\vee \otimes_{\alg^o}
\ch(\Delta)$, a similar computation shows that  ${\sf
  Todd}(\alg)\otimes_\alg \Xi = (\Xi^\vee\otimes_\alg
\Xi)\otimes_{\alg^o} \ch(\Delta) = \ch(\Delta)$ and so 
\[
\begin{split}
(\ch(p)\otimes_\alg \sqrt{\sf Todd}(\alg), \ch(p)\otimes_\alg 
\sqrt{\sf Todd}(\alg))& = 
 (\ch(p)\otimes\ch(q))\otimes_{\alg\otimes\alg^o}\ch(\Delta) \\
& = \ch((p\otimes q)\otimes_{\alg\otimes\alg^o}\Delta) \\
& = \ch((p,q))
\end{split}
\]
Finally, the Chern character $\bbz= \KK(\bbc, \bbc) \rightarrow
\HE(\bbc,\bbc) = \bbc$  is injective, which gives the desired result.
\end{proof}

\begin{corollary}[{\bf D-brane charge formula for noncommutative spacetimes}]
\label{thm:Dbrane-charge}
Suppose that  the noncommutative spacetime $\alg$ satisfies the hypotheses
of Theorem \ref{thm:chNCisometry1} above. Then there is a noncommutative
analogue of the well-known formula {\rm (1.1)} in  \cite{MM} 
for the charge associated to a D-brane $\balg$ in a noncommutative 
spacetime $\alg$ with given weakly $\K$-oriented morphism $f\colon \alg 
\to \balg$ and Chan-Paton bundle $\xi \in \K_\bullet(\balg)$,
\beq\label{eqn:Dbrane-charge}
{\sf Q}_\xi = \ch(f_!(\xi))  \otimes_\alg \sqrt{{\sf Todd}(\alg)} \,.
\eeq
\end{corollary}

This is still not quite the most general situation. Corollary
\ref{thm:Dbrane-charge} deals with charges coming from a (weakly)
$\K$-oriented morphism $f\colon \alg \to \balg$, when a Chan-Paton
bundle, i.e., a $\K$-theory class, is given on $\balg$.  This is
the obvious translation of the situation coming from a flat
D-brane in the commutative case, but one can imagine more general
noncommutative D-branes where the algebra $\balg$ is missing, i.e.,
one simply has a Fredholm module for $\alg$ representing a class
in $\K^\bullet(\alg)$.  (In Corollary \ref{thm:Dbrane-charge},
the associated class in $\K^\bullet(\alg)$ is $\PD_\alg(f_!(\xi))
= f^*(\PD_\balg(\xi))$.) The final version of the charge formula
is the following:

\begin{proposition}[{\bf D-brane charge formula, dual version}]
\label{thm:Dbrane-charge1}
Suppose that  the noncommutative spacetime $\alg$ satisfies the hypotheses
of Theorem \ref{thm:chNCisometry1} above. Then there is a noncommutative
analogue of formula \eqref{dualchargeexpl} of Proposition 
\ref{propdualRR} above for the dual
charge associated to a D-brane in the noncommutative
spacetime $\alg$ represented by a class $\mu\in \K^\bullet(\alg)$:
\beq\label{eqn:Dbrane-charge1}
{\sf Q}_\mu= \sqrt{{\sf Todd}(\alg)}^{\,-1} \otimes_\alg \ch(\mu)\,.
\eeq
This formula satisfies the isometry rule:
\beq\label{eqn:Dbrane-charge2}
\Xi^\vee  \otimes_{\alg \otimes \alg^\op} \bigl(
{\sf Q}_\mu \times {\sf Q}_\nu^\op \bigr)
~=~
\Delta^\vee \otimes_{\alg \otimes \alg^\op} (\mu \times \nu^\op)\,.
\eeq
\end{proposition}
\begin{proof}
We need to check \eqref{eqn:Dbrane-charge2}. By multiplicativity of
the Chern character, the right-hand side is equal to
$\ch(\Delta^\vee) \otimes_{\alg \otimes \alg^\op} \bigl( \ch(\mu)
\times \ch(\nu)^\op \bigr)$. The left-hand side is
\[
\begin{aligned}
& \Xi^\vee  \otimes_{\alg \otimes \alg^\op} \bigl(
{\sf Q}_\mu \times {\sf Q}_\nu^\op \bigr)\\
& = \Xi^\vee  \otimes_{\alg \otimes \alg^\op} \bigl(
 \sqrt{{\sf Todd}(\alg)}^{\,-1} \times  \sqrt{{\sf
     Todd}(\alg)^\op}^{\,-1} \bigr) 
 \otimes_{\alg \otimes \alg^\op} \bigl(\ch(\mu) \times \ch(\nu)^\op \bigr)\\
& = \text{ (by symmetry of }\Xi^\vee \text{ as in the proof of
Theorem \ref{thm:chNCisometry1}})\\
&\phantom{=}\  \Xi^\vee  \otimes_{\alg \otimes \alg^\op}  \bigl(
{\sf Todd}(\alg)^{\,-1} \times 1_{\alg^\op} \bigr) 
\otimes_{\alg \otimes \alg^\op} \bigl(\ch(\mu) \times \ch(\nu)^\op \bigr)\,.
\end{aligned}
\]
But $\ch(\Delta^\vee) =  \Xi^\vee  \otimes_{\alg} {\sf Todd}(\alg)^{\,-1}$,
since (by the diagram calculus)
\[
 \Xi^\vee  \otimes_{\alg} {\sf Todd}(\alg)^{\,-1} = 
 \Xi^\vee  \otimes_{\alg} \bigl(\ch(\Delta^\vee) \otimes_{\alg^\op} \Xi\bigr)
 =\ch(\Delta^\vee) \otimes_{\alg^\op} \bigl( \Xi^\vee  \otimes_{\alg} \Xi\bigr)
 = \ch(\Delta^\vee)\,.
\]
\end{proof}

\begin{remark}
Although our noncommutative formulas for D-brane charges have been derived
under the assumption that $\alg$ is a strong PD algebra, they hold more
generally for any algebra $\alg$ belonging to the class
$\underline{\mathfrak{D}}$ introduced in Section 7.1. This allows us to
write down charge formulas in a variety of very general situations. For
instance, one can in this way obtain a bilinear pairing on twisted
$\K$-theory, $\K^\bullet(X,H)\times\K^\bullet(X,-H)\to\zed$, and
hence an isometric pairing between twisted $\K$-theory and twisted
cohomology, recovering the charge formula (1.13) of [8] for twisted
D-branes.
\end{remark}

\setcounter{section}{0}
\setcounter{theorem}{0}
\setcounter{figure}{0}
\addcontentsline{toc}{section}{Appendix A. The Kasparov Product}
\appendix{The Kasparov Product\label{KaspApp}}

In this appendix, we will summarize the main properties of the
intersection product.  
If $\alg$ is a separable algebra then the exterior (or cup) product exists
and defines a bilinear pairing,  \cite[Thm 2.11]{Kas88}: 
\beq
\KK_i(\alg, \balg_1)\otimes_{\balg_1} \KK_j(\balg_1,\balg_2)
\rightarrow \KK_{i+j}(\alg,\balg_2). 
\eeq
In  \cite[Def.~2.12]{Kas88}, Kasparov also defines the intersection
product (which he calls the cap-cup product) 
$$
\KK_i(\alg_1, \balg_1\otimes \dalg) \otimes_\dalg\KK_j(\dalg\otimes
\alg_2, \balg_2) \rightarrow  
\KK_{i+j}(\alg_1\otimes \alg_2, \balg_1\otimes \balg_2)
$$
by the formula
$$
x_1\otimes _\dalg x_2 = (x_1\otimes 1_{\alg_2})
\otimes_{\balg_1\otimes \dalg\otimes \alg_2} 
(x_2\otimes 1_{\balg_1}).
$$
The exterior (or cup) product is obtained when $\dalg=\bbC$. 
This exterior product has the following properties  \cite[Thm.~2.14]{Kas88}. 
\begin{theorem}\label{properties-Kasparov-product}
Let $\alg_1$ and $\alg_2$ be separable algebras. Then the intersection
(cup-cap) product exists and is:  
\begin{enumerate}
\item bilinear; 
\item contravariant in $\alg_1$ and $\alg_2$; 
\item covariant in $\balg_1$ and $\balg_2$; 
\item functorial in $\dalg$: for any morphism $f: \dalg_1\rightarrow
  \calg$ one has
$$
f(x_1) \otimes _{\dalg_2} x_2 = x_1\otimes _{\dalg_1}f(x_2) \ ;
$$
\item associative: for any $x_1\in \KK_i(\alg_1, \balg_1\otimes \dalg_1)$, 
$x_2\in \KK_j(\dalg_1\otimes \alg_2, \balg_2\otimes \dalg_2)$ and 
$x_3 \in \KK_k(\dalg_2\otimes \alg_3, \balg_3)$, where $\alg_1$,
  $\alg_2$, $\alg_3$ and $\dalg_1$ are  
assumed separable, the following formula holds: 
$$
(x_1 \otimes _{\dalg_1} x_2) \otimes _{\dalg_2} x_3 =
x_1\otimes_{\dalg_1} (x_2\otimes_{\dalg_2} x_3); 
$$
\item For any $x_1\in \KK_i(\alg_1, \balg_1\otimes \dalg_1\otimes \dalg)$, 
$x_2\in \KK_j(\dalg\otimes \dalg_2\otimes \alg_2, \balg_2)$, 
where $\alg_1, \alg_2, \dalg_2$ are separable and $\dalg_1$ is
$\sigma$-unital, the following formula 
holds:
$$
x_1\otimes_\dalg x_2 = (x_1\otimes 1_{\dalg_2}) \otimes_{\dalg_1\otimes
  \dalg\otimes \dalg_2}(1_{\dalg_1} \otimes x_2); 
$$
\item For $x_1\in \KK_i(\alg_1, \balg_1\otimes \dalg)$, $x_2\in
  \KK_j(\dalg\otimes \alg_2, \balg_2)$, for separable algebras
  $\alg_1, \alg_2$ and $\dalg$, the following formula holds: 
$$
(x_1\otimes_{\dalg} x_2)\otimes_{\bbc} 1_{\dalg_1}= (x_1\otimes
  1_{\dalg_1})\otimes _{\dalg\otimes \dalg_1}(x_2\otimes 1_{\dalg_1})
  \ ;
$$
\item the cup product is commutative \textup{(}over $\bbc$\textup{)}: 
$$
x_1\otimes_{\bbc} x_2 = x_2\otimes_{\bbc} x_1;
$$
and
\item the element $1_\bbc \in \KK_0(\bbc, \bbc)$ is a unit for this product: 
$$
1_\bbc\otimes _\bbc x = x\otimes_\bbc 1_\bbc = x
$$
for all $x\in \KK_i(\alg,\balg)$, where $\alg$ is assumed to be separable. 
\end{enumerate}
\end{theorem}

\addcontentsline{toc}{section}{Appendix B. A Diagram Calculus for the Kasparov Product}
\appendix{A Diagram Calculus for the Kasparov Product\label{KaspDiag}}

Keeping track of Kasparov products and
the associativity formulae in the general case described
above in Appendix A can be quite complicated.  In this appendix
we describe a pictorial calculus for keeping track of these things,
which one of us (J.R.) has often found useful as a guide to calculations.
In this appendix we will not write degree labels explicitly on $\KK$-groups
for the sake of notational convenience --- in the most important case,
all elements lie in $\KK_0$ anyway.

The idea is to represent an element of a $\KK$ group by a diagram (which we
read from left to right), with one ``input'' for each tensor factor in
the first argument of $\KK$, and one ``output'' for each tensor factor in
the second argument of $\KK$.  For convenience, we can also
add arrowheads pointing toward the outputs.
Thus, for example, an element of
$\KK(\balg\otimes \alg, \calg\otimes \dalg)$ would be represented by a
diagram like the one in Figure \ref{fig:KK(BA,CD)}.
\begin{figure}[ht]
\[
\xymatrix{\balg \ar '[dr] '[drr] [rrr]& & &\calg\\
&\circ &\circ &\\
\alg \ar '[ur] '[urr] [rrr]& & &\dalg}
\]
\caption{Diagram representing an element of $\KK(\balg \otimes \alg,
  \calg\otimes \dalg)$} 
\label{fig:KK(BA,CD)}
\end{figure}
Note that an element of $\KK(\alg \otimes \balg, \calg\otimes \dalg)$
would be represented 
by an almost identical diagram, having the two input terminals
switched.  The basic rule is that permutation of the input or output
terminals may involve at most the switch of a sign.

The Kasparov product corresponds to concatenation of diagrams, except
that one is only allowed to attach an input to a matching output.
For example, in Figure \ref{fig:KK(BA,CD)}, there are input terminals
corresponding to both $\balg$ and $\alg$, so one can take the product over
a Kasparov class having a $\balg$ or $\alg$ as an output
terminal.  For example, a class in $\KK(\cale,\alg)$ would be represented
by a diagram like Figure \ref{fig:KK(E,A)}, and we can concatenate the
diagrams
as shown in Figure \ref{fig:KK(E,A)KK(BA,CD)} to obtain the
product (over $\alg$) in $\KK(\balg \otimes \cale, \calg
\otimes\dalg)$ shown in Figure \ref{fig:KK(BE,CD)}.

\begin{figure}[ht]
\[
\xymatrix{\cale \ar[r] & \alg}
\]
\caption{Diagram representing an element of $\KK(\cale, \alg)$}
\label{fig:KK(E,A)}
\end{figure}

\begin{figure}[htb]
\[
\xymatrix{&\balg \ar '[dr] '[drr] [rrr]& & &\calg\\
&& \circ & \circ &\\
\cale \ar[r] &\alg \ar '[ur] '[urr] [rrr]&& &\dalg}
\]
\caption{Diagram representing the intersection product,  
$\otimes_\alg\colon
\KK(\cale, \alg)\otimes \KK(\balg\otimes\alg, \calg \otimes\dalg)\to
\KK(\balg\otimes \cale, \calg\otimes\dalg)$} 
\label{fig:KK(E,A)KK(BA,CD)}
\end{figure}

\begin{figure}[htb]
\[
\xymatrix{\balg \ar '[dr] '[drr] [rrr]& & &\calg \\
&\circ &\circ & \\
\cale \ar '[ur] '[urr] [rrr] & & &\dalg}
\]
\caption{Diagram representing an element of $\KK(\balg \otimes \cale,
  \calg \otimes \dalg)$}
\label{fig:KK(BE,CD)}
\end{figure}

The associativity of the Kasparov product corresponds to the principle
that if one has multiple concatenations to do, \emph{the concatenations
can be done in any order}, except perhaps for keeping
track of signs.  For example, if
\[
x\in \KK(\balg \otimes \alg, \calg), \quad y\in \KK(\dalg, \alg), \quad
\text{and } z\in \KK(\ealg, \balg),
\]
then the associativity of the product gives
\[
z\otimes_\balg ( y \otimes_\alg x) = \pm y \otimes_\alg (z\otimes_\balg x),
\]
even though when written this way, it seems to be somewhat
counter-intuitive. But one can ``prove'' this graphically with the
picture in Figure \ref{fig:KKassoc}.

\begin{figure}[htb]
\[
\xymatrix{\ealg \ar[r]^z &\balg \ar '[dr] [drr]^x & & \\
&&\circ & \calg\\
\dalg \ar[r]^y & \alg \ar '[ur] [urr] &&}
\] 
\caption{Diagram showing that
$z\otimes_\balg ( y \otimes_\alg x) = \pm y \otimes_\alg (z\otimes_\balg x)$}
\label{fig:KKassoc}
\end{figure}

Of course, a picture by itself is not a rigorous proof, but it can
be made into one as follows. Here $\times$ is used to denote the
``exterior'' Kasparov product, and for simplicity we
assume that all elements lie in $\KK_0$, so that we
don't have to worry about sign changes (which all have to do with
conventions about orientation of the Bott element).
On the one hand, we have
\beq
\label{eq:KaspprodA}
\begin{aligned}
z\otimes_\balg ( y \otimes_\alg x) &:= (z\times 1_\dalg)
\otimes_{\balg\otimes \dalg} 
( y \otimes_\alg x)\\
&= (z\times 1_\dalg) \otimes_{\balg\otimes \dalg} \bigl(
(1_\balg\times y) \otimes_{\balg\otimes \alg} x \bigr)\\
&= \bigl[ (z\times 1_\dalg) \otimes_{\balg\otimes \dalg}
  (1_\balg\times y) \bigr] 
\otimes_{\balg\otimes \alg} x .
\end{aligned}
\eeq
But on the other hand we have
\beq
\label{eq:KaspprodB}
\begin{aligned}
y \otimes_\alg (z\otimes_\balg x) &:= (1_\ealg\times y)
\otimes_{\ealg\otimes \alg} 
(z\otimes_\balg x) \\
&= (1_\ealg \times y) \otimes_{\ealg \otimes \alg}
\bigl( (z\times 1_\alg) \otimes_{\balg\otimes \alg} x \bigr)\\
&= \bigl[ (1_\ealg\times y) \otimes_{\ealg \otimes \alg}(z\times 1_\alg )
\bigr] \otimes_{\balg \otimes \alg} x .
\end{aligned}
\eeq
So to prove the associativity formula, it suffices to observe that
\beq
\label{eq:KaspprodC}
(z\times 1_\dalg) \otimes_{\balg\otimes \dalg} (1_\balg\times y)
= z \times y = (1_\ealg\times y) \otimes_{\ealg \otimes \alg}(z\times 1_\alg ).
\eeq

We should mention incidentally that essentially everything we said about
the Kasparov  product applies equally well to products in bivariant
cyclic homology, whose formal properties are exactly the same.


\end{document}